\pdfoutput=1

\documentclass[cernpreprint,texlive=2016,UKenglish]{latex/atlasdoc}

\usepackage[subfigure=true,block=none,biblatex=true,backend=bibtex]{latex/atlaspackage}
\usepackage{latex/atlasphysics}

\usepackage{amssymb}
\nolinenumbers
\usepackage[centering,scale=0.75]{geometry}
\usepackage{graphicx}
\usepackage{float}
\floatstyle{plaintop}
\restylefloat{table}
\usepackage[tableposition=top]{caption}
\usepackage{adjustbox,lipsum}
\usepackage{rotating}
\usepackage{pdflscape}

\usepackage{latex/atlascover}

\usepackage{hyperref}
\usepackage{latex/atlascontribute}

\usepackage{latex/atlasbiblatex}
\usepackage{mathrsfs}
\usepackage{xspace}
\usepackage{float}
\usepackage{color, colortbl}
\definecolor{Gray}{gray}{0.9}

\usepackage{TOPQ-2017-01-PAPER-defs}

\AtlasTitle{Measurements of differential cross sections of top quark pair production in association with jets in \boldmath${pp}$ collisions at \boldmath${\sqrt{s}=13\,\TeV}$ using the ATLAS detector}

\author{The ATLAS Collaboration}

\AtlasAbstract{
Measurements of differential cross sections of top quark pair production in association with jets by the ATLAS experiment at the LHC are presented. 
The measurements are performed as functions of the top quark transverse momentum, the transverse momentum of the top quark-antitop quark system and the out-of-plane transverse momentum using data from $pp$ collisions at $\sqrt{s}=13\,\TeV$ collected by the ATLAS detector at the LHC in 2015 and corresponding to an integrated luminosity of \lumitot. The top quark pair events are selected in the lepton (electron or muon) + jets channel. The measured cross sections, which are compared to several predictions, allow a detailed study of top quark production. 
} 
\AtlasRefCode{TOPQ-2017-01}
\PreprintIdNumber{CERN-EP-2017-227}

\author{The ATLAS Collaboration}

\AtlasJournalRef{JHEP 10 (2018) 159}
\AtlasDOI{DOI:10.1007/JHEP10(2018)159}

\AtlasJournal{JHEP}
\date{\today}

\begin{document}

\title{Measurements of differential cross sections of top quark pair production in association with jets in \boldmath${pp}$ collisions at \boldmath${\sqrt{s}=13\,\TeV}$ using the ATLAS detector}

\maketitle
\author{The ATLAS Collaboration}

\section{Introduction}\label{sec:Introduction}
The large number of top quark pair ($\ttbar{}$) events produced at the Large Hadron Collider (LHC) allows detailed studies of the characteristics of $\ttbar{}$ production as a function of different kinematic variables. In this paper, the data collected by the ATLAS experiment in 2015 are used to measure differential cross sections for $\ttbar{}$ production in association with jets. The measurement of differential cross sections in different bins of jet multiplicity provides a better understanding of the effect of gluon radiation on $\ttbar{}$ kinematic variables than differential cross sections inclusive in the number of jets previously published by the ATLAS Collaboration at $\sqrt{s}=13\,\TeV$~\cite{TOPQ-2016-01}. 

Since the top quark decays almost always to a $W$ boson and a $b$-quark, the decay of a top quark pair produces six particles in the final state, whose identity depends on the decays of the intermediate $W$ bosons. The channel considered in this analysis is characterised by the leptonic decay of one $W$ boson and the hadronic decay of the other $W$ boson; this is commonly referred to as semileptonic decay mode or $\ell$+jets channel. The final-state configuration contains one electron or muon, one neutrino giving rise to missing transverse momentum ($E^\mathrm{miss}_\mathrm{T}$) and four jets, two of which originate from $b$-quarks. Events may include additional jets from gluon radiation off initial- or final-state quarks. To study the dependence of this emission on the observables, three configurations are defined depending on the number of additional jets produced within the detector acceptance in association with the top quark pair: the "4-jet exclusive configuration" (no additional jets); the "5-jet exclusive configuration" (only one additional jet); and the "6-jet inclusive configuration" (two or more additional jets). The latter configuration is of particular interest since it provides a similar phase space to the one used by measurements such as Higgs boson production in association with two top quarks and searches with high jet multiplicity. 

The three configurations with increasing number of additional jets are expected to provide a better understanding of the effect of gluon radiation on the kinematic variables of top quark pair production. ATLAS already published differential cross section measurements as a function of the number of additional jets~\cite{TOPQ-2011-21,TOPQ-2015-04,TOPQ-2015-17} and of several kinematic variables ~\cite{TOPQ-2011-07,TOPQ-2012-08,TOPQ-2013-07,TOPQ-2015-06,TOPQ-2016-01}. The results presented in this paper combine the two types of measurements to provide additional information about top quark production and explore the effect of the gluon radiation on $\ttbar{}$ kinematic variables. The CMS Collaboration published similar measurements~\cite{CMS-TOP-11-013, CMS-TOP-12-028, CMS-TOP-12-041, CMS-TOP-16-008}.

The observables studied here are the transverse momentum ($\pt$)\unskip\footnote{ATLAS uses a right-handed coordinate system with its origin at the nominal interaction point (IP) in the centre of the detector and the $z$-axis along the beam pipe. The $x$-axis points from the IP to the centre of the LHC ring, and the $y$-axis points upward. Cylindrical coordinates ($r$,$\phi$) are used in the transverse plane, $\phi$ being the azimuthal angle around the beam pipe. The pseudorapidity is defined in terms of the polar angle $\theta$ as $\eta = - \ln \tan(\theta/2)$ and the angular separation between particles is defined as $\Delta R = \sqrt{(\Delta \phi)^2 + (\Delta \eta)^2}$. The transverse momentum is the projection of the momentum on the transverse plane.} of the top quark-antitop quark system (\ptttbar) and the absolute value of the out-of-plane momentum (|$\Poutttbar$|), defined as the projection of the top quark three-momentum onto the direction perpendicular to a plane defined by the other top quark and the beam axis ($\hat{z}$) in the laboratory frame \cite{PhysRevLett.81.2642}:

\begin{equation*}
\absPoutttbar  = \left | \vec{p}^{~t, {\mathrm{had}}} \cdot \frac{\vec{p}^{~t,{\mathrm{lep}}} \times \hat{z}}{|\vec{p}^{~t,{\mathrm{lep}}}\times \hat{z}|} \right | \,,\\
\end{equation*}

where $\vec{p}^{~t,{\mathrm{lep}}}$ and $\vec{p}^{~t,{\mathrm{had}}}$ are the momenta of the semileptonically and hadronically decaying top quarks, respectively. This observable is complementary to \ptttbar{} since \Poutttbar{} is expected to be more sensitive to the direction of gluon radiation; for example the emission of a low \pt{} jet at a large angle with respect to the plane defined by the two top quarks is expected to be better measured with \Poutttbar{} than \ptttbar{}.
In addition, the differential cross section as a function of the transverse momentum of the hadronic top quark (\ptthad{}) is measured. In previous publications~\cite{TOPQ-2015-06,TOPQ-2016-01}, differences between the data and the predictions by several Standard Model Monte Carlo (MC) event generators were observed. By measuring the differential cross section of this observable in different jet multiplicities it is possible to identify the regions of phase space in which the discrepancy is largest. The measured differential cross sections as functions of these three observables are compared to predictions from several MC event generators, namely \PowhegBox~\cite{Frixione:2007vw}, \mgamcatnlo{}~\cite{ttVSim} and \Sherpa{}~\cite{Gleisberg:2008ta}.

\section{ATLAS detector}\label{sec:Detector}

ATLAS is a multipurpose detector \cite{PERF-2007-01} that provides nearly full solid angle coverage around the interaction point.
Charged-particle trajectories with pseudorapidity $|\eta| <2.5$ are reconstructed in the inner detector, which comprises a silicon pixel detector, a silicon microstrip detector and a~transition radiation tracker. The innermost pixel layer, the insertable B-layer~\cite{ATLAS-TDR-19}, was added before the start of 13 \TeV{} LHC operation at an average~radius of 33~mm around a~new, thinner beam pipe. The inner detector is embedded in a superconducting solenoid generating a~2~T axial magnetic field, allowing precise measurements of charged-particle momenta. Sampling calorimeters with several different designs span the pseudorapidity range up to $|\eta| = 4.9$. High-granularity liquid argon (LAr) electromagnetic (EM) calorimeters are used up to $|\eta| = 3.2$. Hadronic calorimeters based on scintillator-tile
active material cover $|\eta| < 1.7$ while LAr technology is used for hadronic calorimetry in the region $1.5 < |\eta| < 4.9$. The calorimeters are surrounded by a~muon spectrometer within a~magnetic field provided by air-core toroid magnets with a~bending integral of about 2.5~Tm in the barrel and up to 6~Tm in the end-caps.
Three stations of precision drift tubes and cathode-strip chambers provide 
an accurate measurement of the muon track curvature in the region  $|\eta| < 2.7$. 
Resistive-plate and thin-gap chambers provide muon triggering capability up to $|\eta| = 2.4$.

Data are selected from inclusive $pp$ interactions using a~two-level trigger system~\cite{TRIG-2016-01}. A~hardware-based trigger uses custom-made hardware and coarser-granularity detector data to initially reduce the trigger rate to approximately $75\,$kHz from the original 40 MHz LHC bunch crossing rate. 
Next, a software-based high-level trigger, which has access to full detector granularity, is applied to further reduce the event rate to $1\,$kHz.

\section{Data and simulation} \label{sec:DataSimSamples}
The differential cross sections are measured using a data set collected during the 2015 LHC $pp$ run at $\rts =13$~\TeV{} and with 25 ns bunch spacing. The average number of $pp$ interactions per bunch crossing ranged from approximately 5 to 25, with a~mean of 14. 
After applying data-quality assessment criteria based on beam, detector and data-taking quality, the available data correspond to a~total integrated luminosity of $3.2$~\ifb. 

The data were collected using a combination of multiple single-muon and single-electron triggers. For each lepton type, multiple trigger conditions are combined to maintain good efficiency in the full momentum range, while controlling the trigger rate.  For electrons, the \pt{} thresholds are 24~\GeV{}, 60~\GeV{} and 120~\GeV{}, while for muons the thresholds are 20~\GeV{} and 50~\GeV{}. Isolation requirements are applied to the triggers with the lowest \pt{} thresholds.

The signal and background processes are modelled with various MC event generators described below and summarised in Table~\ref{tab:MC}. Multiple overlaid $pp$ collisions were simulated with the soft QCD processes of \PythiaEightP{}~\cite{Sjostrand:2007gs} using parameter values from tune A2~\cite{ATL-PHYS-PUB-2012-003} and the MSTW2008LO~\cite{Martin:2009iq} set of parton distribution functions (PDFs). The EvtGen v1.2.0 program~\cite{EvtGen} was used to simulate the decay of bottom and charm hadrons, except for the \Sherpa event generator. The detector response was simulated~\cite{SOFT-2010-01} in \textsc{GEANT\textrm4}~\cite{Geant4}.
 
\subsection{Signal simulation samples} \label{sec:SignalMC}
In this section the MC samples used for the generation of \ttb{} events are described for the nominal sample, the alternative samples used to estimate systematic uncertainties and the other samples used in the post-unfolding comparison. The top quark mass ($m_{t}$) was set to 172.5 \GeV{} in all MC event generators. 

For the generation of \ttb{} events, the \PowhegBox v2 event generator~\cite{Frixione:2007vw,Nason:2004rx,POWHEGBOX}, from now on called \Powheg, with the CT10 PDF set~\cite{CT10} was used for the matrix element calculations. The factorisation and hadronisation scales are set to $\sqrt{m_{t}^{2} + p^{2}_{\mathrm{T,t}}}$ where $m_{t}$ and $p_{\mathrm{T,t}}$ are the top quark mass and the transverse momentum of the top quark, respectively, evaluated for the underlying Born configuration. Events in which both $W$ bosons decay hadronically were not included. For this process, the top quarks were decayed using MadSpin~\cite{Artoisenet:2012st} to preserve all spin correlations, while parton shower, hadronisation, and the underlying event were simulated using \PythiaSixP{}~\cite{Sjostrand:2006za} with the CTEQ6L1 PDF set~\cite{CTEQ6L1} and the Perugia2012 tune~\cite{perugia}. 
The $h_{\mathrm{damp}}$ parameter, which controls the $\pT$ of the first gluon or quark emission beyond the Born configuration in \Powheg, was set to the mass of the top quark~\cite{ATL-PHYS-PUB-2016-004}. The main effect of this parameter is to regulate the high-$\pT$ emission against which the \ttbar{} system recoils. Signal \ttb{} events generated with those settings are referred to as the nominal signal sample.

To estimate the effect of the parton shower algorithm, a \Powheg{}+\herwigpp{} sample was generated with the same \Powheg{} settings as for the nominal sample. The parton shower, hadronisation and underlying event simulation were produced with \herwigpp~\cite{Bahr:2008pv} (version 2.7.1) using the UE-EE-5 tune~\cite{ATL-PHYS-PUB-2014-021} and the CTEQ6L1 PDF set. 

The impact of the matrix element (ME) event generator choice is evaluated using events generated with \mgamcatnlo{}+\herwigpp{} with the UE-EE-5 tune. The events were generated with version 2.1.1 of \mgamcatnlo{}. \NLO{} matrix elements and the CT10 PDF set were used for the \ttbar{} hard-scattering process. These events were passed through a~fast simulation using a~parametrisation of the performance of the ATLAS electromagnetic and hadronic calorimeters~\cite{ATL-SOFT-PUB-2014-01} and full simulation of the response in the inner detector and muon spectrometer.

The effects of different levels of gluon radiation are evaluated using two samples with different factorisation and hadronisation scales relative to the nominal sample, as well as a different $h_{\mathrm{damp}}$ parameter value. 
Specifically, in one sample the factorisation and hadronisation scales were reduced by a~factor of 0.5, the $h_{\mathrm{damp}}$ parameter was increased to $2 m_{t}$ and the `radHi' tune variation from the Perugia2012 tune set is used. 
In the second sample, the factorisation and hadronisation scales were increased by a~factor of two, the $h_{\mathrm{damp}}$ parameter was unchanged and the `radLo' tune variation from the Perugia2012 tune set was used.

The measured differential cross sections are compared to several additional \ttbar{} MC samples~\cite{ATL-PHYS-PUB-2016-004,ATL-PHYS-PUB-2016-020,ATL-PHYS-PUB-2017-007}. 
\begin{itemize}
\item Two~\mgamcatnlo{}+\PythiaEight{} samples having two different hard-scattering scales, $H_{\mathrm{T}}/2$~\footnote{$H_{\mathrm{T}}$ is defined as the scalar sum of the transverse momenta of the two top quarks.} and $\sqrt{m_{t}^{2} + p^{2}_{\mathrm{T,t}}}$ and using the same A14 tune.
\item Two~\Powheg{}+\PythiaEight{} samples simulated with different values of the $h_{\mathrm{damp}}$ parameter ($h_{\mathrm{damp}} = m_{t}$ and $h_{\mathrm{damp}} =1.5m_{t}$) also using the A14 tune. 
\item Two additional~\Powheg{}+\PythiaEight{} samples with alternative radiation settings: the factorisation and renormalisation scales are coherently varied by a factor of 2.0 (0.5) and the A14 tune `Var3c Down' (`Var3c Up') variation is used.
\item A~\Powheg{}+\HerwigSeven{} sample generated with the $h_{\mathrm{damp}}$ parameter set to $1.5m_{t}$ and using the H7-UE-MMHT tune which use the NNPDF3.0 PDF~\cite{Ball:2014uwa} for the ME.
\item A $\textsc{Sherpa 2.2.1}$ sample in which events were generated with a $\ttbar{}$ matrix element and up to one additional parton simulated at NLO and two, three and four partons at LO. The CT10 PDF set was used. 
\end{itemize}

The $t\bar{t}$ samples are normalised using $\sigma_{t\bar{t}} = 832^{+20}_{-29}(\mathrm{scale})~\pm 35~(\mathrm{PDF})$~pb as calculated with the Top++2.0 program to next-to-next-to-leading order (NNLO) in perturbative QCD, including soft-gluon resummation to next-to-next-to-leading-log order (NNLL) (see Ref.~\cite{Czakon:2011xx} and references therein), and assuming a top quark mass $m_{t}$ = 172.5 \GeV{}. The first uncertainty comes from the independent variation of the factorisation and renormalisation scales, $\mu_{\mathrm{F}}$ and $\mu_{\mathrm{R}}$, while the second one is associated with variations in the PDF and $\alpha_{\mathrm{S}}$, following the PDF4LHC prescription with the MSTW2008 68\% CL NNLO, CT10 NNLO and NNPDF2.3 5f FFN PDF sets see Refs.~\cite{Martin:2009bu,CT10,PDF4LHC,NNPDF}. 

\subsection{Background simulation samples} \label{sec:BackgroundMC}
Several processes can produce the same final state as the $\ttbar{}$ semileptonic channel. The events produced by these backgrounds need to be estimated and subtracted from data to calculate the top quark pair cross sections. They are fully estimated using MC simulation with the exception of the $W$+jets background, for which data-driven techniques complement the MC simulation prediction. The processes considered are single-top quark production, $W$+jets and $Z$+jets production, diboson final states and top quark pairs produced in association with weak bosons ($t\bar{t}$ + $W/Z/WW$, denoted by $t\bar{t}V$).  

The simulation of single-top quark events from $Wt$ and $s$-channel production was performed using the configuration described above for the nominal $t\bar{t}$ sample. The overlap between the $Wt$ and \ttb{}~samples was handled using the diagram-removal scheme \cite{SingleTopWt}. Electroweak $t$-channel single-top quark events were generated using the \PowhegBox v1 event generator. The single-top quark cross sections for the $t$- and $s$-channels are normalised using their \NLO{} predictions, while for the $Wt$ channel it is normalised using its \NLO+\NNLL{} prediction~\cite{Kidonakis:2011wy,Aliev:2010zk,Kant:2014oha}. 

Inclusive samples containing single $W$ or $Z$ bosons in association with jets were simulated using the \Sherpa 2.1.1 event generator~\cite{Gleisberg:2008ta}. Matrix elements were calculated with up to two partons at \NLO{} and four partons at leading-order (\LO)~using the Comix~\cite{Gleisberg:2008fv} and OpenLoop~\cite{Cascioli:2011va} matrix element event generators and merged with the \Sherpa parton shower~\cite{Schumann:2007mg} using the ME+PS@NLO prescription~\cite{Hoeche:2012yf}. The CT10 PDF sets were used in conjunction with dedicated parton shower tuning developed by the authors of \Sherpa{}. The $Z+$jets events are normalised using the \NNLO~cross sections~\cite{ATL-PHYS-PUB-2016-003} while the normalisation for the $W+$jets events is obtained with a data-driven method described in Section~\ref{sec:BackgroundDetermination}.

Diboson processes, with one of the bosons decaying hadronically and the other leptonically, were simulated using the \Sherpa 2.1.1 event generator~\cite{Gleisberg:2008ta,ATL-PHYS-PUB-2016-002}. They are calculated for up to one ($ZZ$) or zero ($WW$, $WZ$) additional partons at \NLO{} and up to three additional partons at LO using the Comix and OpenLoops matrix element event generators and merged with the \Sherpa parton shower using the ME+PS@NLO prescription. The CT10 PDF sets were used in conjunction with dedicated parton shower tuning developed by the authors of \Sherpa. The event generator cross sections, which are already at the NLO accuracy, are used in this case.

The $\ttbar$$V$ events were simulated using the \mgamcatnlo{} event generator at LO interfaced to the \Pythia 8.186 parton shower model~\cite{ATL-PHYS-PUB-2016-005}. The matrix elements were simulated with up to two ($t\bar{t}$ + $W$), one ($t\bar{t}$ + $Z$) or no ($t\bar{t}$ + $WW$) extra partons. The ATLAS underlying-event tune A14 was used together with the NNPDF2.3LO PDF sets. The events are normalised using their respective NLO cross sections~\cite{ttVSim}. 

\begin{table*}[t] 
\scriptsize
\centering
\begin{tabular}{|l|l|c|c|c|c|} \hline
Physics process                  & Generator             & PDF set for  & Parton shower   & Tune         & Cross section   \\
                                 &                       & hard process &                 &              & normalisation   \\ \hline
$\ttbar$ signal                  & \PowHegBox v2         & CT10         & \Pythia 6.428   & Perugia2012  & \NNLO+\NNLL     \\ 
$\ttbar$ PS syst.                & \PowHegBox v2         & CTEQ6L1      & \herwigpp 2.7.1 & UE-EE-5      & \NNLO+\NNLL     \\
$\ttbar$ ME syst.                & {\textsc MadGraph5}\_ & CT10         & \herwigpp 2.7.1 & UE-EE-5      & \NLO            \\
                                 & a{\textsc MC@NLO}     &              &                 &              &                 \\
$\ttbar$ rad. syst.              & \PowHegBox v2         & CT10         & \Pythia 6.428   & 'radHi/Lo'   & \NNLO+\NNLL     \\ 
Single top:  $t$-channel         & \PowHegBox v1         & CT10f4       & \Pythia 6.428   & Perugia2012  & \NLO            \\ 
Single top:  $s$-channel         & \PowHegBox v2         & CT10         & \Pythia 6.428   & Perugia2012  & \NLO            \\ 
Single top: $Wt$-channel         & \PowHegBox v2         & CT10         & \Pythia 6.428   & Perugia2012  & \NLO+\NNLL      \\ 
$\ttbar$+$W/Z/WW$                & {\textsc MadGraph5}\_ & NNPDF2.3LO   & \Pythia 8.186   & A14          & \NLO            \\ 
                                 & a{\textsc MC@NLO}     &              &                 &              &                 \\
$W(\to \ell \nu) $+ jets         & \Sherpa 2.1.1         & CT10         & \Sherpa         & \Sherpa      & \NNLO           \\ 
$Z(\to \ell {\bar \ell}) $+ jets & \Sherpa 2.1.1         & CT10         & \Sherpa         & \Sherpa      & \NNLO           \\ 
$WW, WZ, ZZ$                     & \Sherpa 2.1.1         & CT10         & \Sherpa         & \Sherpa      & \NLO            \\ \hline
\end{tabular}
\caption{Summary of MC samples, showing the event generator for the hard-scattering process, cross section normalisation precision, PDF choice, as well as the parton shower generator and the corresponding tune used in the analysis.}
\label{tab:MC}
\end{table*}

\section{Object reconstruction and event selection} \label{sec:EventReco}

The following sections describe the reconstruction- and particle-level objects used to characterise the final-state event topology and to define the fiducial phase space regions for the measurements. The reconstruction level is applied to data and MC samples.

\subsection{Detector-level object reconstruction}\label{sec:ObjectDef}

Primary vertices are formed from reconstructed tracks which are spatially compatible with the interaction region. The hard-scatter primary vertex is chosen to be the one with at least two associated tracks and the highest $\sum \pt^2$, where the sum extends over all tracks with $\pt > 0.4\,\mathrm{\GeV{}}$ matched to the vertex. 

Electron candidates are reconstructed by matching tracks in the inner detector to energy deposits in the EM calorimeter. They must satisfy a~``tight'' likelihood-based identification criterion based on shower shapes in the EM calorimeter, track quality and detection of transition radiation produced in the transition radiation tracker detector~\cite{ATL-PHYS-PUB-2011-006}. The reconstructed EM clusters are required to have a~transverse energy $\ET>$ 25~\GeV{} and a pseudorapidity $|\eta| < 2.47$, excluding the transition region between the barrel and the end-cap calorimeters ($1.37 < |\eta| < 1.52$). The associated track must have a~longitudinal impact parameter $|z_0\, {\sin} \theta|<0.5$~mm and a~transverse impact parameter significance $|d_0|/\sigma(d_0)<5$, where $d_0$ is measured with respect to the beam line. 
Isolation requirements based on calorimeter and tracking quantities are used to reduce the background from non-prompt and fake electrons~\cite{TOPQ-2015-09}. The isolation criteria are \pT{}- and $\eta$-dependent, and ensure an efficiency of 90\% for electrons with \pT{} of 25~\GeV{} and 99\% efficiency for electrons with $\pt{}$ of 60~\GeV{}. The identification, isolation and trigger efficiencies are measured using electrons from $Z$ boson decays~\cite{PERF-2016-01}. 

Muon candidates are identified by matching tracks in the muon spectrometer to tracks in the inner detector~\cite{PERF-2015-10}. The track $\pt$ is determined through a~global fit of the hits which takes into account the energy loss in the calorimeters. Muons are required to have \pt $>$ 25~\GeV{} and $|\eta|<2.5$.
To reduce the background from muons originating from heavy-flavour decays inside jets, muons are required to be isolated using track quality and isolation criteria similar to those applied to electrons. If a muon shares a track with an electron, it is likely to have undergone bremsstrahlung and hence the electron is not selected. Muon efficiencies are reconstruction and isolation efficiencies and for muon candidates with \pt $>$ 25~\GeV{} these efficiencies are of 99\% and are obtained using muons from $J$/$\psi$ and $Z$ decays.

Jets are reconstructed using the anti-$k_{t}$ algorithm~\cite{akt1} with radius parameter $R = 0.4$ as implemented in the {\textsc FastJet} package \cite{Fastjet}. Jet reconstruction in the calorimeter starts from topological clustering of individual calorimeter cell signals calibrated to be consistent with electromagnetic or hadronic cluster shapes using corrections determined in simulation and inferred from test-beam data~\cite{PERF-2014-07}. Jet four-momenta are then corrected for pile-up effects using the jet-area method \cite{Cacciari:2008gn}. To reduce the number of jets originating from pile-up, an additional selection criterion based on a~jet-vertex tagging technique is applied. The jet-vertex tagging is a~likelihood discriminant that combines information from several track-based variables~\cite{PERF-2014-03} and the criterion is only applied to jets with $\pT < 60\GeV{}$ and $|\eta|< 2.4$.
Jets are calibrated using an energy- and $\eta$-dependent simulation-based calibration scheme with \textit{in situ} corrections based on data~\cite{PERF-2012-01, PERF-2016-04}, and are accepted if they have \pT $>$ 25~\GeV{} and $|\eta| < 2.5$. 

For objects satisfying both the jet and lepton selection criteria, a procedure called "overlap removal" is applied to assign objects to a unique hypothesis. To prevent double-counting of electron energy deposits as jets, the jet closest to a reconstructed electron is discarded if they are \mbox{$\Delta R < 0.2$} apart.
Subsequently, to reduce the impact of non-prompt electrons, if an electron is \mbox{$\Delta R < 0.4$} from a jet, then that electron is removed. 
If a jet has fewer than three tracks and is \mbox{$\Delta R < 0.4$} from a~muon, the  jet is removed. Finally, the muon is removed if it is \mbox{$\Delta R < 0.4$} from a jet with at least three tracks.

The purity of the selected \ttbar{} sample is improved by identifying jets containing $b$-hadrons, so called $b$-tagged jets. This identification exploits the long lifetime of $b$-hadrons and the invariant mass of tracks from the corresponding reconstructed secondary vertex, which is on average several \GeV{} larger than that in jets originating from gluons or light-flavour quarks. Information from the track impact parameters, secondary-vertex location and decay topology are combined in a multivariate algorithm (MV2c20)~\cite{ATL-PHYS-PUB-2015-022}.
The operating point corresponds to an overall 77\% $b$-tagging efficiency in \ttbar{} events, with a corresponding rejection of charm-quark jets (light-flavour and gluon jets) by a factor of 4.5 (140)~\cite{ATL-PHYS-PUB-2015-022}. Jets that pass this selection are identified as $b$-tagged jets.

The $\met$ vector is computed from the sum of the transverse momenta of the reconstructed calibrated physics objects (electrons, photons, hadronically decaying $\tau$-leptons, jets and muons) together with the transverse energy deposited in the calorimeter cells, calibrated using tracking information, not associated with these objects \cite{PERF-2014-04}. 
To avoid double-counting of energy, the muon energy loss in the calorimeters is subtracted in the $\met$ calculation. This variable is not used in the selection but is used in the top quark reconstruction described below.

\subsection{Particle-level object definition}\label{sec:TruthObjectDef}
Particle-level objects are defined in simulated events using only stable particles, i.e. particles with a mean lifetime $\tau$ > 30 ps. The fiducial phase space for the measurements presented in this paper is defined using a series of requirements applied to particle-level objects analogous to those used in the selection of the reconstruction-level objects, described above. 

Electrons and muons must not originate, either directly or through a~$\tau$ decay, from a~hadron in the MC event record. This ensures that the lepton is from the decay of a real $W$ boson without requiring a direct match to it. The four-momenta of leptons are modified by adding the four-momenta of all photons within $\Delta R=0.1$ and not originating from hadron decays, to take into account final-state photon radiation. Such leptons are then required to have \pT $>$ 25~\GeV{} and $|\eta| < 2.5$. 

Particle-level jets are reconstructed using the same anti-$k_{t}$ algorithm used at reconstruction level. The jet-reconstruction procedure takes as input all stable particles, except for leptons not from hadron decay as described above, inside a radius $R = 0.4$. Particle level jets are required to have \pT $>$ 25~\GeV{} and $|\eta| < 2.5$.
A jet is identified as a $b$-jet if a hadron containing a $b$-quark is matched to the jet through a ghost-matching technique described in Ref.~\cite{Cacciari:2008gn}; the hadron must have \pT $>$ 5~\GeV{}. 
No overlap removal criteria are applied to particle-level objects. 
Neutrinos and charged leptons from hadron decays are included in particle-level jets. 

\subsection{Event selection and fiducial phase space definition}\label{sec:EventSelection}
Events at both reconstruction and particle levels are required to contain exactly one electron or muon and at least four jets, with at least two tagged as $b$-jets. Each event is then unequivocally assigned to the 4-jet, 5-jet or 6-jet-inclusive configurations, depending on the number of reconstructed jets.

Dilepton $\ttbar$ events, where only one lepton satisfies the fiducial selection, are included by definition in the fiducial measurement.
In the fiducial phase space definition, semileptonic \ttbar{} decays into $\tau$-leptons are considered as signal only if the $\tau$-lepton decays leptonically.

\section{Background determination and event yields} 
\label{sec:BackgroundDetermination}

After the event selection, various backgrounds still contribute to the event yields. The different background contributions are estimated by using MC simulations or data-driven techniques as detailed below for each source. The latter are used when the accuracy of the MC simulation is not adequate, as in the case of $W$ boson production in association with multiple jets and the background originating from jets mimicking the signature of charged leptons.

The single-top quark background is the largest background contribution in all considered regions, amounting to~5\% of the total event yield and~30\% of the total background estimate. This background is modelled with a MC simulation, and the event yields are normalised using calculations of their cross sections, as described in Section \ref{sec:DataSimSamples}. 

Multijet production processes, including \ttbar production{} with all hadronic decay and \ttbar decays into $\tau$-leptons which then decay hadronically, have a large cross section and can mimic the lepton+jets signature due to hadrons misidentified as prompt leptons (fake leptons), conversion of photons for the electron channel or semileptonic decays of heavy-flavour hadrons (non-prompt real leptons). The multijet background is estimated directly from data by using a matrix method~\cite{TOPQ-2011-02} in which signal and control regions are defined using lepton identification criteria. The method depends on the probability of a real (fake) lepton to pass the tight selection criteria, which is referred to as the real (fake) efficiency. These efficiencies are measured in data control regions dominated by real or fake lepton events. In the $e$+jets channel, the fake efficiency is parametrised as a function of $p_{\mathrm{T}}$ and $\eta$, as well as the azimuthal angle difference between the lepton and the $\met$ vector, $\Delta\phi$. In the $\mu$+jets channel, the fake efficiency is calculated for low and high lepton $p_{\mathrm{T}}$. The low $p_{\mathrm{T}}$ parametrisation depends on $\Delta\phi$, $p_{\mathrm{T}}$ and $\met$ , whereas the high $p_{\mathrm{T}}$ parametrisation only uses $p_{\mathrm{T}}$. The real efficiencies are measured with the $Z \rightarrow \ell \ell$ events using the tag-and-probe method. In the $e$+jets channel, the efficiency is parametrised as a function of $p_{\mathrm{T}}$, whereas in the $\mu$+jets channel the parametrisation depends on $\Delta\phi$ and $p_{\mathrm{T}}$. The multijet background contributes to the total event yield at the level of approximately 4$\%$ and~30\% of the total background estimate.

The $W$+jets background represents the third largest background, amounting to $2$--$3$\% of the total event yield and~20\% of the total background estimate.
The estimation of this background is performed using a combination of MC simulations and data-driven techniques; the \Sherpa MC event generator is used to estimate the contribution from the $W$+jets process. The normalisation and the heavy-flavour fractions of this process, which are affected by large theoretical uncertainties, are determined from data. The overall $W$+jets normalisation is obtained by exploiting the expected charge asymmetry in the production of $W^+$ and $W^-$ bosons in $pp$ collisions. This asymmetry is predicted by theory~\cite{Halzen:2013bqa} and evaluated using MC simulations, assuming other processes are symmetric in charge except for a small contamination from single-top quark, $t\bar{t}V$ and $WZ$ events, which is subtracted using MC simulations. The total number of $W$+jets events with a positively or negatively charged $W$ boson ($N_{W^+} + N_{W^-}$) in the sample is thus estimated using the following equation:

\begin{equation}
N_{W^+} + N_{W^-} = \left(\frac{r_{\mathrm{MC}} + 1}{r_{\mathrm{MC}} - 1}\right)(D_{\mathrm+} - D_{\mathrm-})\,,
\label{eq:Wchargeasymm}
\end{equation}

where $r_{\mathrm{MC}}$ is the ratio of the number of events with positively charged leptons to the number of events with negatively charged leptons in the MC simulations, and $D_{\mathrm+}$ and $D_{\mathrm-}$ are the numbers of events with positive and negative leptons in the data, respectively, corrected for the aforementioned non-$W+$jets charge-asymmetric contributions using simulation. The corrections due to event generator mis-modelling of $W$ boson production in association with jets of different flavour ($W+$$b\bar{b}$, $W+$$c\bar{c}$, $W+$$c$, $W+$light flavours) are estimated using a dedicated control sample in data which uses the same lepton as for the signal but requiring exactly two jets. In their determination, the overall normalisation scaling factor obtained using Eq.~(\ref{eq:Wchargeasymm}) is applied first. Then heavy-flavour scaling factors obtained in the two-jet control region are extrapolated to the signal region using MC simulations, assuming constant relative rates for the signal and control regions. Taking into account the heavy-flavour scale factors, the overall normalisation factor is calculated again using Eq.~(\ref{eq:Wchargeasymm}).
This iterative procedure is repeated until the total predicted $W+$jets yield in the two-jet control configuration agrees with the data yield.
The procedure is explained in detail in Ref.~\cite{TOPQ-2012-18}.

The background contributions from $Z$+jets, $t\bar{t}V$ and diboson events are obtained from MC simulation, and the event yields are normalised using the theoretical calculations of their cross sections, as described in Section \ref{sec:DataSimSamples}. The total contribution from these processes is 1--2\% of the total event yield or 11--14\% of the total background.

Dilepton top quark pair events can satisfy the event selection if one lepton does not satisfy the requirements listed above and at least two additional jets are produced. Events with at least a top quark decaying to a $\tau$-lepton which subsequently decays leptonically, can also pass the event selection. These events contribute 3-5\% to the total event yield, and are considered in the analysis at both reconstruction and particle levels. Cases where both top quarks decay semileptonically into $\tau$-leptons, and where both $\tau$-leptons decay hadronically, are accounted for in the multijet background.

The event yields in the three configurations are displayed in Table~\ref{Tab:EventYields} for data, simulated signal, and backgrounds. 
Figure~\ref{fig:controls_4je2bi_detector} shows\footnote{All data as well as theory points are plotted at the graphical bin centre on the $x$-axis throughout this paper.} the comparison between data and predictions for the 4-jet configuration for different distributions. All of the distributions are shown for the combined \ljets{} channel (combining electron and muon channels). The background contributions in the other configurations are similar, as shown in Figure~\ref{fig:controls_jets_detector}. The event selection results in a total background contamination of $10$--$15$\%, depending on the configuration. 
A constant difference between data and prediction is observed in Figures ~\ref{fig:controls_jets_detector}\subref{fig:jet_pt_bilog_5je2bi} and~\ref{fig:controls_jets_detector}\subref{fig:jet_pt_bilog_6ji2bi}, the same effect is also seen in the distribution of the number of jets, shown in Figure~\ref{fig:jet_n_co_4ji2bi}. This discrepancy has also been observed in studies of associated production of jets with top quark pairs~\cite{TOPQ-2015-17}. Nevertheless, the predictions obtained using the nominal sample are compatible with the data within the uncertainties; this level of agreement allows to carry out the unfolding described in Section~\ref{sec:unfolding}.

\begin{table}[t]
\resizebox{1 \textwidth}{!}{ 
   \begin{tabular}{ccc}
      {\renewcommand{\arraystretch}{1.2}
\begin{tabular}{lr}
\hline \hline
4-jet exclusive & \multicolumn{1}{r}{} \\
\hline \hline
Sample & Yield \vspace{0.01cm}\\ 
\hline 
\vspace{0.13cm}
$t\bar{t}$ & $61400^{+3300}_{-3400}$ \\
\vspace{0.13cm}
$W$+jets & $2200^{+1400}_{-1600}$ \\
\vspace{0.13cm}
$Z$+jets & $840^{+630}_{-620}$ \\
\vspace{0.13cm}
Diboson & $140^{+100}_{-100}$ \\
\vspace{0.13cm}
Single top & $3600^{+360}_{-360}$ \\
\vspace{0.13cm}
Multijet & $3300^{+1700}_{-1800}$ \\
\vspace{0.13cm}
$t\bar{t}$~$V$ & $103^{+17}_{-17}$ \\
\vspace{0.13cm}
Total prediction & $71600^{+4800}_{-5000}$ \\
\hline
Data &  75768 \\
\hline
Data/prediction &  $1.06\pm0.07$\\
\hline \hline
\end{tabular}
}

      {\renewcommand{\arraystretch}{1.2}
\begin{tabular}{lr}
\hline \hline
5-jet exclusive & \multicolumn{1}{r}{} \\
\hline \hline
Sample & Yield  \vspace{0.01cm}\\ 
\hline 
\vspace{0.13cm}
$t\bar{t}$ & $36900^{+3700}_{-3700}$ \\
\vspace{0.13cm}
$W$+jets & $890^{+600}_{-680}$ \\
\vspace{0.13cm}
$Z$+jets & $340^{+330}_{-330}$ \\
\vspace{0.13cm}
Diboson & $100^{+100}_{-100}$ \\
\vspace{0.13cm}
Single top & $1730^{+240}_{-240}$ \\
\vspace{0.13cm}
Multijet & $1460^{+770}_{-780}$ \\
\vspace{0.13cm}
$t\bar{t}$~$V$ & $132^{+21}_{-21}$ \\
\vspace{0.13cm}
Total prediction & $41600^{+4000}_{-4300}$ \\
\hline
Data &  46243 \\
\hline
Data/prediction &  $1.11\pm0.11$ \\
\hline \hline
\end{tabular}
}

      {\renewcommand{\arraystretch}{1.2}
\begin{tabular}{lr}
\hline \hline
6-jet inclusive & \multicolumn{1}{r}{} \\
\hline \hline
Sample & Yield \vspace{0.01cm}\\ 
\hline 
\vspace{0.13cm}
$t\bar{t}$ & $25400^{+4700}_{-4400}$ \\
\vspace{0.13cm}
$W$+jets & $540^{+400}_{-450}$ \\
\vspace{0.13cm}
$Z$+jets & $160^{+100}_{-100}$ \\
\vspace{0.13cm}
Diboson & $110^{+57}_{-57}$ \\
\vspace{0.13cm}
Single top & $980^{+210}_{-200}$ \\
\vspace{0.13cm}
Multijet & $920^{+500}_{-500}$ \\
\vspace{0.13cm}
$t\bar{t}$~$V$ & $224^{+40}_{-40}$ \\
\vspace{0.13cm}
Total prediction & $28400^{+4900}_{-4900}$ \\
\hline
Data &  33582 \\
\hline
Data/prediction &  $1.2\pm0.2$  \\
\hline \hline
\end{tabular}
}

   \end{tabular}
}
\caption{Event yields in the 4-jet exclusive (left), 5-jet exclusive (centre) and 6-jet inclusive (right) configurations. The uncertainties include the combined statistical and systematic uncertainties, excluding the systematic uncertainties related to the modelling of the \ttbar{} system.
}
\label{Tab:EventYields}
\end{table}

\begin{figure*}[p]
\centering
\subfigure[]{\includegraphics[width=0.45\textwidth]{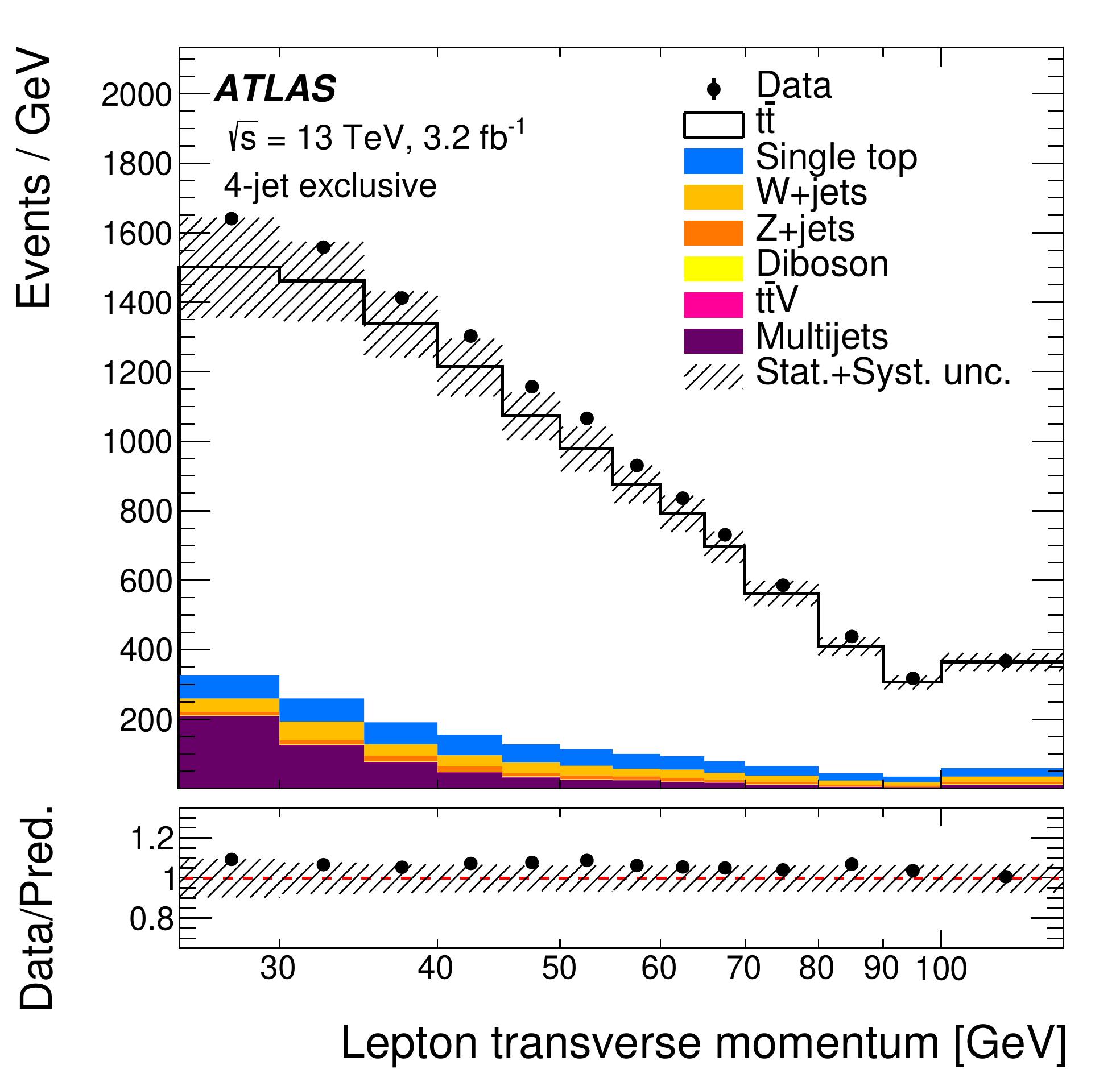}\label{fig:lep_pt_logx_4je2bi}}
\subfigure[]{\includegraphics[width=0.45\textwidth]{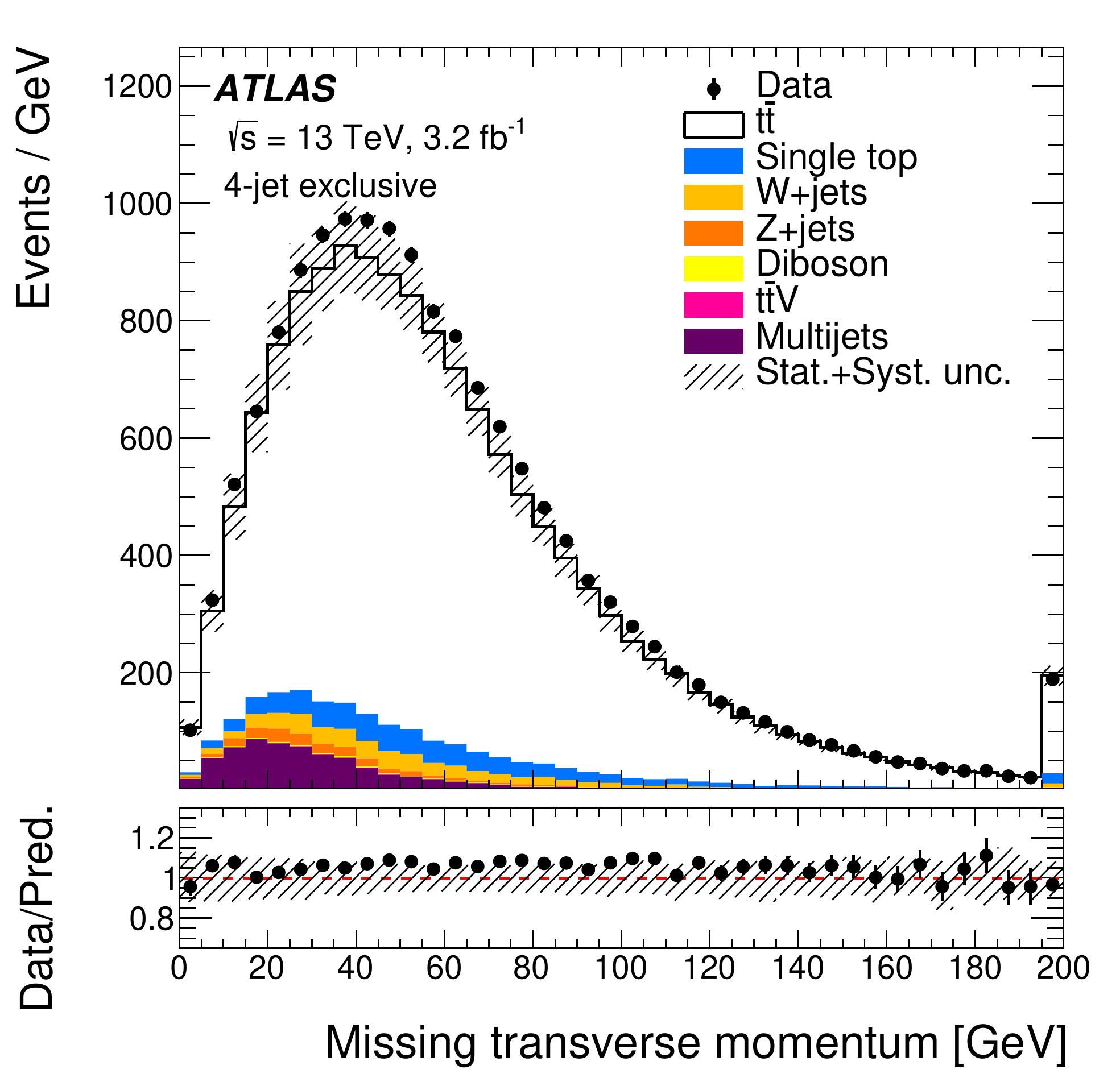}\label{fig:met_pt_2_4je2bi}}
\subfigure[]{\includegraphics[width=0.45\textwidth]{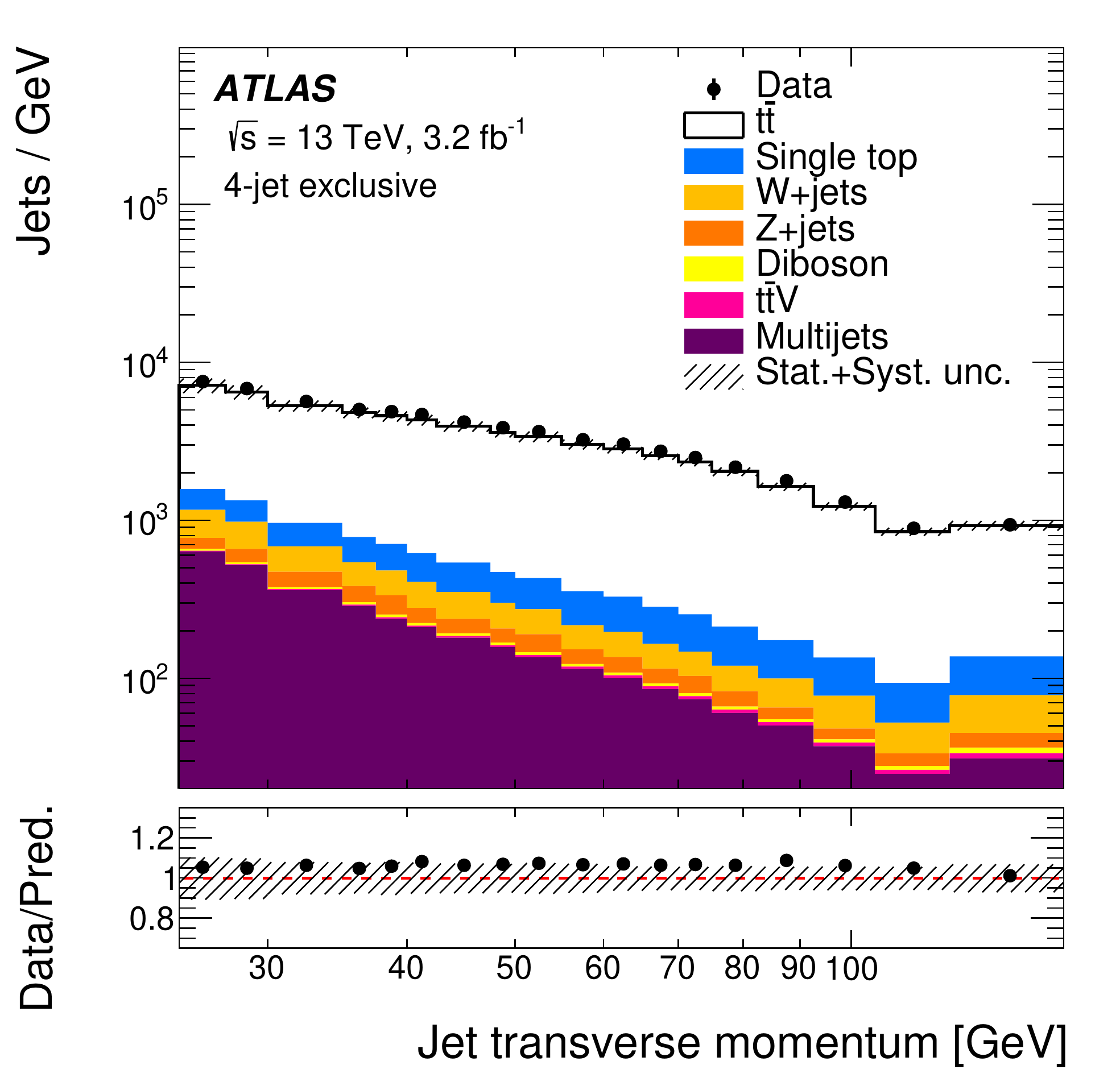}\label{fig:jet_pt_bilog_4je2bi}}
\subfigure[]{\includegraphics[width=0.45\textwidth]{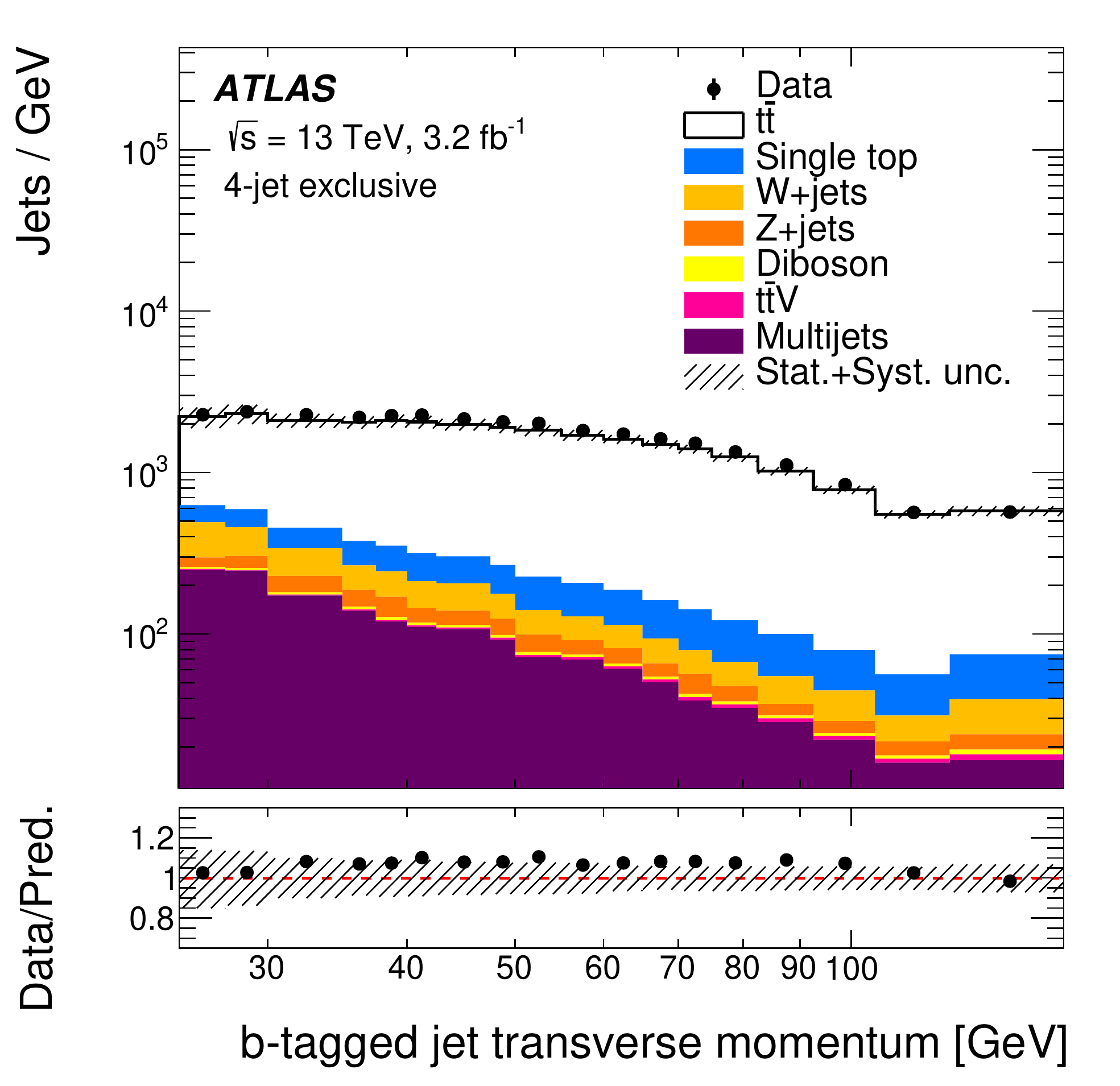}\label{fig:bjet_pt_bilog_4je2bi}}
\caption{Kinematic distributions in the 4-jet exclusive configuration at reconstruction level:~\subref{fig:lep_pt_logx_4je2bi}~lepton transverse momentum,~\subref{fig:met_pt_2_4je2bi}~missing transverse momentum, transverse momentum of~\subref{fig:jet_pt_bilog_4je2bi} the selected jets and~\subref{fig:bjet_pt_bilog_4je2bi} of the selected $b$-tagged jets. Data are compared to the sum of signal and background predictions using the nominal sample as the \ttbar{} signal model. The hatched area indicates the combined statistical and systematic uncertainties in the total prediction, excluding systematic uncertainties related to the modelling of the \ttbar{} production. Events beyond the range of the horizontal axis are included in the last bin.}
\label{fig:controls_4je2bi_detector}
\end{figure*}

\begin{figure*}[p]
\centering
\subfigure[]{ \includegraphics[width=0.45\textwidth]{figures/control/co_reco_cutflow_4je2bi_jet_pt_bilog.pdf}\label{fig:jet_pt_co_4je2bi}}
\subfigure[]{ \includegraphics[width=0.45\textwidth]{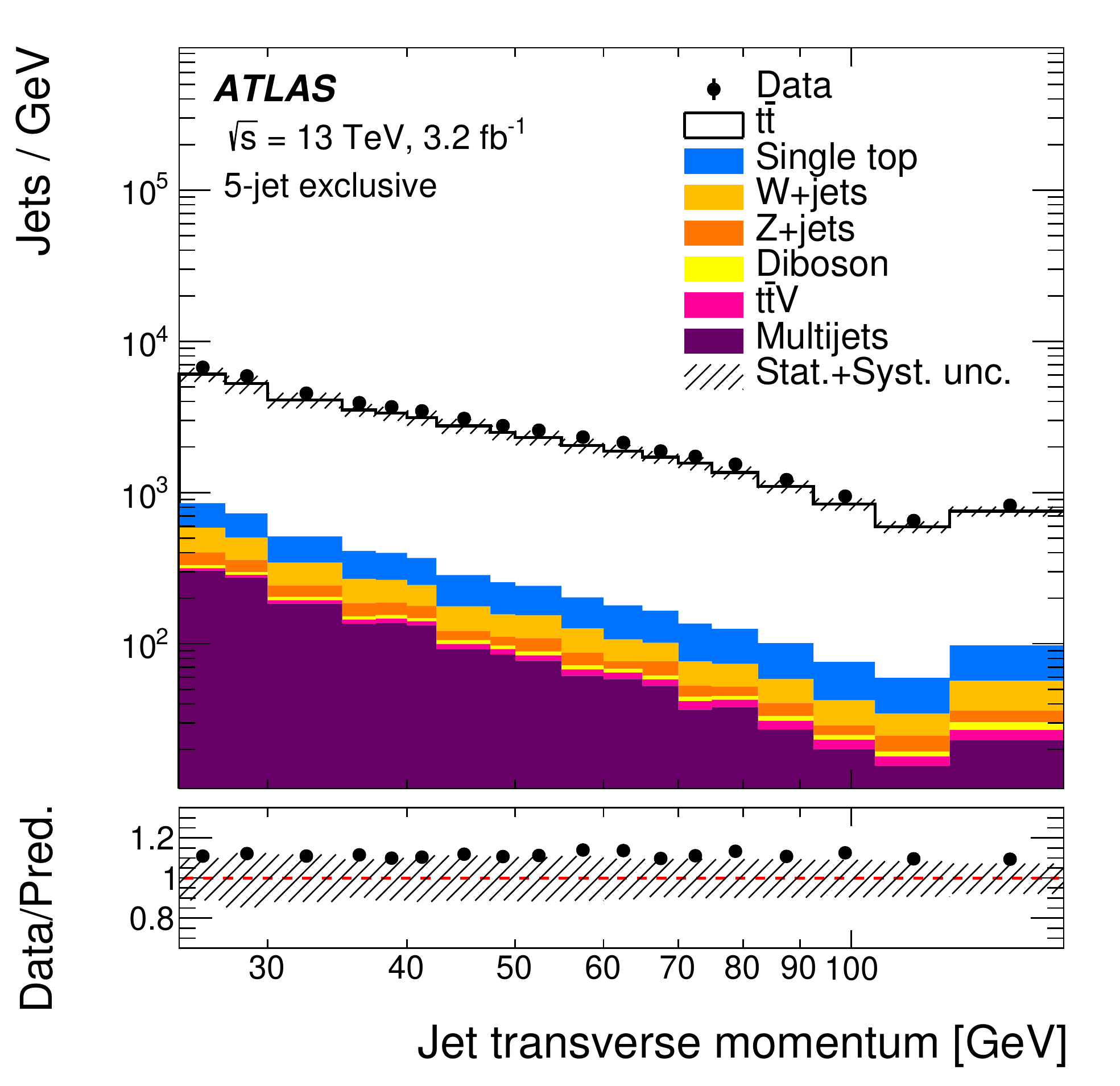}\label{fig:jet_pt_bilog_5je2bi}}
\subfigure[]{ \includegraphics[width=0.45\textwidth]{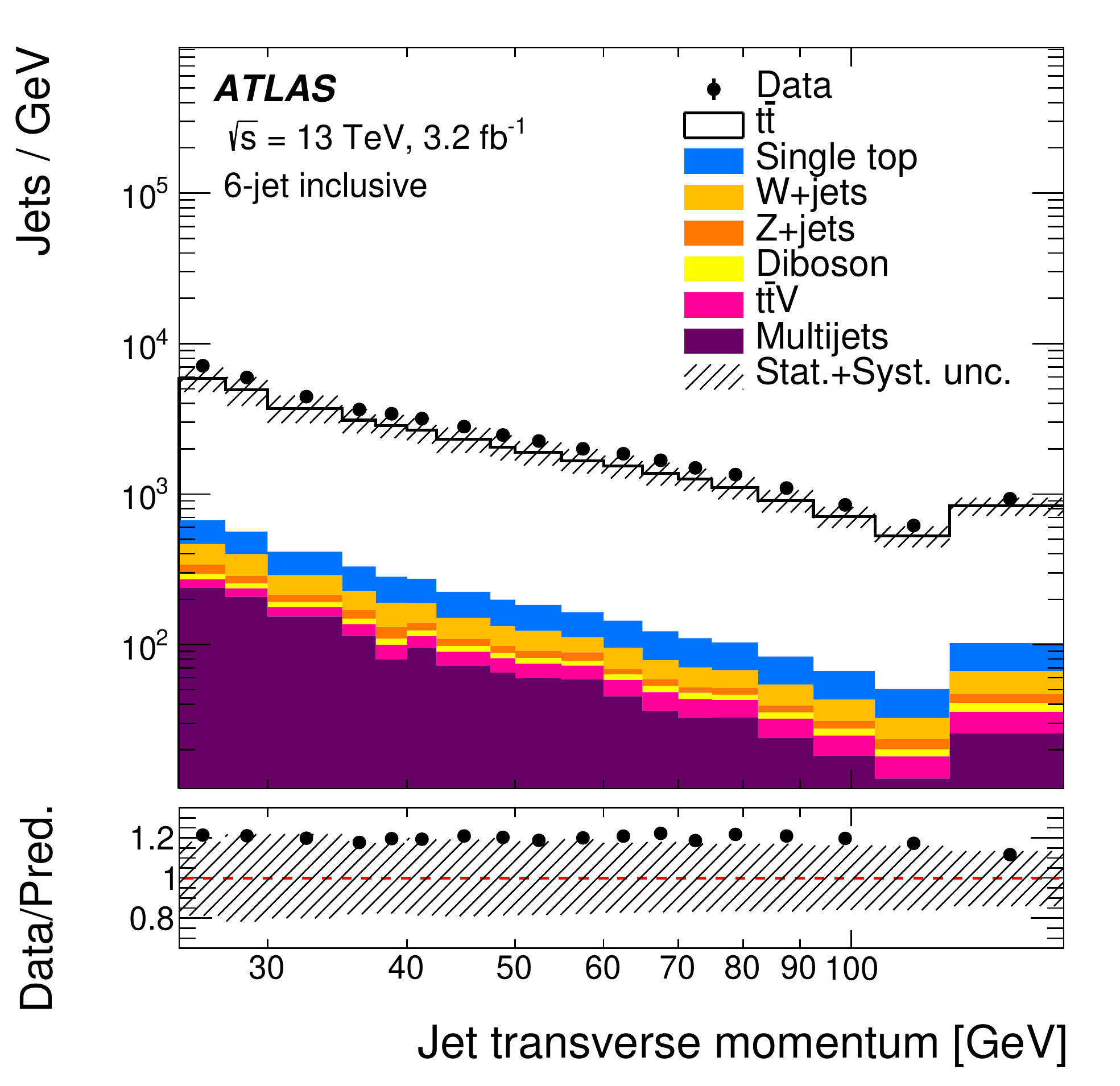}\label{fig:jet_pt_bilog_6ji2bi}}
\caption{Distribution of the transverse momentum of selected jets in the~\subref{fig:jet_pt_co_4je2bi}~4-jet exclusive,~\subref{fig:jet_pt_bilog_5je2bi}~5-jet exclusive and~\subref{fig:jet_pt_bilog_6ji2bi}~6-jet inclusive configurations at reconstruction level. Data are compared to the sum of signal and background predictions using the nominal sample as the \ttbar{} signal model. The hatched area indicates the combined statistical and systematic uncertainties in the total prediction, excluding systematic uncertainties related to the modelling of the \ttbar{} production. Events beyond the range of the horizontal axis are included in the last bin.}
\label{fig:controls_jets_detector}
\end{figure*}

\begin{figure*}[p]
\centering
\includegraphics[width=0.45\textwidth]{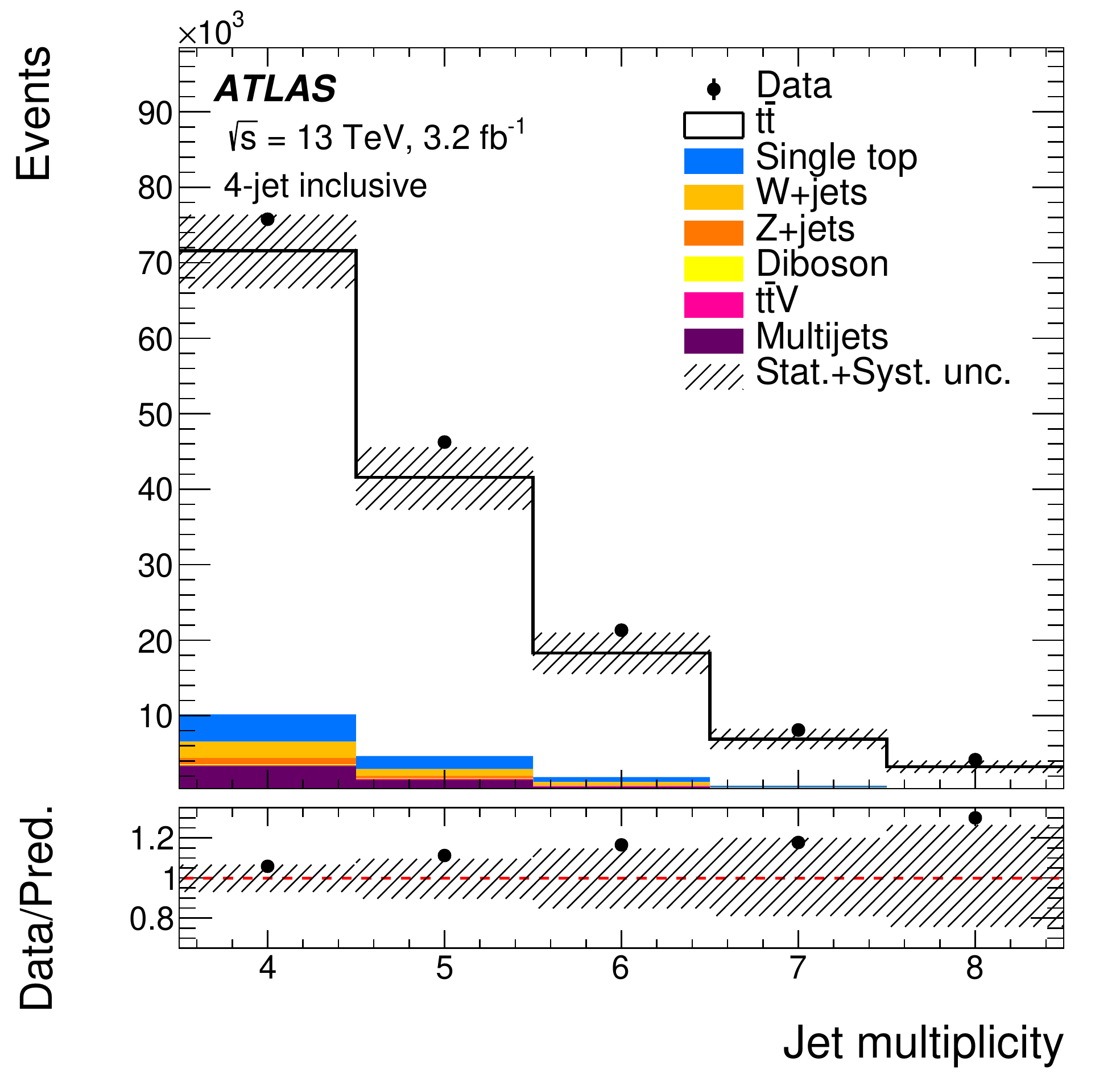}
\caption{Distribution of the jet multiplicity in the 4-jet inclusive configuration. Data are compared to the sum of signal and background predictions using the nominal sample as the \ttbar{} signal model. The hatched area indicates the combined statistical and systematic uncertainties in the total prediction, excluding systematic uncertainties related to the modelling of the \ttbar{} production. Events beyond the range of the horizontal axis are included in the last bin.}
\label{fig:jet_n_co_4ji2bi}
\end{figure*}

\FloatBarrier
\section{Reconstruction of top quark kinematic properties} \label{sec:PseudoTop}

The two top quarks are reconstructed from their decay products so that the differential cross sections can be measured as functions of observables involving the top quark and the \ttbar{} system. In the following, the leptonic (hadronic) top quark refers to the one that decays into a leptonically (hadronically) decaying $W$ boson.

The pseudo-top algorithm~\cite{TOPQ-2013-07} reconstructs the four-momenta of the top quarks and their complete decay chain from final-state objects, namely the charged lepton (electron or muon), missing transverse momentum, and four jets, two of which are $b$-tagged. Only about 14\% of the selected events contain more than two $b$-tagged jets, in which case the two with the highest transverse momentum are considered as coming from the top quarks, while the others are considered for the $W$ reconstruction. The same algorithm is used to reconstruct the kinematic properties of top quarks at reconstruction level and particle level in the three configurations.

The algorithm starts with the reconstruction of the neutrino four-momentum. While the $x$ and $y$ components of the neutrino momentum are set to the corresponding components of the missing transverse momentum, the $z$ component is calculated by imposing a $W$ boson mass constraint on the invariant mass of the charged-lepton--neutrino system. If the resulting quadratic equation has two real solutions, the one with the smaller value of $|p_z|$ is chosen. If the discriminant of the equation is negative, only the real part is considered. 

The leptonically decaying $W$ boson is reconstructed from the charged lepton and the reconstructed neutrino. The leptonic top quark is reconstructed from the leptonic $W$ boson and the \btagged jet closest in $\Delta R$ to the charged lepton. The hadronic $W$ boson is reconstructed from the two jets whose invariant mass is closest to the mass of the $W$ boson; only jets that do not pass the $b$-tagging requirements are considered. Finally, the hadronic top quark is reconstructed from the hadronic $W$ boson and the other \bjet. This choice yields the best performance of the algorithm in terms of the correspondence between the reconstruction level and particle level.

The performance of the algorithm was studied in each of the three configurations. The algorithm reconstructs the masses of the hadronic $W$ boson and the top quark with similar performances in all three configurations. Hence, the presence of additional jets in the 5- and 6-jet configurations, where different combinations in the jet assignment to the $W$ boson are possible, does not impact the reconstruction significantly.

\section{Measured observables}\label{sec:YieldsAndPlots}

The goal of this analysis is to measure differential cross sections for observables in regions of phase space sensitive to gluon radiation. Three observables are chosen because they are shown to be sensitive to radiation or other effects correlated with the number of jets: $\ptthad$, $\ptttbar$ and $\absPoutttbar$. 

Figure~\ref{Fig:reco_level_variables_tt_pt} shows the $\ptttbar$ distributions for the three configurations. The $\ptttbar$ distribution is expected to depend strongly on gluon radiation; if no additional jets beyond those of the $\ttbar$ decay are produced, the $\ttbar$ system should have small $\ptttbar$. If an additional jet is produced, the $\ttbar{}$ system recoils against it, hence it should take larger $\pt{}$ values. This effect is more pronounced with more additional jets, as observed in Figure~\ref{Fig:reco_level_variables_tt_pt}. The $\ptthad$ distributions for the three configurations are shown in Figure~\ref{Fig:reco_level_variables_topH}. The predictions tend to underestimate (overestimate) the data at low (high) $\ptthad$. This effect is most clearly observed in the 4-jet configuration; a stress test performed on the unfolding (described in Section~\ref{sec:unfolding}) demonstrated that the difference between data and prediction does not affect the results. The $\absPoutttbar$ distributions are shown in Figure~\ref{Fig:reco_level_variables_pout}; the shape of the measured distribution displays a small dependence on the number of additional jets.

\begin{figure*}[ht]
\centering
\subfigure[]{ \includegraphics[width=0.45\textwidth]{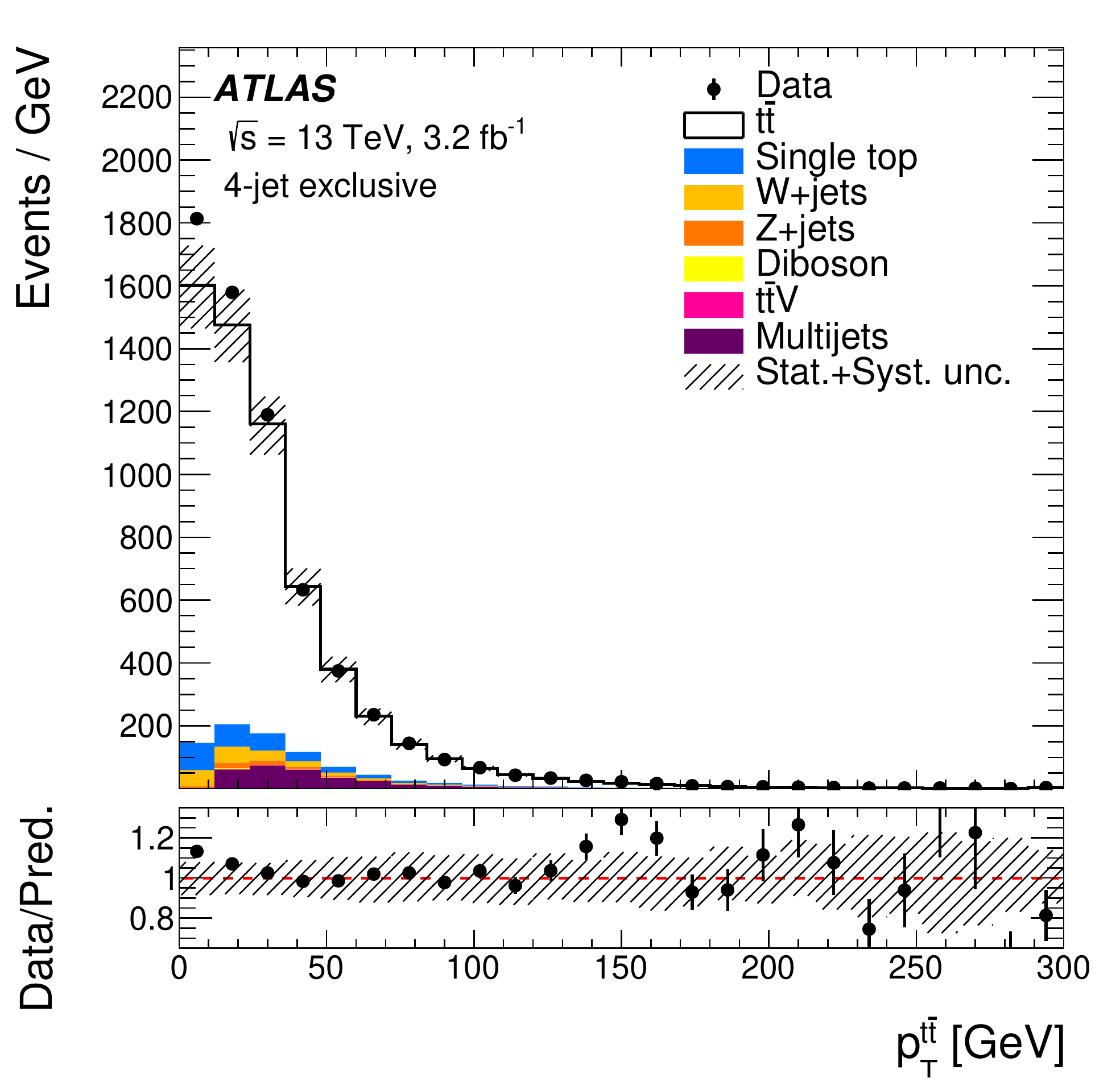}\label{fig:4j_tt_pt_co}}
\subfigure[]{ \includegraphics[width=0.45\textwidth]{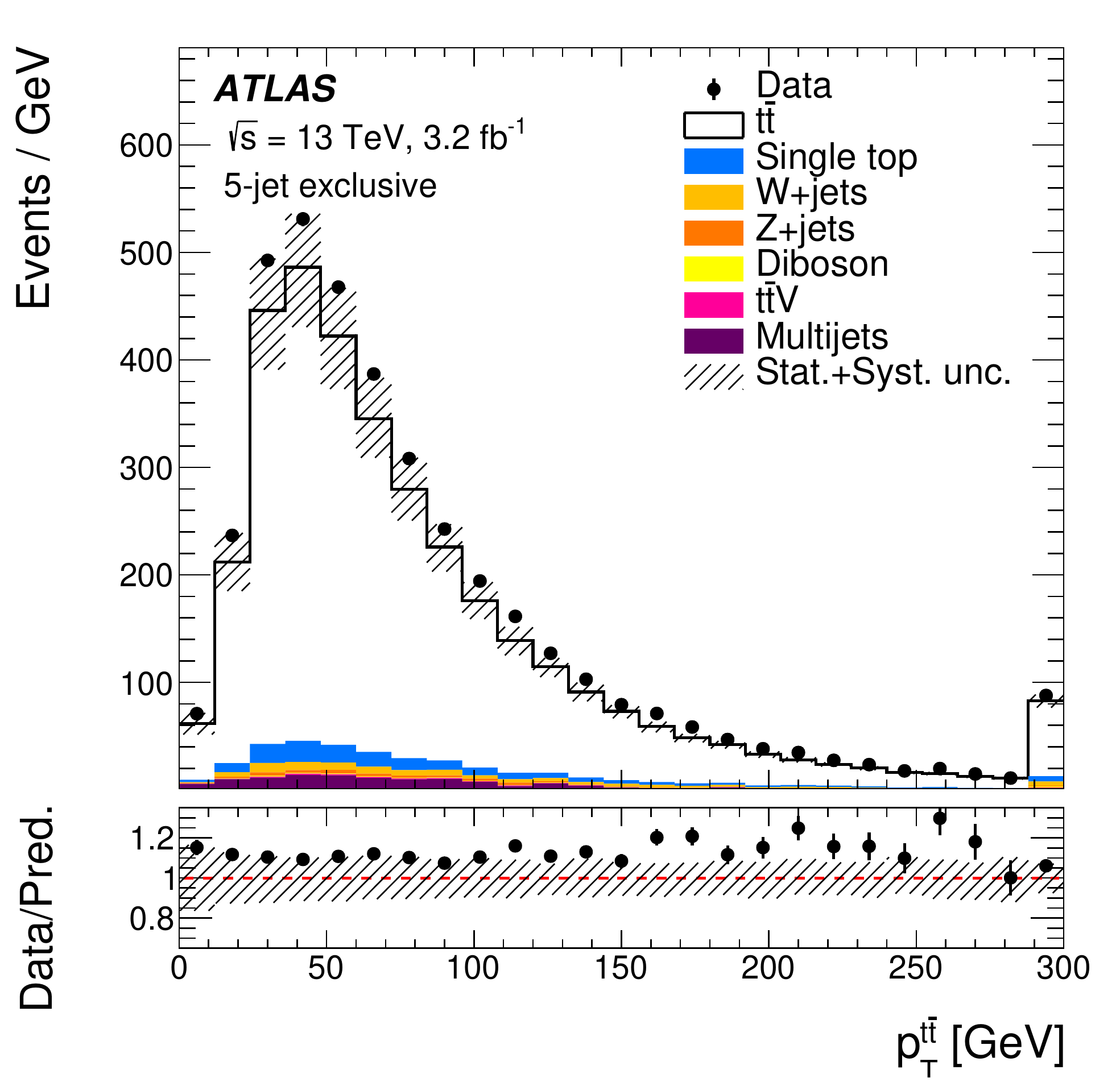}\label{fig:5j_tt_pt_co}}
\subfigure[]{ \includegraphics[width=0.45\textwidth]{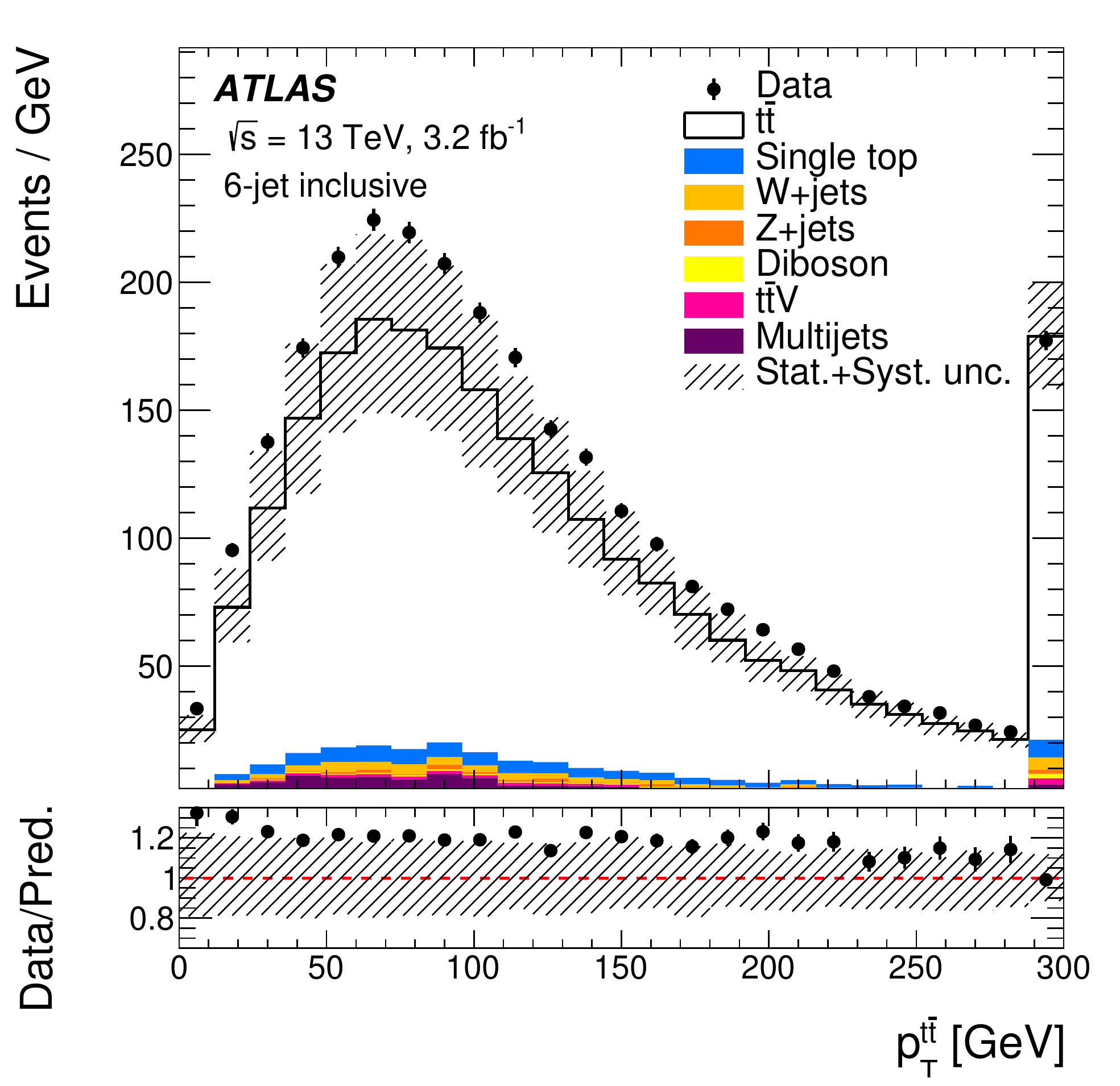}\label{fig:6j_tt_pt_co}}
\caption{Distributions of $\ptttbar$ at reconstruction level: \subref{fig:4j_tt_pt_co} 4-jet exclusive, \subref{fig:5j_tt_pt_co} 5-jet exclusive and \subref{fig:6j_tt_pt_co} 6-jet inclusive configurations. Data are compared to the sum of signal and background predictions using the nominal sample as the \ttbar{} signal model. The hatched area indicates the combined statistical and systematic uncertainties in the total prediction, excluding systematic uncertainties related to the modelling of the $\ttbar$ production.  Events beyond the range of the horizontal axis are included in the last bin.}
\label{Fig:reco_level_variables_tt_pt}
\end{figure*}

\begin{figure*}[ht]
\centering
\subfigure[]{ \includegraphics[width=0.45\textwidth]{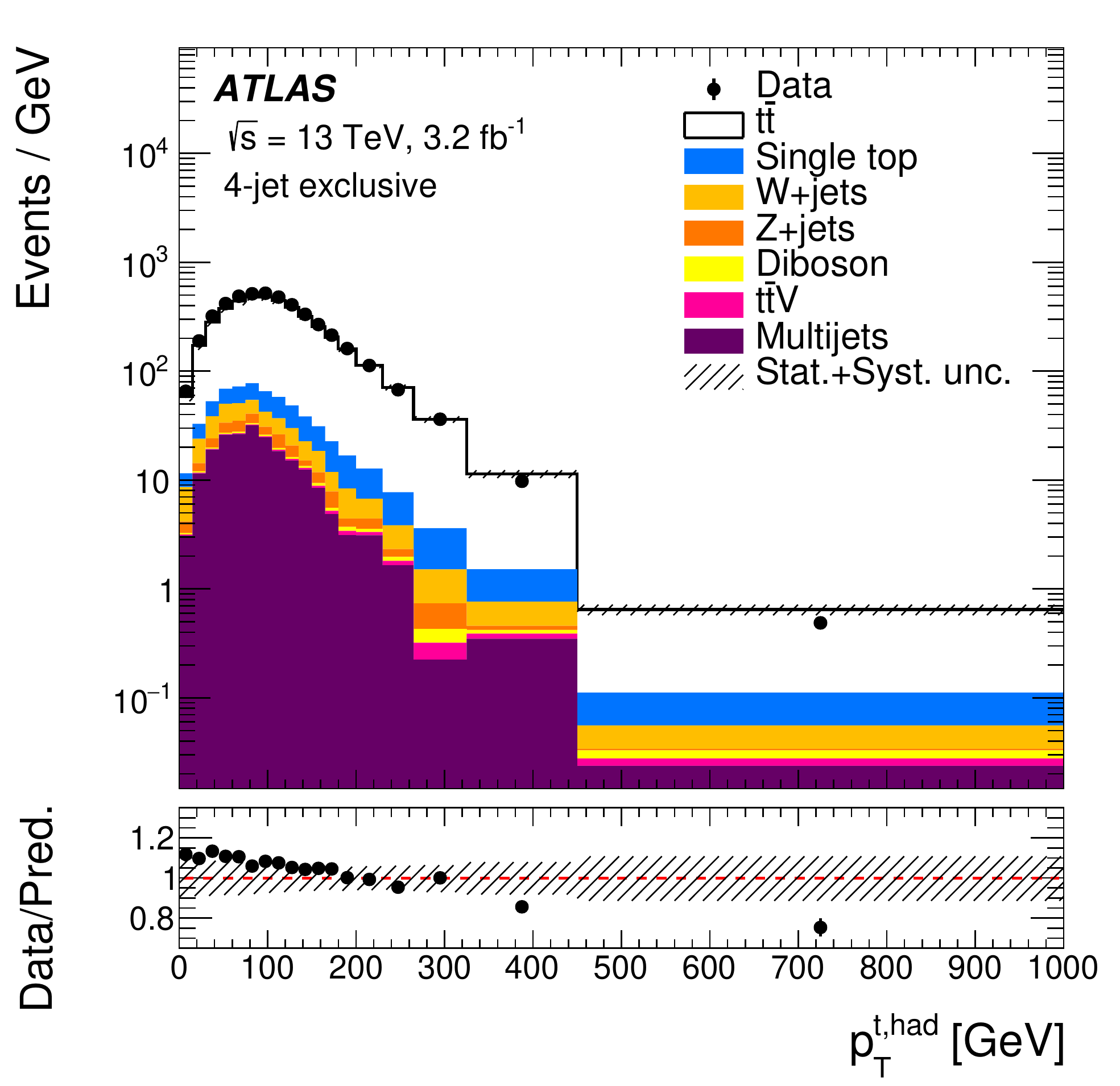}\label{fig:4j_topH_pt_co}}
\subfigure[]{ \includegraphics[width=0.45\textwidth]{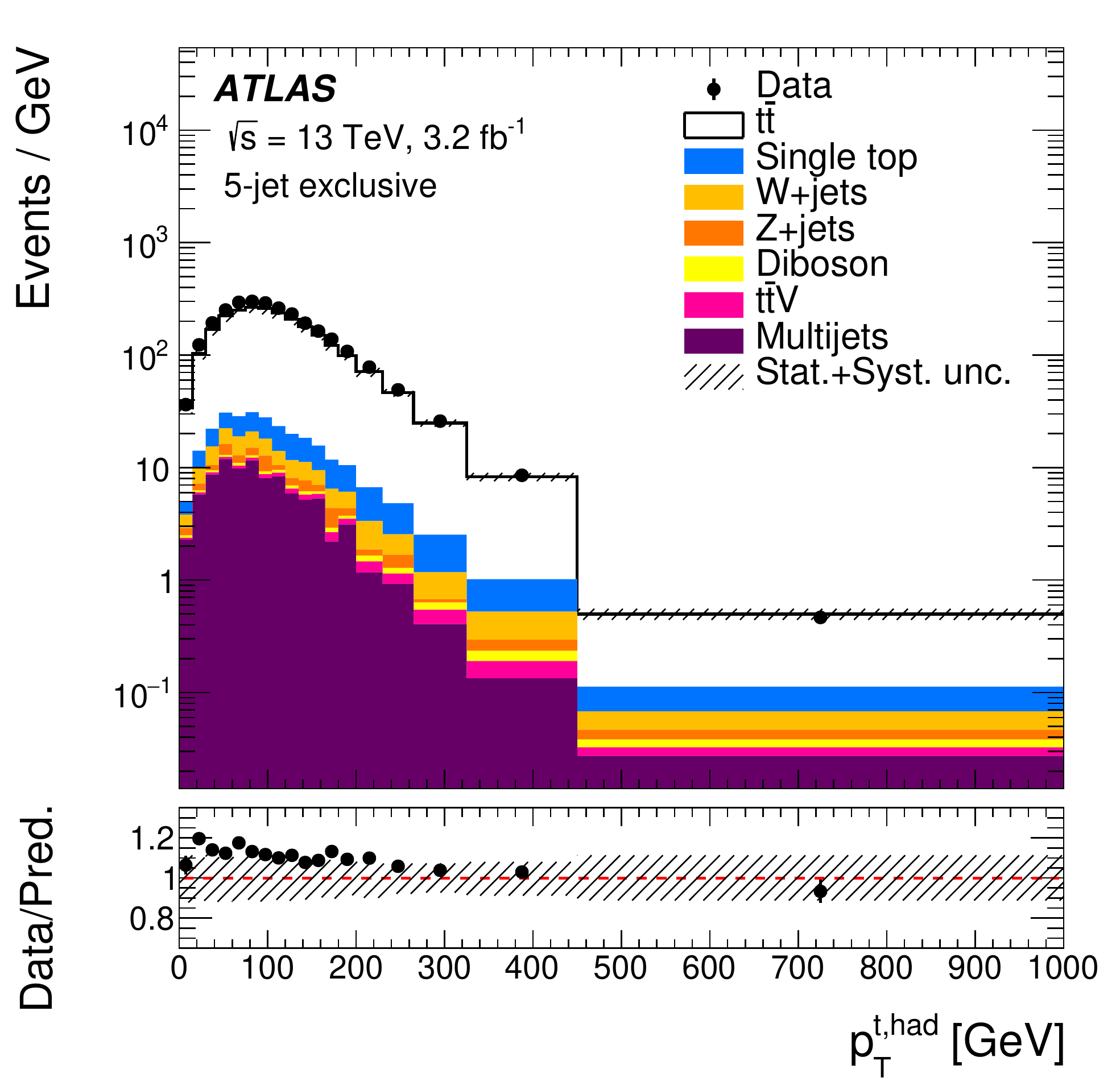}\label{fig:5j_topH_pt_co}}
\subfigure[]{ \includegraphics[width=0.45\textwidth]{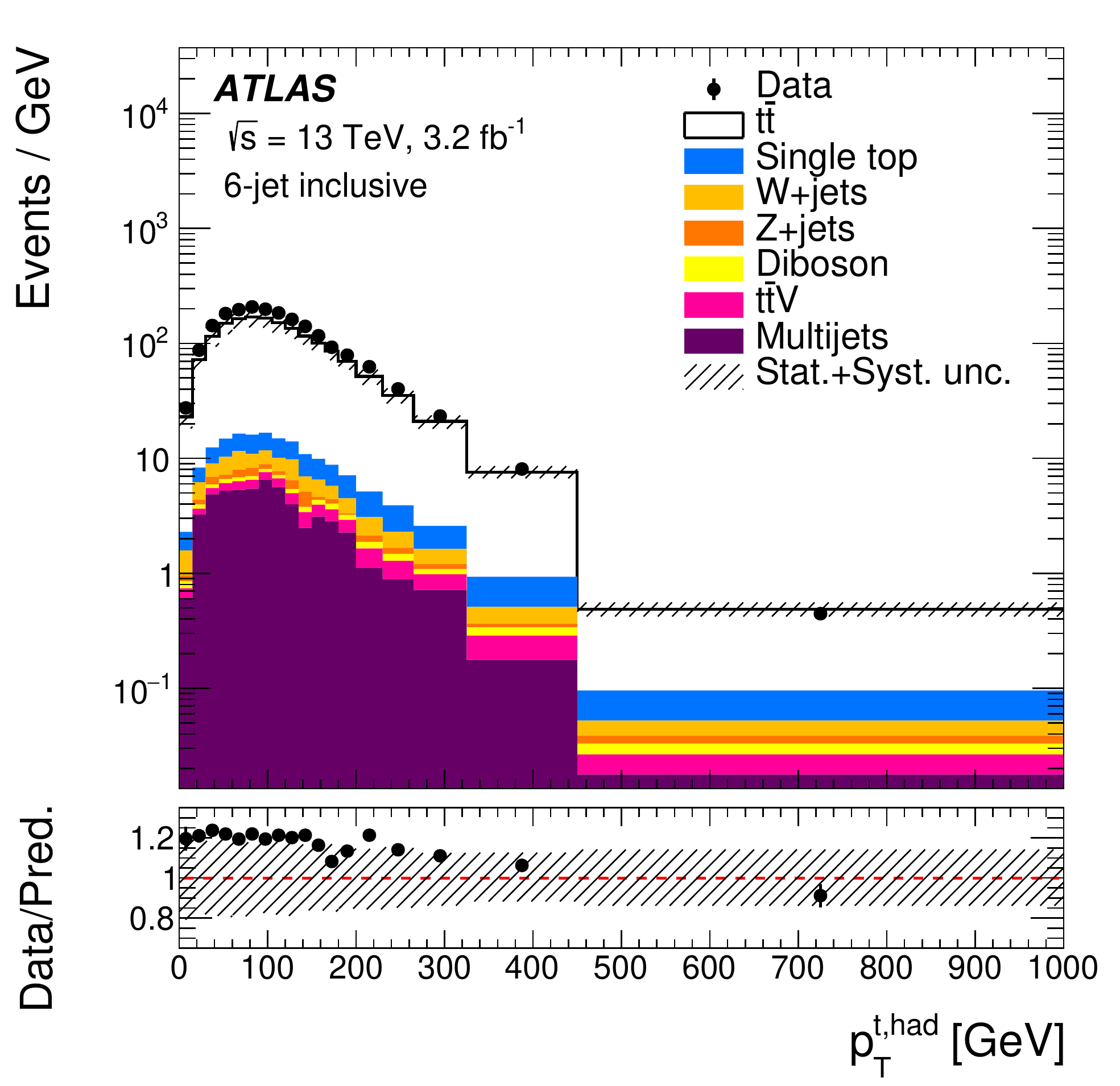}\label{fig:6j_topH_pt_co}}
\caption{Distributions of \ptthad{} at reconstruction level: \subref{fig:4j_topH_pt_co} 4-jet exclusive, \subref{fig:5j_topH_pt_co} 5-jet exclusive and \subref{fig:6j_topH_pt_co} 6-jet inclusive configurations. Data are compared to the sum of signal and background predictions using the nominal sample as the \ttbar{} signal model. The hatched area indicates the combined statistical and systematic uncertainties in the total prediction, excluding systematic uncertainties related to the modelling of the $\ttbar$ production.  Events beyond the range of the horizontal axis are included in the last bin.}
\label{Fig:reco_level_variables_topH}
\end{figure*}

\begin{figure*}[ht]
\centering
\subfigure[]{ \includegraphics[width=0.45\textwidth]{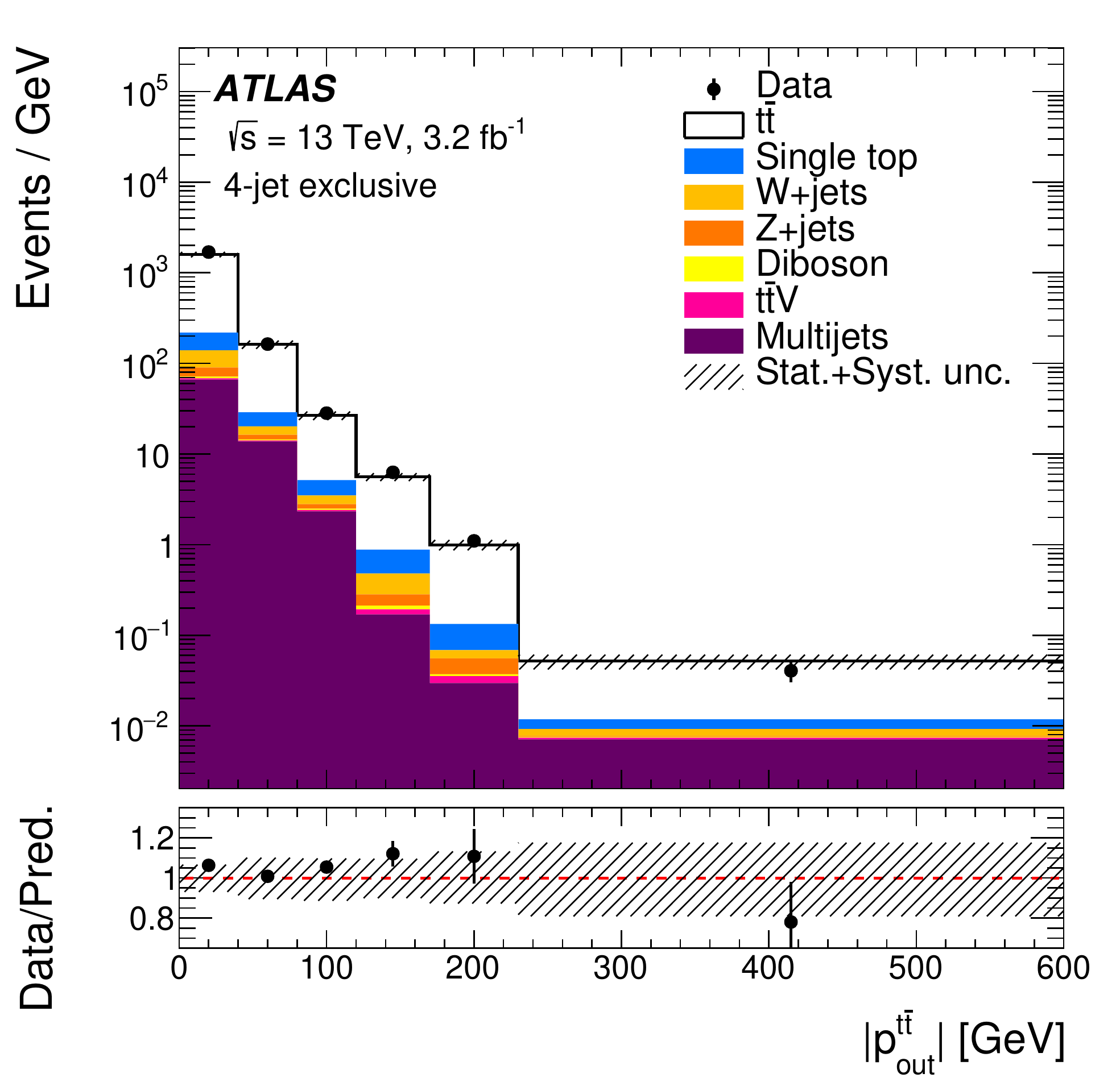}\label{fig:4je_pout_co}}
\subfigure[]{ \includegraphics[width=0.45\textwidth]{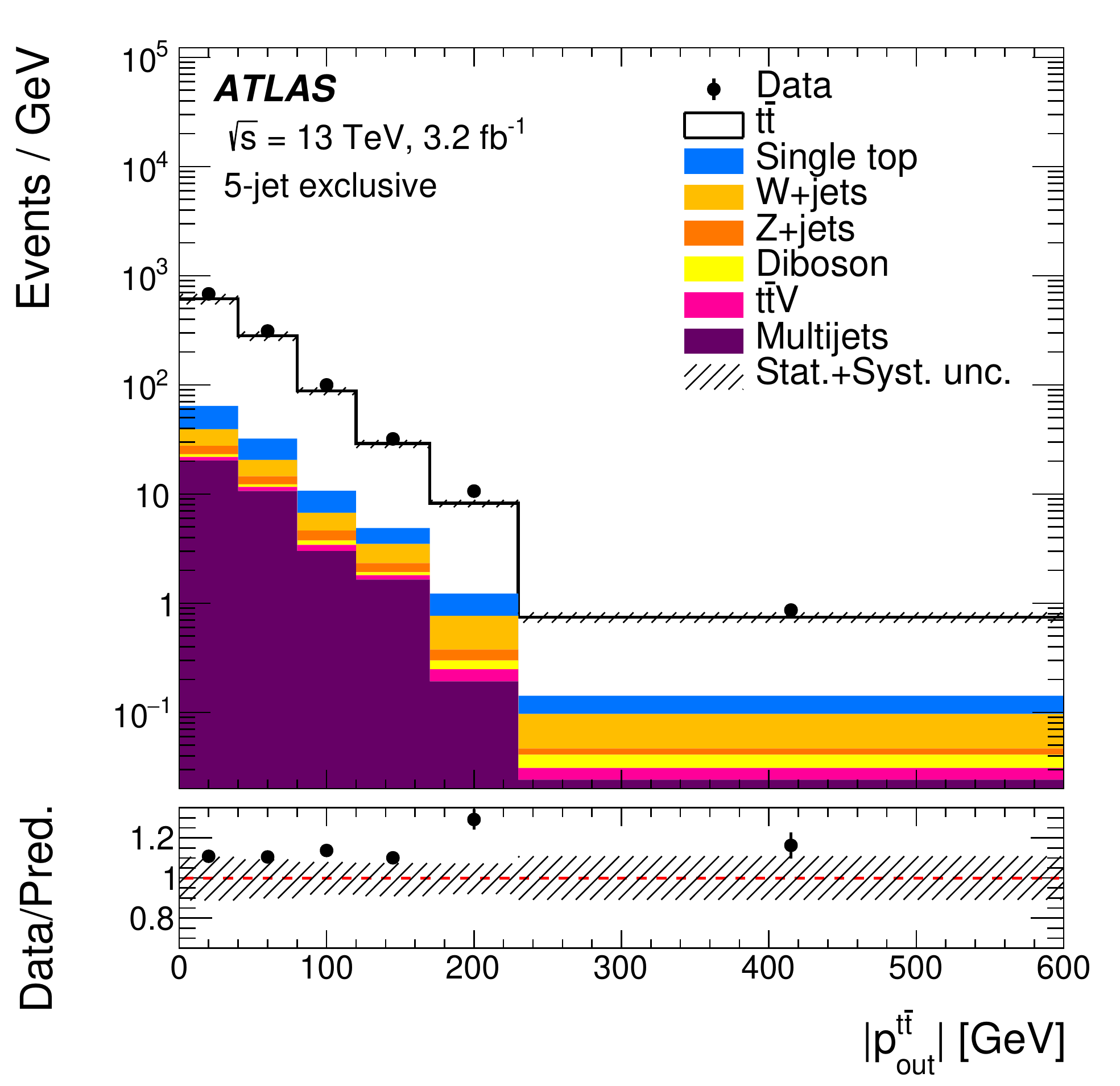}\label{fig:5j_pout_co}}
\subfigure[]{ \includegraphics[width=0.45\textwidth]{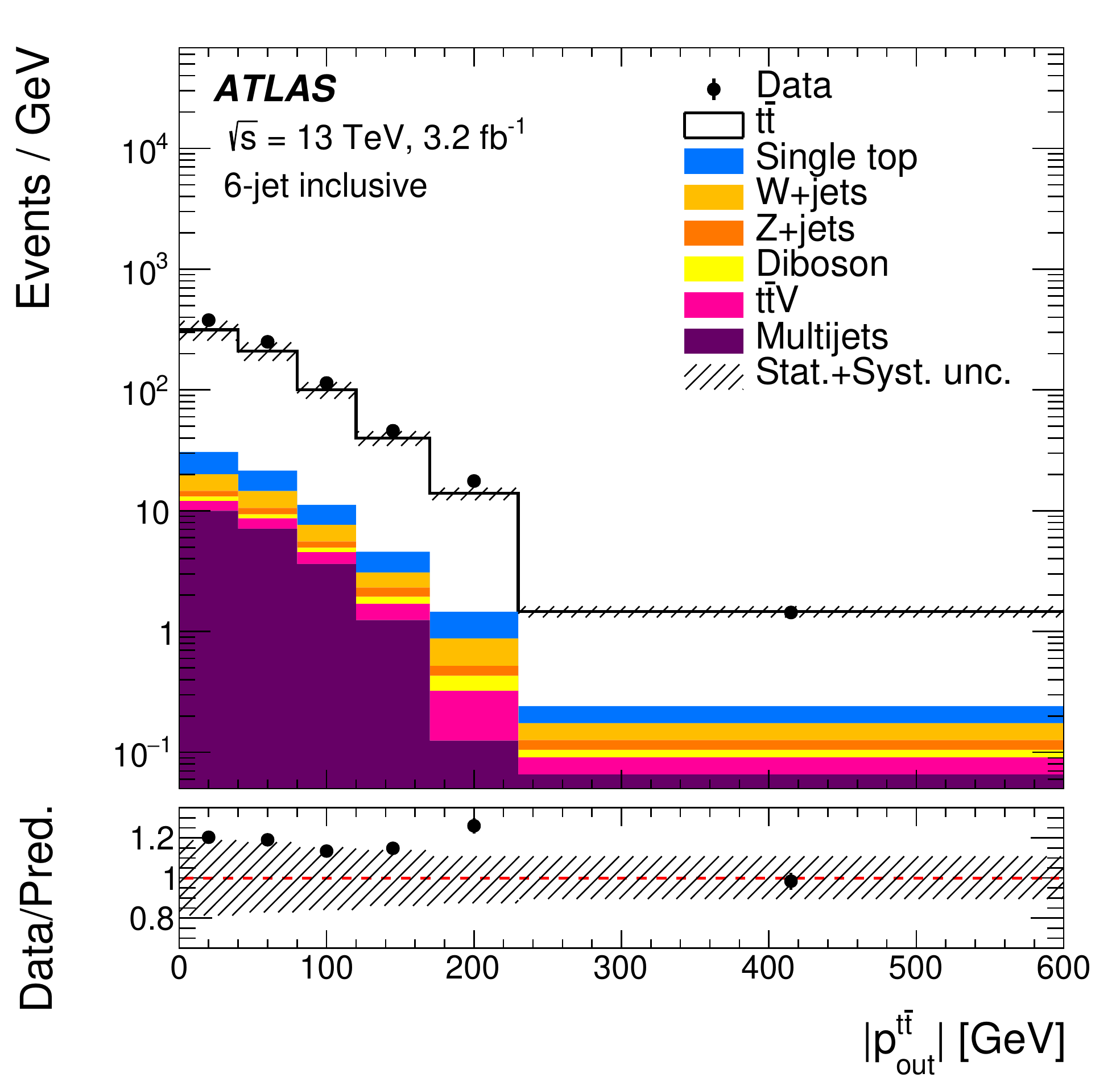}\label{fig:6j_pout_co}}
\caption{Distributions of $\absPoutttbar$ at reconstruction level: \subref{fig:4je_pout_co} 4-jet exclusive, \subref{fig:5j_pout_co} 5-jet exclusive and \subref{fig:6j_pout_co} 6-jet inclusive configurations. Data are compared to the sum of signal and background predictions using the nominal sample as the \ttbar{} signal model. The hatched area indicates the combined statistical and systematic uncertainties in the total prediction, excluding systematic uncertainties related to the modelling of the $\ttbar$ production.  Events beyond the range of the horizontal axis are included in the last bin.}
\label{Fig:reco_level_variables_pout}
\end{figure*}

\FloatBarrier
\section{Unfolding procedure}\label{sec:unfolding}

The measured differential cross sections are obtained from the reconstruction-level distributions using an unfolding technique which corrects for detector and reconstruction effects. The iterative Bayesian method~\cite{unfold:bayes} as implemented in RooUnfold~\cite{Adye:2011gm} is used.

The individual electron and muon channels have very similar corrections and give compatible results at reconstruction level. They are therefore combined by summing the distributions before the unfolding procedure.
 
For each observable, the unfolding procedure starts from the number of events at reconstruction level in bin $j$ of the distribution ($N_{\mathrm{reco}}^j$), after subtracting the background events estimated as described in Section~\ref{sec:BackgroundDetermination} ($N_{\mathrm{bg}}^j$). 
Next, the acceptance correction $f_{\mathrm{acc}}^j$ is defined as the ratio of the number of events passing both the particle- and reconstruction-level selections to the number of events passing the reconstruction-level selection. This factor corrects for events  that  are  generated outside the fiducial phase space region but pass the reconstruction-level selection.

The reconstruction-level objects used to reconstruct the top quarks are required to be angularly matched to the corresponding particle-level object as assigned by the pseudo-top algorithm. The jets assigned to the $W$ boson can be swapped. This requirement leads to a better correspondence between the particle and reconstruction levels. The matching requirement for the lepton, using the direction given by its associated track, is $\Delta R < 0.02$  while jets are required to be within $\Delta R < 0.35$.
The matching correction $f_{\mathrm{match}}^j$ is defined as the ratio of events matched among the events passing both the particle-level and reconstruction-level selections for the same number of jets; it corrects for events in which a match is not found.

The unfolding step uses a migration matrix ($\mathcal{M}$) derived from simulated \ttbar{} events which maps the binned particle-level events to the binned reconstruction-level events.
The probability for particle-level events to be reconstructed in the same bin is therefore represented by the elements on the diagonal, and the off-diagonal elements describe the fraction of particle-level events that migrate into other bins. Therefore, the elements of each row add up to unity (within rounding). The number of bins is optimised for maximum information extraction under stable unfolding conditions. This is achieved by requiring that closure and stress tests are satisfied without introducing any bias. The unfolding is performed using four iterations to balance the unfolding stability with respect to the previous iteration (below 0.1\%) and the growth of the statistical uncertainty. The effect of varying the number of iterations by one was found to be negligible. Finally, the efficiency $\epsilon$ is defined as the ratio of the number of matched events to the number of events passing the particle-level selection. This factor corrects for the inefficiency of the reconstruction.

The unfolding procedure for an observable $X$ at particle level is summarised by the following expression for the absolute differential cross section:

\begin{equation*}
\frac{{\mathrm{d}}\sigma^{\mathrm{fid}}}{{\mathrm{d}}X^i} \equiv \frac{1}{\mathcal{L} \cdot \Delta X^i} \cdot  \frac{1}{\epsilon^i} \cdot \sum_j \mathcal{M}_{ij}^{-1} \cdot f_{\mathrm{match}}^j \cdot  f_{\mathrm{acc}}^j \cdot \left(N_{\mathrm{reco}}^j - N_{\mathrm{bg}}^j\right)\hbox{,}
\end{equation*}

where the index $j$ labels bins at reconstruction level while the $i$ index labels bins at particle level; $\Delta X^i$ is the bin width while $\mathcal{L}$ is the integrated luminosity, and the Bayesian unfolding is symbolised by $\mathcal{M}_{ij}^{-1}$.
The integrated fiducial cross section is obtained by integrating the unfolded cross section over the bins, and its value is used to compute the normalised differential cross section: 

\begin{equation*}
\frac{1}{\sigma^{\mathrm{fid}}}\cdot\frac{{\mathrm{d}}\sigma^{\mathrm{fid}}}{{\mathrm{d}}X^i}\hbox{.}
\end{equation*}

The unfolding of the observables is carried out independently in each configuration taking into account the bin-to-bin correlations within the distributions but not across jet multiplicity bins or among different observables within one jet multiplicity. Events that have a different number of jets at particle level and reconstruction level do not enter any migration matrix but are considered by the acceptance correction.

\FloatBarrier
\section{Systematic uncertainties} \label{sec:Uncertainties}

This section describes the estimation of systematic uncertainties related to object reconstruction and calibration, MC event generator modelling and background estimation. The uncertainty in the unfolded distribution is evaluated as follows. The considered distribution is varied at reconstruction level, unfolded using corrections from the nominal $\ttbar$ signal sample, and the unfolded distribution is compared to the particle-level distribution. All reconstruction- and background-related systematic uncertainties are evaluated using the nominal event generator, while alternative event generators are employed to assess uncertainties in the $\ttbar$ system modelling as discussed in Sec.~\ref{sec:SigMod}. In these cases, the corrections derived from the event generator are used to unfold the reconstruction-level spectra of the alternative event generator.

The covariance matrix incorporating statistical and systematic uncertainties is obtained for each observable by summing two covariance matrices.
The first covariance matrix includes statistical and systematic uncertainties from detector effects and background estimation by using pseudo-experiments to combine the sources.  The second covariance matrix is derived by adding four separate covariance matrices corresponding to the effects of the signal modelling: event generator, parton shower and hadronisation, initial- and final-state radiation (ISR/FSR) and PDF uncertainties. The bin-to-bin correlation values are set to unity for all these matrices. 

The covariance matrices due to the statistical and systematic uncertainties are obtained for each observable by evaluating the covariance between the kinematic bins using pseudo-experiments. In particular, the correlations due to statistical fluctuations from the size of both the data sample and the simulated signal samples are evaluated by varying the event counts independently in every bin before unfolding, and then propagating the resulting variations through the unfolding. The full description of the method is provided in Ref.~\cite{TOPQ-2014-15}.

\subsection{Experimental uncertainties}\label{sec:DetMod}
The jet energy scale (JES) uncertainty is estimated using a combination of simulations, test-beam data and \textit{in situ} measurements~\cite{PERF-2012-01,PERF-2011-03,PERF-2011-05}. Additional contributions from jet-flavour composition, $\eta$-intercalibration, hadrons passing through the calorimeter without interacting (punch-through), single-particle response, calorimeter response to different jet flavours, and pile-up are considered, resulting in 19 eigenvector uncertainty components. The uncertainty in the jet energy resolution (JER) is obtained with an \textit{in situ} measurement of the jet response in dijet events~\cite{PERF-2011-04}. 

The efficiency to tag jets containing $b$-hadrons is corrected in simulated events by applying scale factors, extracted from a $\ttbar$ dilepton sample, to account for the residual difference between data and simulation. Scale factors are also applied for jets originating from light or charm quarks that are misidentified as $b$-jets.
The associated flavour-tagging uncertainties, split into eigenvector components, are computed by varying the scale factors within their uncertainties~\cite{ATLAS-CONF-2014-004,ATLAS-CONF-2012-043,ATLAS-CONF-2012-040}. 

The lepton reconstruction efficiency in simulated events is corrected by scale factors derived from measurements of these efficiencies in data using a control region enriched in $Z \to \ell^+ \ell^-$ events.
The lepton-trigger and reconstruction-efficiency scale factors, energy scale and resolution are varied within their uncertainties \cite{PERF-2014-04,PERF-2015-10}.

The uncertainty associated with $\met$ is calculated by propagating the energy scale and resolution uncertainties to all jets and leptons in the 
$\met$ calculation. Additional $\met$ uncertainties arising from energy deposits not associated with any reconstructed objects are also included \cite{PERF-2014-04}.

\subsection{Signal modelling uncertainties}\label{sec:SigMod}
Uncertainties in the signal modelling affect the kinematic properties of simulated $\ttbar$ events as well as reconstruction- and particle-level efficiencies. To assess the uncertainty related to the matrix-element model and matching algorithm used in the MC event generator for the $\ttbar$ signal process, events simulated with \mgamcatnlo{} + \herwigpp{} are unfolded using the migration matrix and correction factors derived from an alternative \PowHeg{}+\herwigpp{} sample. The difference between the unfolded distribution and the known particle level distribution of the \mgamcatnlo{}+\herwigpp{} sample is assigned as the uncertainty, which is then symmetrised. 

To assess the impact of different parton-shower models, events simulated with \PowHeg interfaced to \herwigpp{} are unfolded using the migration matrix and correction factors derived with the nominal sample. The difference between the unfolded distribution and the known particle-level distribution of the \PowHeg+\herwigpp{} sample is assigned as the relative uncertainty, which is then symmetrised.

To evaluate the uncertainties related to the modelling of the initial- and final-state gluon radiation (ISR/FSR), $\ttbar$ MC samples with modified ISR/FSR modelling are used. The MC samples used for the evaluation of this uncertainty are generated using the \PowHeg event generator interfaced to the \Pythia shower model, where the parameters are varied as described in Section~\ref{sec:DataSimSamples}. 
The impact of the uncertainty related to the PDF is assessed using the $\ttbar$ sample generated with a \mgamcatnlo{} interfaced to \herwigpp. PDF-varied corrections and response matrix for the unfolding procedure are obtained by reweighting the central PDF4LHC15 PDF set to the full set of its 30 eigenvectors as described in Ref.~\cite{PDF4LHC}. Using these corrections, the central \mgamcatnlo{}+\herwigpp{} distribution is unfolded, the relative difference is computed with respect to the expected central particle-level spectrum, and the total uncertainty is obtained by adding these relative differences in quadrature. In addition, the difference between the central PDF4LHC15 and CT10 is evaluated in a similar way and added in quadrature to the PDF uncertainty.

\subsection{Background modelling uncertainties}\label{sec:BkgMod}
Systematic uncertainties affecting the backgrounds evaluated with MC simulation are estimated using an alternative background MC sample produced by rescaling the nominal background sample. The alternative sample, instead of the nominal one, is subtracted from data. The uncertainty is evaluated as the difference between the unfolded distribution using the alternative background MC sample and the nominal one.

A $15$\% normalisation uncertainty is applied to the single-top quark background. This includes the uncertainty associated with the emission of additional radiation which is evaluated to be smaller than $15$\%. The $5$\% theoretical uncertainty in the normalisation is also included. 

In the case of the $Z$+jets and diboson backgrounds, the uncertainties include a contribution from the overall cross section normalisation as well as an additional 24\% uncertainty per additional jet~\cite{Alwall:2007fs}: 48\%, 72\% and 96\% in the 4-jet, 5-jet and 6-jet configurations, respectively. 

The systematic uncertainties due to the overall normalisation and the heavy-flavour fractions of $W$+jets events are obtained by varying the data-driven scale factors. The overall impact of these uncertainties is less than $2$\%. Each detector systematic uncertainty includes the impact of those on the $W$+jets estimate. In addition, a $24$\% uncertainty per radiated jet, as described for the $Z$+jets and diboson samples, is applied to the $W$+jets background uncertainty. 

The uncertainty in the background from non-prompt and fake leptons is evaluated by changing the selection used to define the control region and propagating the statistical uncertainty of the parametrisations of the efficiency to pass the tighter lepton requirements for real and fake leptons. 

In addition, an extra 50\% normalisation uncertainty is applied to this background to account for the remaining mis-modelling observed in various control regions. This systematic uncertainty also includes the impact of the normalisation on the estimation of the $W$+jets background.

\subsection{Size of the simulated samples and luminosity uncertainty}
Test distributions, created with independent Poisson fluctuations of the event count in each bin, are unfolded to account for the size of the simulated samples. The uncertainty is the standard deviation given by all unfolded distributions. 

The uncertainty in the integrated luminosity is 2.1\% and is derived, following techniques similar to those described in Ref.~\cite{DAPR-2013-01}, from the luminosity scale calibration using a~pair of $x$--$y$ beam-separation scans performed in August 2015.

\subsection{Summary plots of systematic uncertainties}

Figure~\ref{fig:unc_results:fiducial:topH} presents the uncertainties as a function of $\ptthad{}$ in the \ttbar{} fiducial phase space differential cross sections. The uncertainties are between 8\% and 25\% for the absolute cross sections and between 4\% and 9\% in almost the full range of the normalised cross sections. In all configurations the uncertainties are larger at the low and high ends of the spectrum; this shape is due to the combination of different components. The background and JES uncertainties are bigger at low value in $\pt$ and decrease with the $\pt$ while the signal uncertainties have the opposite behaviour. Comparing Figures~\ref{fig:unc_particle:topH_pt:abs:4je2bi},~\ref{fig:unc_particle:topH_pt:abs:5je2bi} and ~\ref{fig:unc_particle:topH_pt:abs:6ji2bi} shows that the JES uncertainty increases with the number of jets and is the dominant uncertainty in the 6-jet configuration. The uncertainties for the other observables have similar values and behaviour. In the 4-jet configuration, the dominant uncertainty is due to flavour-tagging. The total uncertainties are reduced for the normalised cross sections because of the cancelling out of correlated uncertainties, such as the flavour-tagging and the JES uncertainties as seen by comparing Figures~\ref{fig:unc_particle:topH_pt:rel:4je2bi} and ~\ref{fig:unc_particle:topH_pt:abs:4je2bi}.

\begin{figure*}[htbp]
\centering
\subfigure[]{  \includegraphics[width=0.45\textwidth]{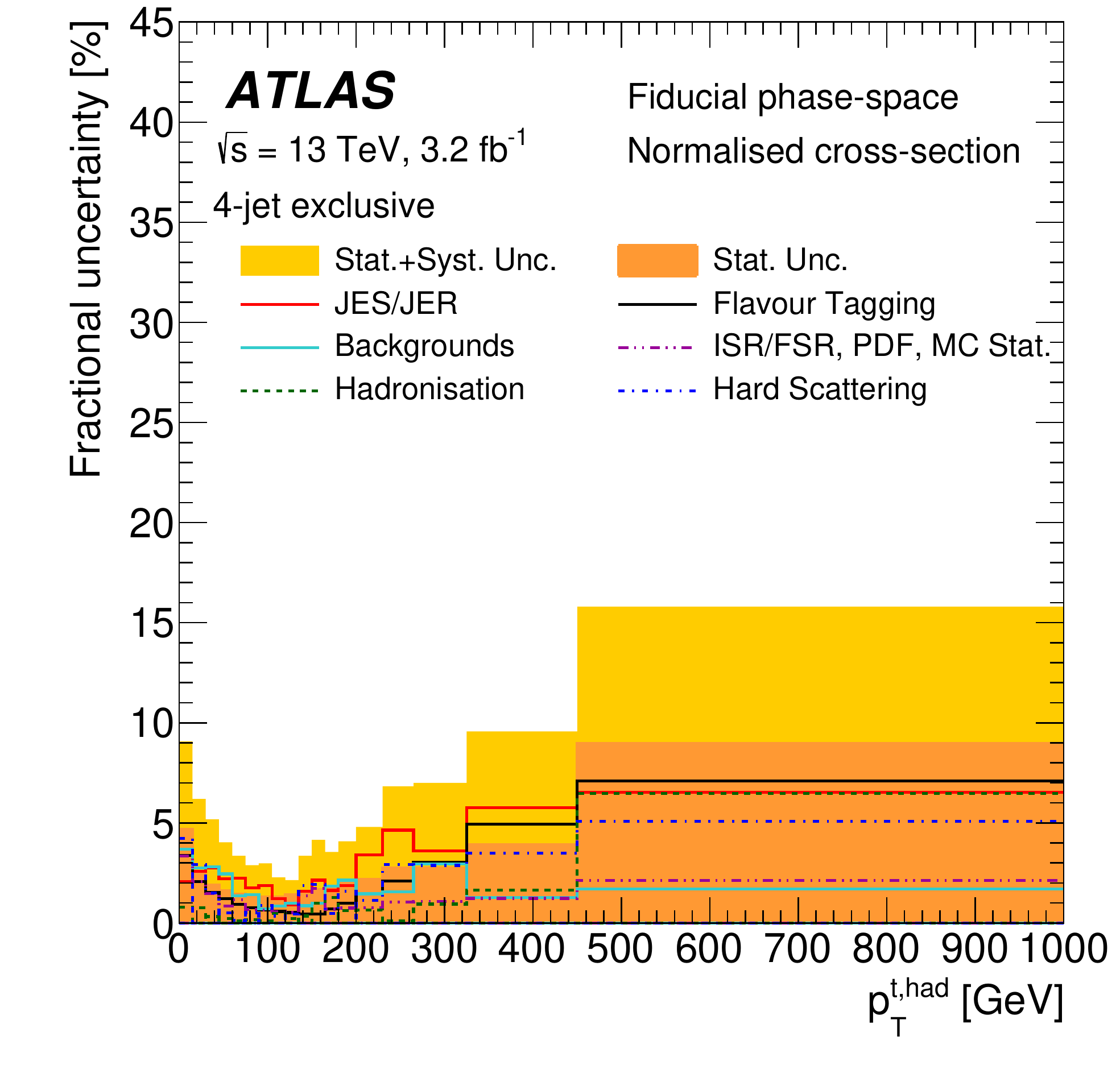}\label{fig:unc_particle:topH_pt:rel:4je2bi}}
\subfigure[]{  \includegraphics[width=0.45\textwidth]{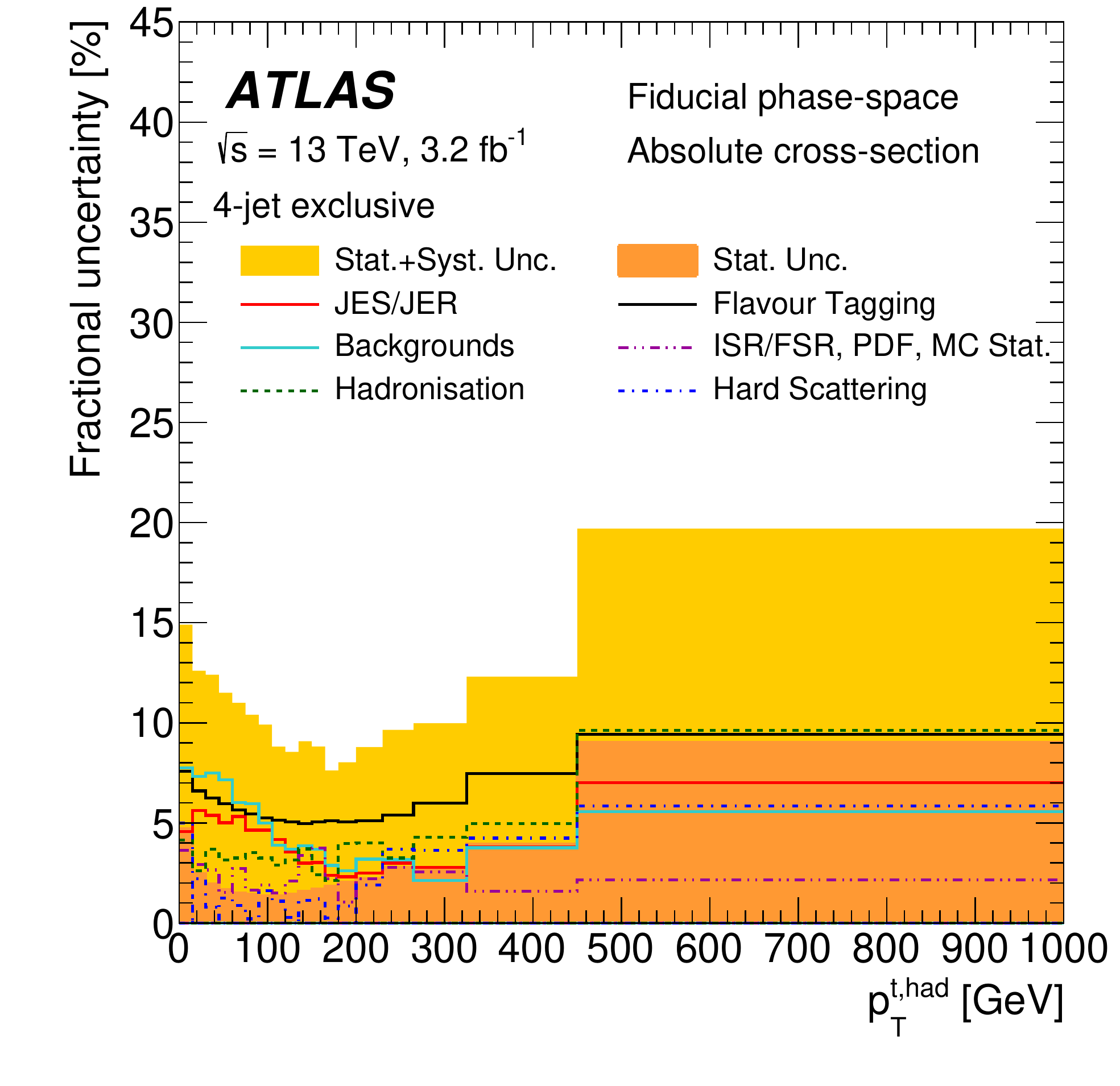}\label{fig:unc_particle:topH_pt:abs:4je2bi}}
\subfigure[]{  \includegraphics[width=0.45\textwidth]{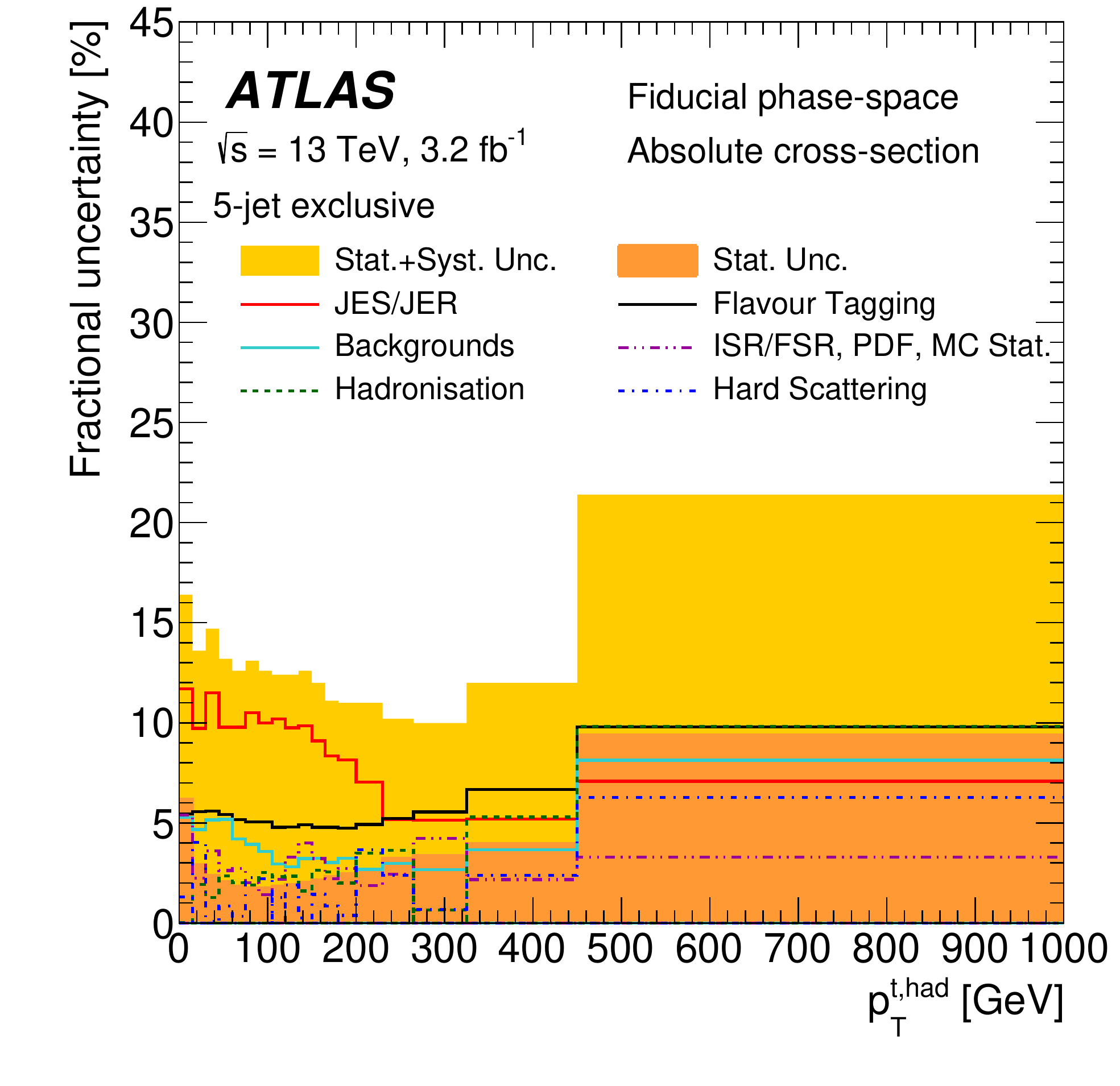}\label{fig:unc_particle:topH_pt:abs:5je2bi}}
\subfigure[]{  \includegraphics[width=0.45\textwidth]{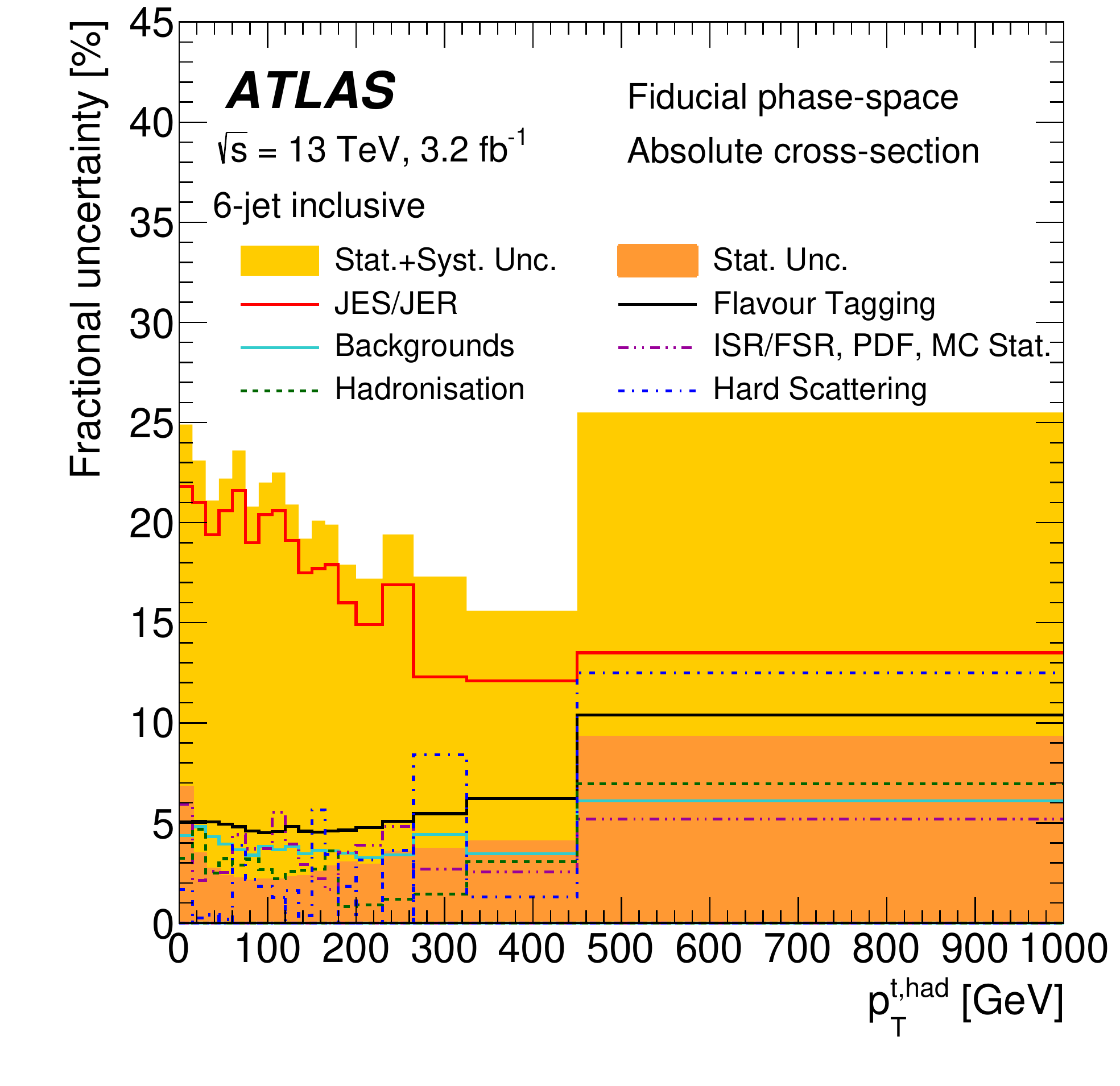}\label{fig:unc_particle:topH_pt:abs:6ji2bi}}
\caption{Uncertainties in the fiducial phase space differential cross sections as a function of $\ptthad{}$: normalised~\subref{fig:unc_particle:topH_pt:rel:4je2bi} in the 4-jet exclusive configuration; absolute~\subref{fig:unc_particle:topH_pt:abs:4je2bi} in the 4-jet exclusive,~\subref{fig:unc_particle:topH_pt:abs:5je2bi} 5-jet exclusive, and~\subref{fig:unc_particle:topH_pt:abs:6ji2bi} 6-jet inclusive configurations. The yellow bands indicate the total uncertainty in each bin. 
}
\label{fig:unc_results:fiducial:topH}
\end{figure*}

\FloatBarrier
\section{Results and comparisons with predictions} \label{sec:Results}

The measured differential cross sections as functions of $\ptthad$, $\ptttbar$ and $\absPoutttbar$ are shown in Figures~\ref{fig:unfolding:particle:topH_pt}--\ref{fig:unfolding:particle:pout} for the three configurations. All absolute differential cross sections are presented while only a selection of the normalised results is presented in which shape effects are more visible (this includes the $\ptthad$ results in the 4-jet configuration and the $\ptttbar$ and $\absPoutttbar$ results in the 6-jet configuration). Several MC predictions are compared to data; a subset of the most relevant predictions is shown in the figures while the compatibility to data is tested for a comprehensive list of MC predictions and shown in Tables~\ref{tab:modelcomparison_top_pt_abs}--\ref{tab:modelcomparison_pout_rel}. 

The level of agreement between the measured differential cross sections and the predictions is quantified using $\chi^2$ values which are evaluated employing the full covariance matrices of the uncertainties; the uncertainties in the theoretical predictions are not included in this evaluation. The $p$-values (probabilities that the $\chi^2$ is larger than or equal to the observed value) are then evaluated from the $\chi^2$ and the number of degrees of freedom (NDF). The detailed procedure for the calculation of the $\chi^2$ and $p$-values is described in Ref.~\cite{TOPQ-2016-01}.

The differential cross section as a function of $\ptthad$ is shown in Figure~\ref{fig:unfolding:particle:topH_pt}. All MC predictions underestimate (overestimate) the data at low (high) values of $\ptthad$; this tendency is reduced at higher jet multiplicity. This is consistent with the CMS results for the same observable and various jet multiplicities~\cite{CMS-TOP-16-008}. In addition, these results obtained here improve the understanding of similar effects observed in previous ATLAS analyses~\cite{TOPQ-2015-06,TOPQ-2016-01}; the effect is mainly due to events with exactly four jets.
The $\chi^2$ values for all predictions and all configurations are shown in Tables~\ref{tab:modelcomparison_top_pt_abs} and \ref{tab:modelcomparison_top_pt_rel} for the absolute and normalised differential cross sections, respectively. In general, all predictions are compatible with the data in the 5- and 6-jet configurations for both the absolute and normalised differential cross sections while there is tension for the 4-jet configuration, especially for the absolute differential cross section. 
The main exception is the prediction obtained from the \Powheg{}+\herwigpp{} calculation, which is inconsistent with the measured differential cross sections for the 4- and 6-jet configurations. 

The differential cross sections as a function of $\ptttbar$ for different jet multiplicities are shown in Figure~\ref{fig:unfolding:particle:tt_pt} and the $\chi^2$ values are presented in Tables~\ref{tab:modelcomparison_tt_pt_abs} and \ref{tab:modelcomparison_tt_pt_rel}. In general, good agreement is observed in the 4- and 5-jet configurations while there is some tension in the 6-jet configuration. However, the $\chi^2$ values show that the \mgamcatnlo{} event generator is not compatible with data in the 4- and 5-jet configurations in both the absolute and normalised differential cross sections. This was not observed in the measurement inclusive in the number of jets~\cite{TOPQ-2016-01} because different configurations are dominant at different values of $\ptttbar$. Indeed, the absolute cross section in the first two bins of the 4-jet configuration is larger than in the other configurations while the cross section in the last two bins is largest in the 6-jet configuration. Since the mis-modelling is observed in regions of \pt in which the cross section in that configuration is subdominant, it could not be observed in the previous measurement.
The \Powheg{}+\herwigpp{} prediction does not model the data well in all configurations. Furthermore, both \Powheg{}+\Pythia{} calculations with additional radiation (`radHi' and `Var3c Up') are not compatible with the data for the 4- and 6-jet configurations for the absolute differential cross sections as shown in Table~\ref{tab:modelcomparison_tt_pt_abs}. 

The differential cross sections as functions of $\absPoutttbar$ are shown in Figure~\ref{fig:unfolding:particle:pout} and confirm the mis-modelling of the \mgamcatnlo{} prediction for the 4- and 5-jet configurations observed for $\ptttbar$. The $p$-values shown in Tables~\ref{tab:modelcomparison_pout_abs} and ~\ref{tab:modelcomparison_pout_rel} drop significantly at higher jet multiplicity for all predictions. Several predictions are not compatible with the absolute cross sections in the 6-jet configuration but have better agreement with the normalised cross sections; nevertheless, some discrimination is still observed with the normalised cross sections. As before, the \Powheg{}+\herwigpp{} prediction is not compatible with data in the 5-jet configuration.

The complementarity of $\absPoutttbar$ and $\ptttbar$ is highlighted by the different agreement with data of the \Powheg{}+\HerwigSeven{} prediction in the 6-jet configuration; in $\absPoutttbar$ the agreement is poor ($p$-value of 0.06) while it is good in the $\ptttbar$ observable ($p$-value of 0.54). Contrariwise, in the 4-jet configuration the \Powheg{}+\PythiaSixP{} `radHi' prediction has a poor agreement in the $\ptttbar$ variable ($p$-value 0.01) while it is in good agreement in the $\absPoutttbar$ observable ($p$-value 0.32).

An example of the discriminating power of the analysis is given in Figure~\ref{fig:unfolding:special:hdamp}; several predictions with different values of the fragmentation and renormalisation scales and of the $h_{\mathrm{damp}}$ parameter are compared to the measured differential cross sections for the 6-jet configuration. From the comparison shown in Figure~\ref{fig:unfolding:special:hdamp}(a), it can be seen that among the three \PythiaSix{} predictions the best agreement is obtained by the `radLo' calculation which is tuned to yield a lower amount of gluon radiation. This sample has an $h_{\mathrm{damp}} = m_{t}$ and the factorisation and renormalisation scales increased by a factor of two compared to their nominal value. Since the $h_{\mathrm{damp}}$ parameter in the  `radLo' calculation is the same as the one in the nominal sample, it is possible to conclude that the reason for the different behaviour is due to the scale variation. A similar conclusion can be drawn for the comparison of the \Powheg{}+\PythiaEight{} sample in Figure~\ref{fig:unfolding:special:hdamp}(c) where the `Var3c Down' calculation shows the best agreement. Changing $h_{\mathrm{damp}}$ has a small impact as shown in the comparison of the two \Powheg{}+\PythiaEight{} predictions with different $h_{\mathrm{damp}}$ presented in Figure~\ref{fig:unfolding:special:hdamp}(b). The relative levels of agreement between data and the `radHi' and `radLo' predictions in the 6-jet configuration is opposite to what was observed in Ref.~\cite{TOPQ-2015-17} where the `radHi' prediction was observed to have a better agreement with data, e.g. for the jet multiplicity spectrum. This is not the only difference between the results of the two analyses; for example, in Ref.~\cite{TOPQ-2015-17} \mgamcatnlo{}+\herwigpp{} was compatible with data while it is not compatible in some of the combinations of variables and jet multiplicity considered in this paper. It is clear that MC models have difficulty describing the sets of observables listed in the two papers simultaneously, but it is hoped that the sensitivity of the measurements shown here with respect to various MC parameters will provide constraints for any future MC models.

Figure~\ref{fig:fit} shows the ratio of the data to the nominal prediction for the normalised $\ptthad$ and $\ptttbar$ differential cross sections for the three configurations. It can be seen that the differences between the data and the prediction are largest for the 4-jet configuration. The description of the 5- and 6-jet configurations by the prediction is slightly better. For $\ptttbar$, the conclusions are less clear, and a reduction of the uncertainties would help to discriminate between the different predictions.  
\clearpage

\begin{figure}[htb]
\begin{center}
\subfigure[]{ \includegraphics[width=0.45\textwidth]{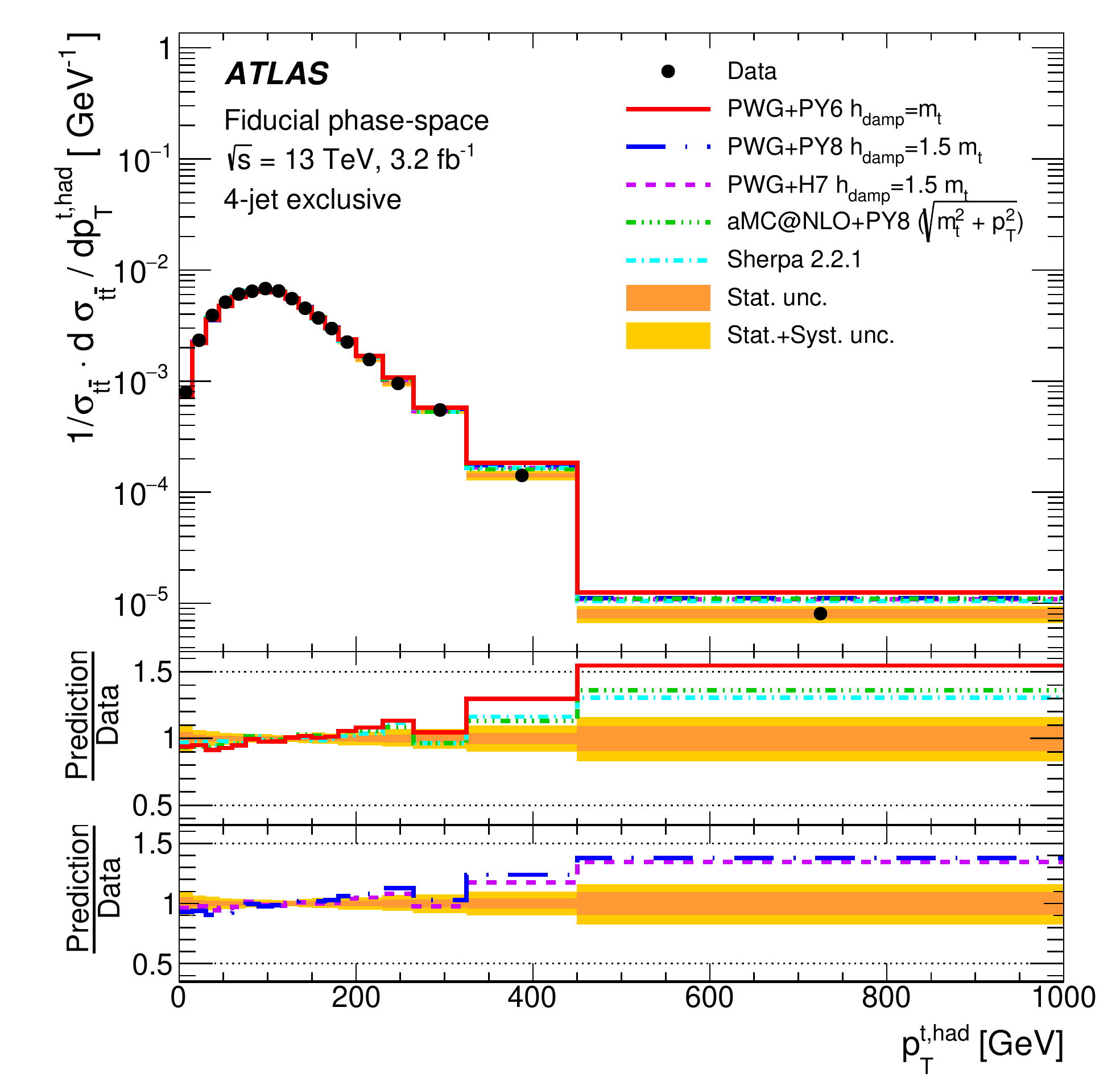}\label{fig:4je2bi_top_had_rel}}
\subfigure[]{ \includegraphics[width=0.45\textwidth]{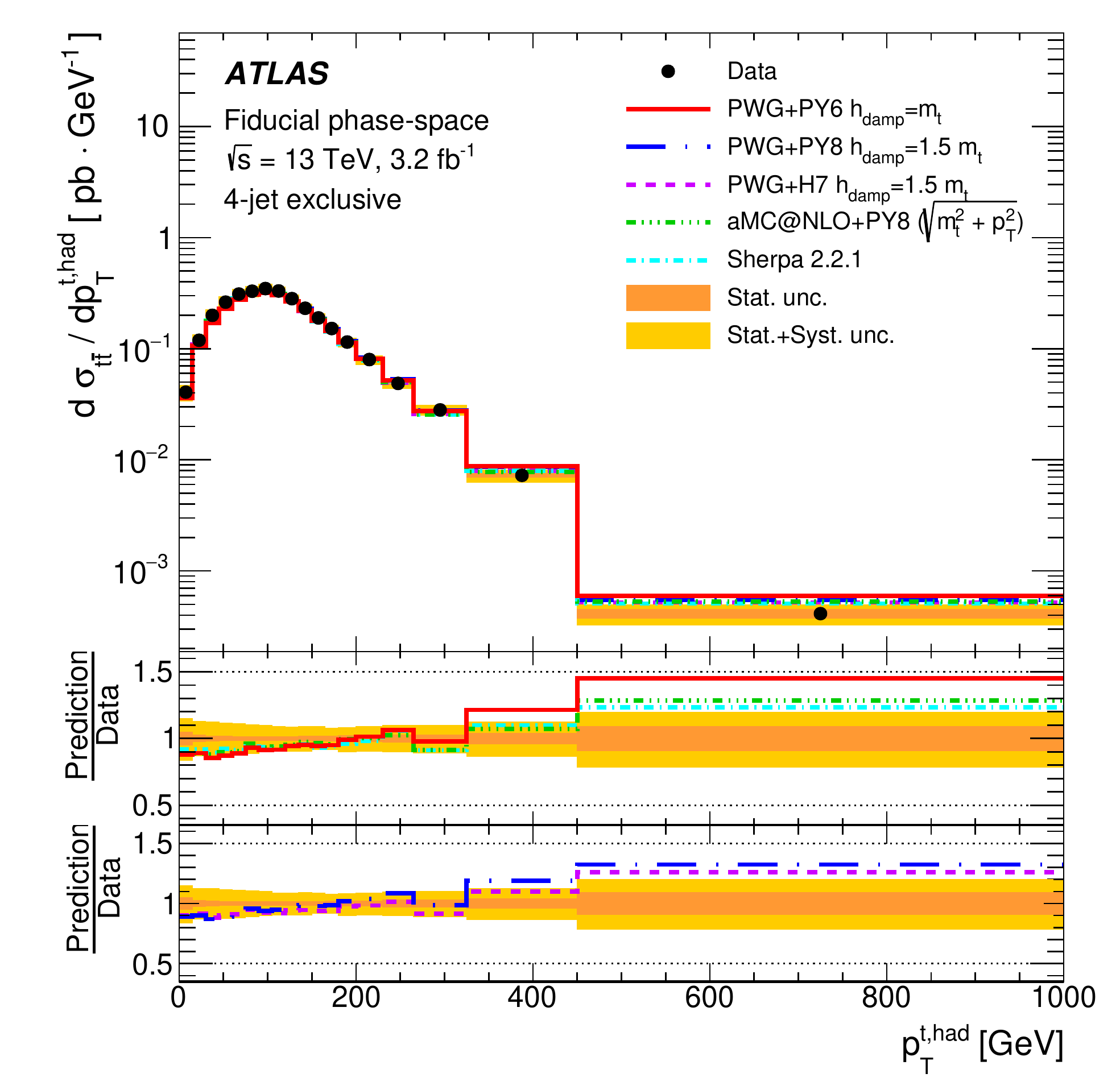}\label{fig:4je2bi_top_had_abs}}
 \subfigure[]{ \includegraphics[width=0.45\textwidth]{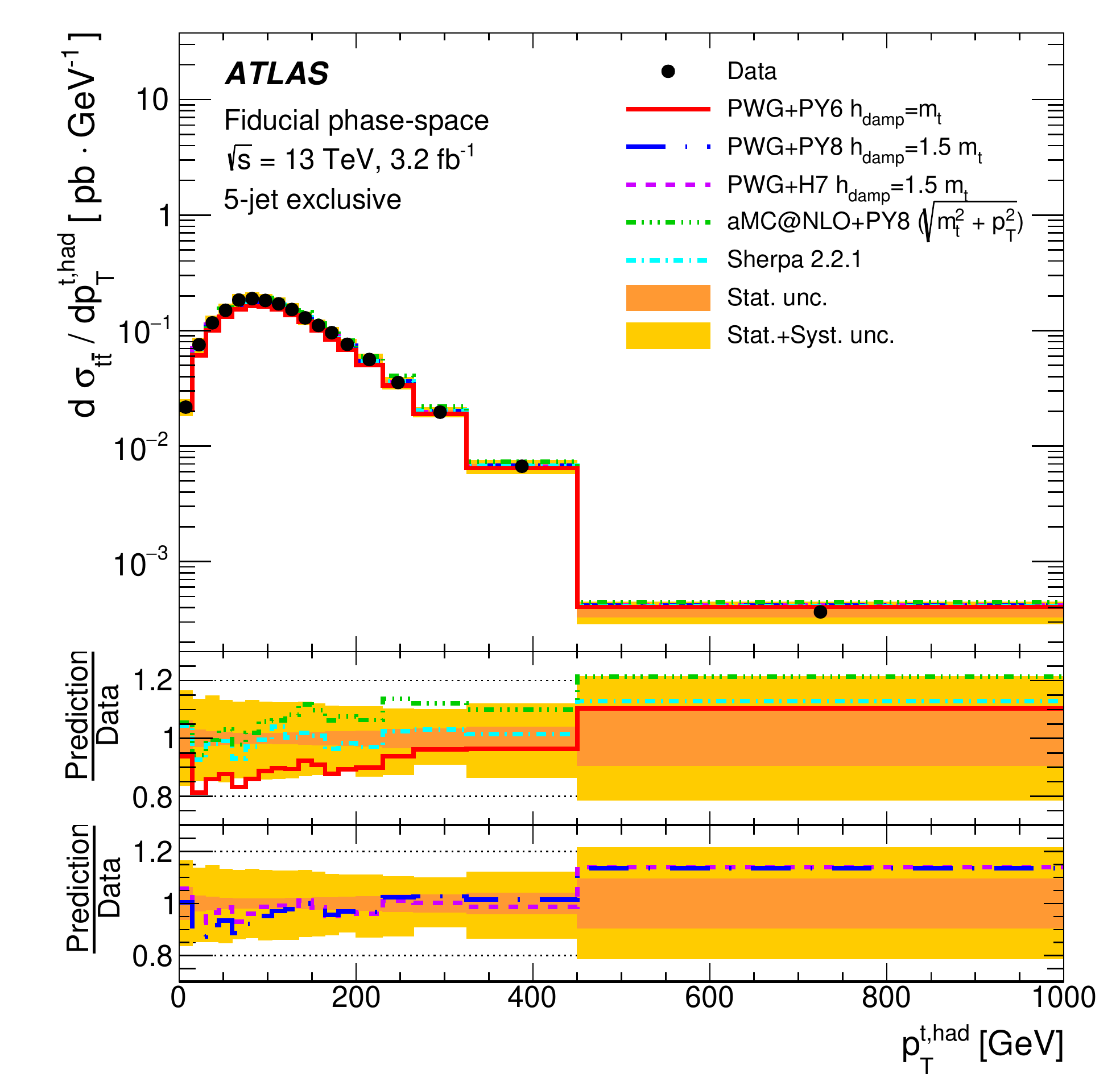}\label{fig:5je2bi_top_had_abs}}
 \subfigure[]{ \includegraphics[width=0.45\textwidth]{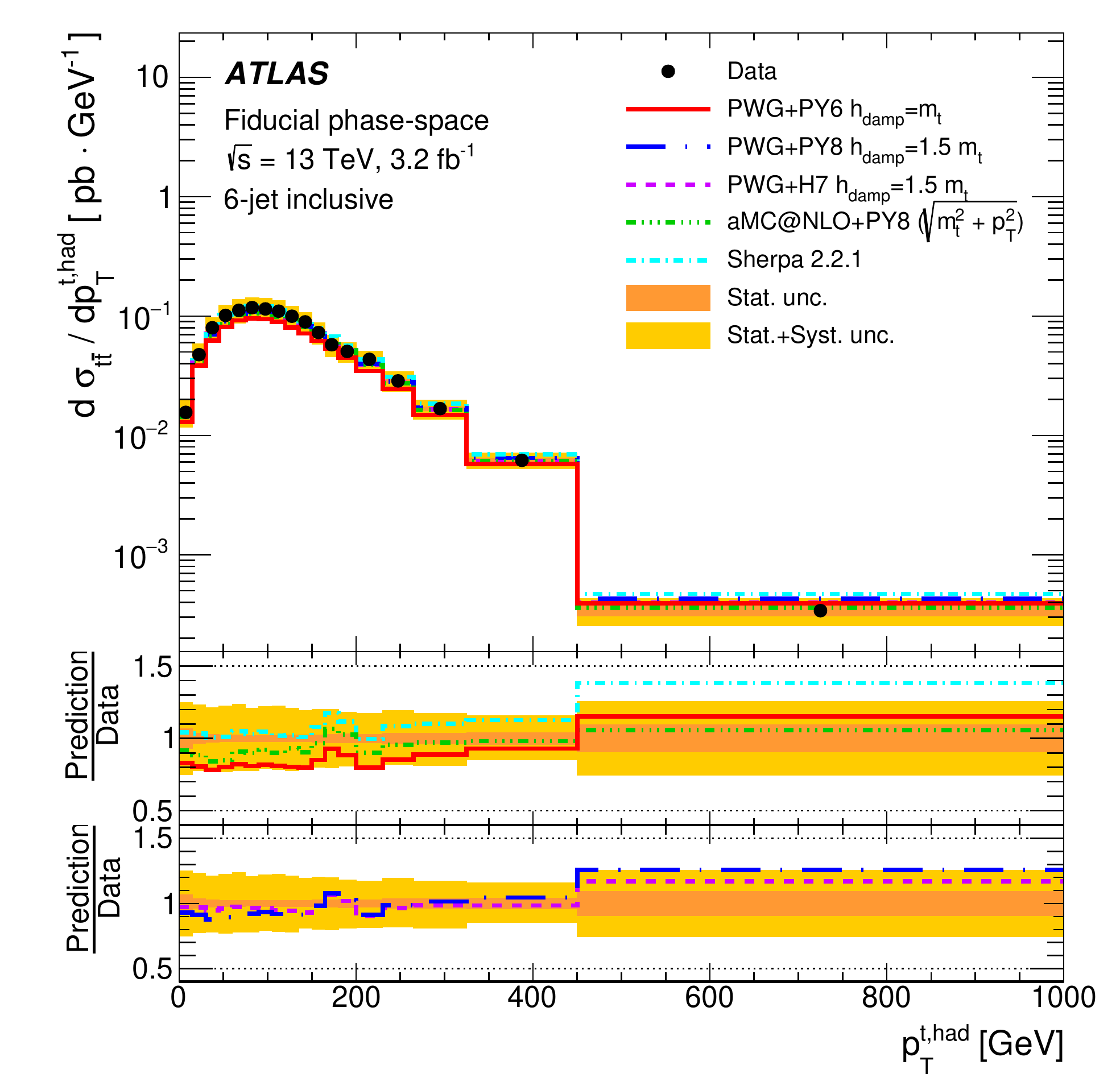}\label{fig:6ji2bi_top_had_abs}}
\caption{\small {Differential cross sections in the fiducial phase space as a function of $\ptthad{}$: normalised~\subref{fig:4je2bi_top_had_rel} in the 4-jet exclusive configuration, absolute~\subref{fig:4je2bi_top_had_abs} in the 4-jet exclusive,~\subref{fig:5je2bi_top_had_abs} 5-jet exclusive and~\subref{fig:6ji2bi_top_had_abs} 6-jet inclusive configurations. The shaded area represents the total statistical and systematic uncertainties.}}
\label{fig:unfolding:particle:topH_pt}
\end{center}
\end{figure}

\begin{figure}[htb]
\begin{center}
\subfigure[]{ \includegraphics[width=0.45\textwidth]{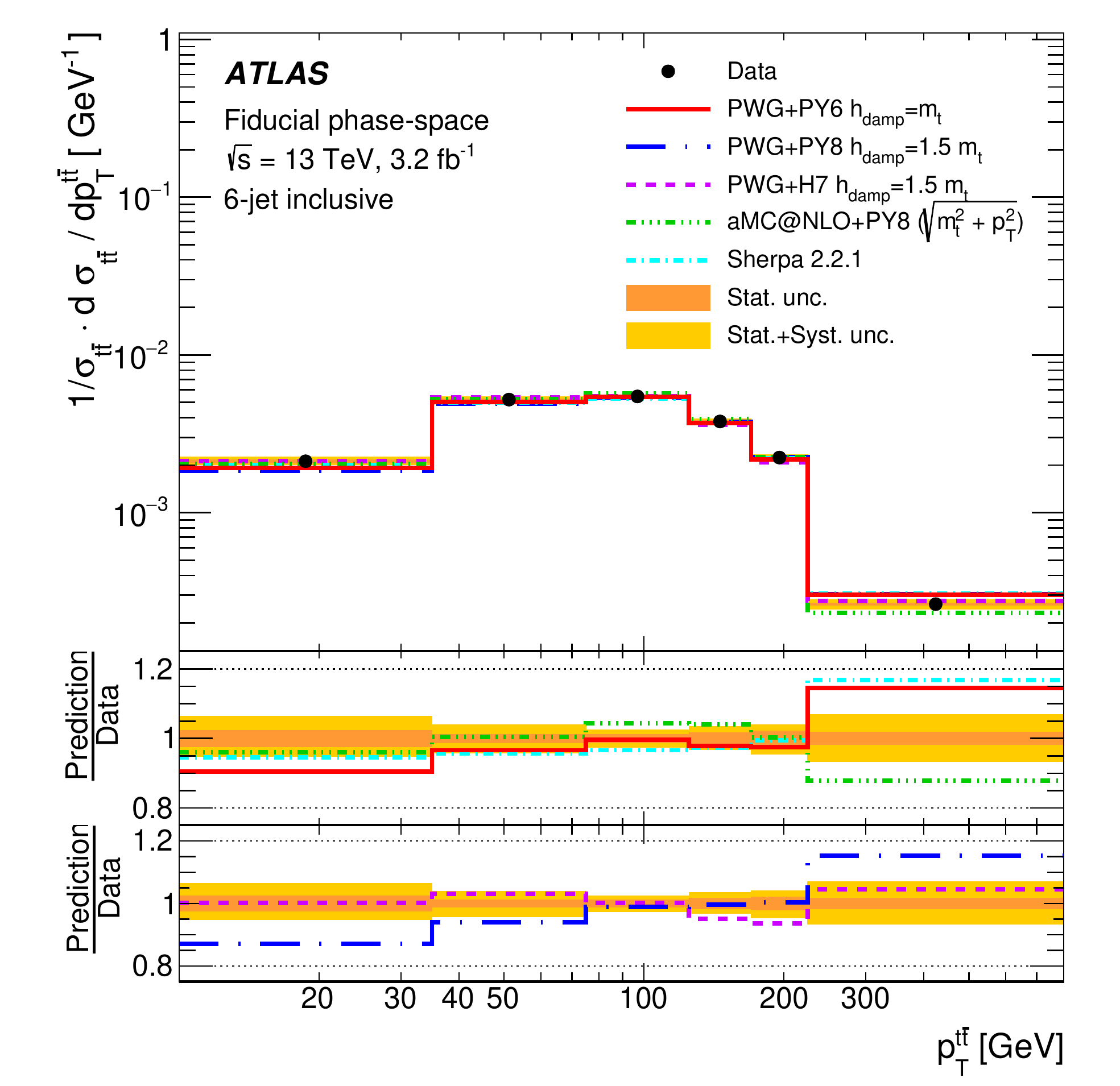}\label{fig:6ji2bi_tt_pt_rel}}
\subfigure[]{ \includegraphics[width=0.45\textwidth]{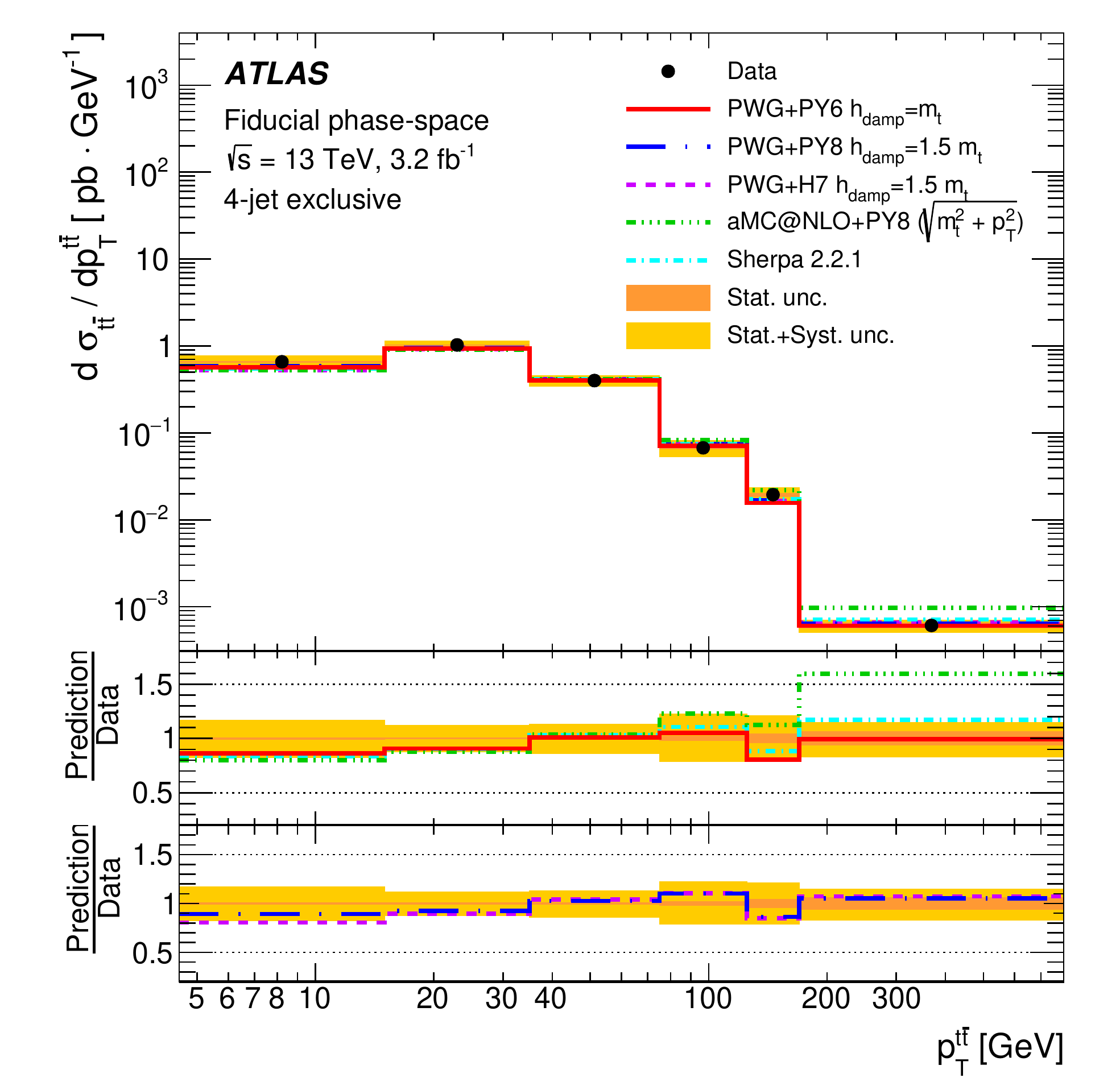}\label{fig:4je2bi_tt_pt_abs}}
 \subfigure[]{ \includegraphics[width=0.45\textwidth]{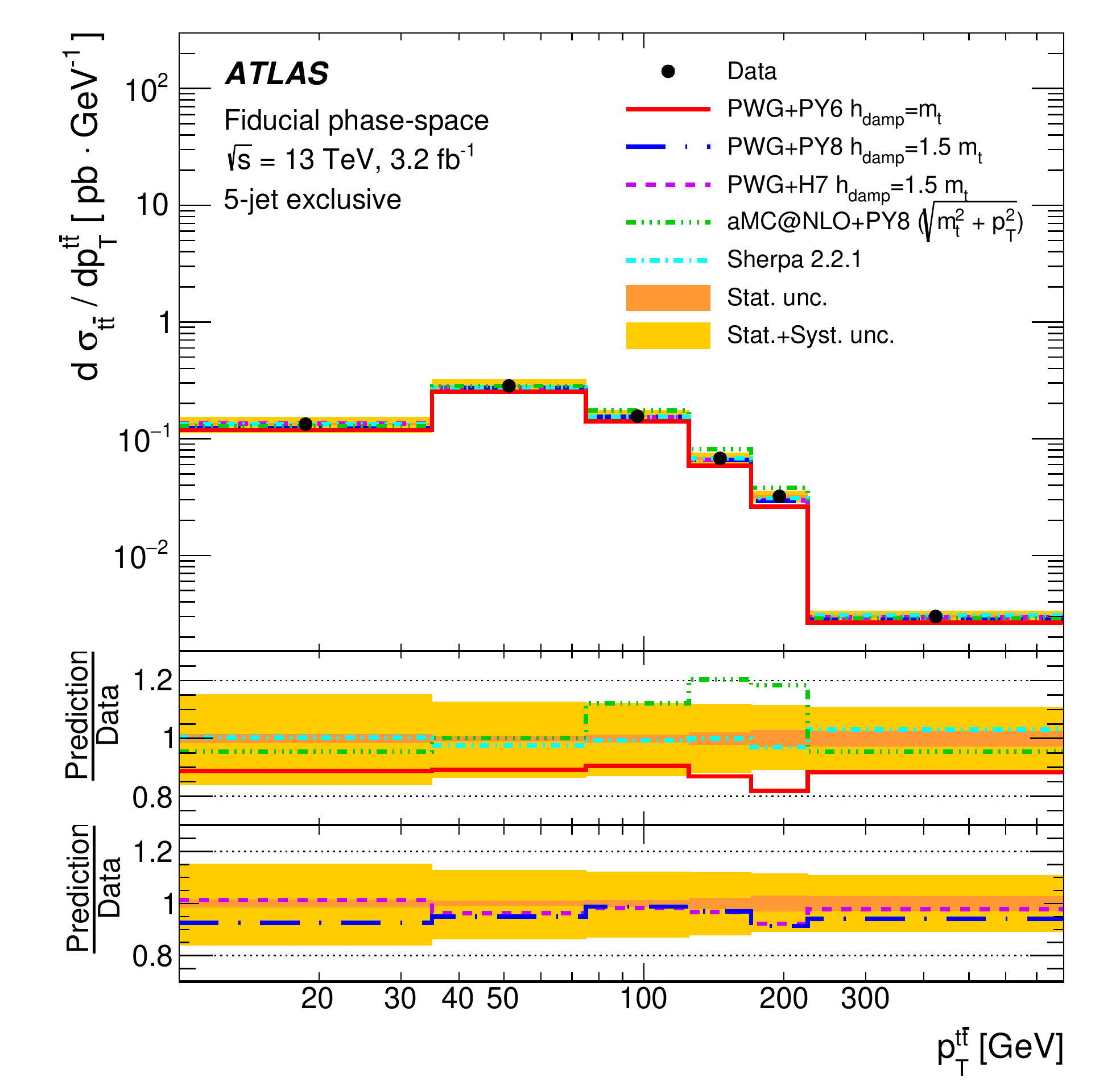}\label{fig:5je2bi_tt_pt_abs}}
 \subfigure[]{ \includegraphics[width=0.45\textwidth]{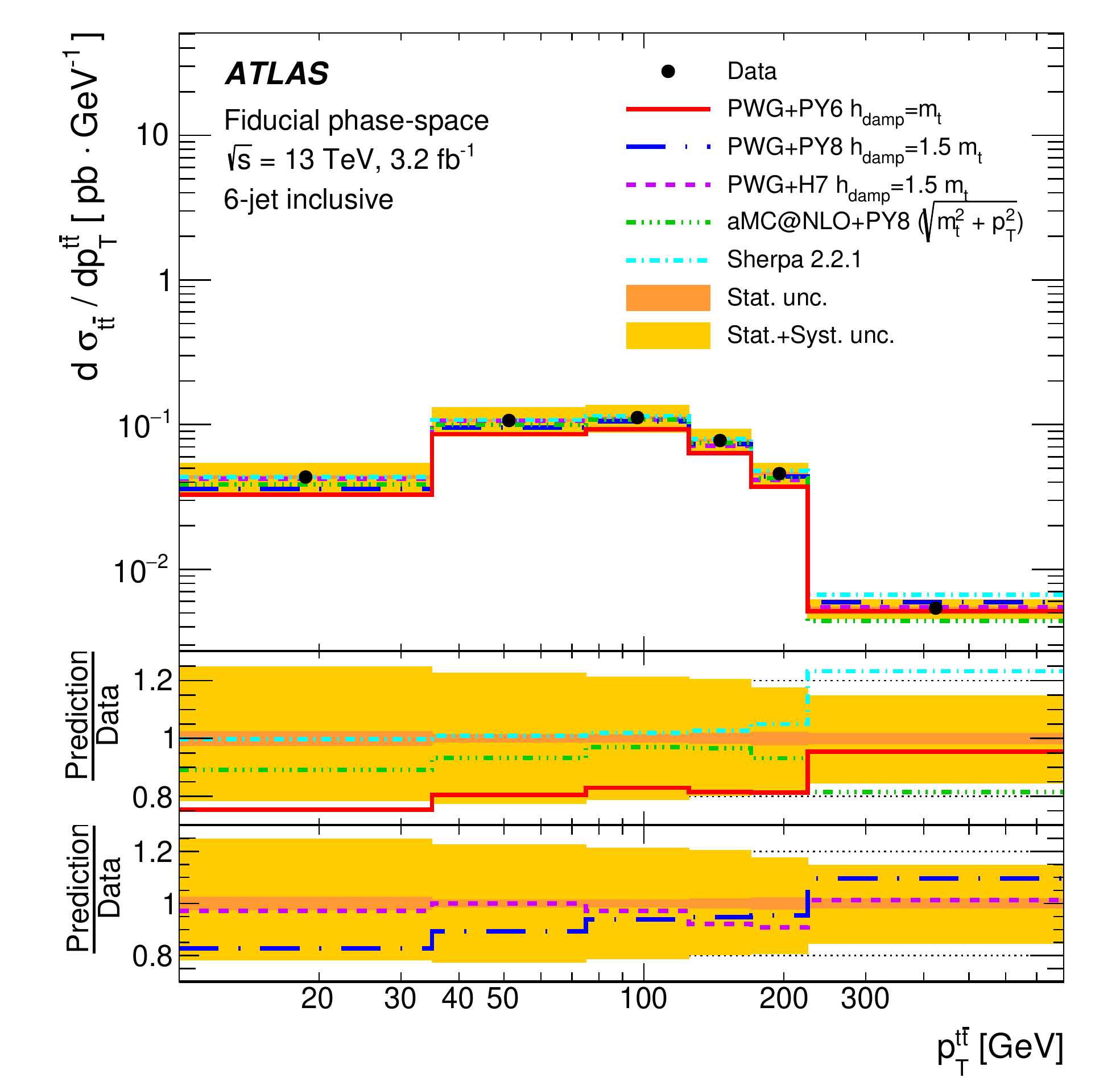}\label{fig:6ji2bi_tt_pt_abs}}
\caption{\small {Differential cross sections in the fiducial phase space as a function of $\ptttbar{}$: normalised~\subref{fig:6ji2bi_tt_pt_rel} in the 6-jet inclusive configuration, absolute~\subref{fig:4je2bi_tt_pt_abs} in the 4-jet exclusive,~\subref{fig:5je2bi_tt_pt_abs} 5-jet exclusive and~\subref{fig:6ji2bi_tt_pt_abs} 6-jet inclusive configurations. The shaded area represents the total statistical and systematic uncertainties.}}
\label{fig:unfolding:particle:tt_pt}
\end{center}
\end{figure}

\begin{figure}[htb]
\begin{center}
\subfigure[]{\includegraphics[width=0.45\textwidth]{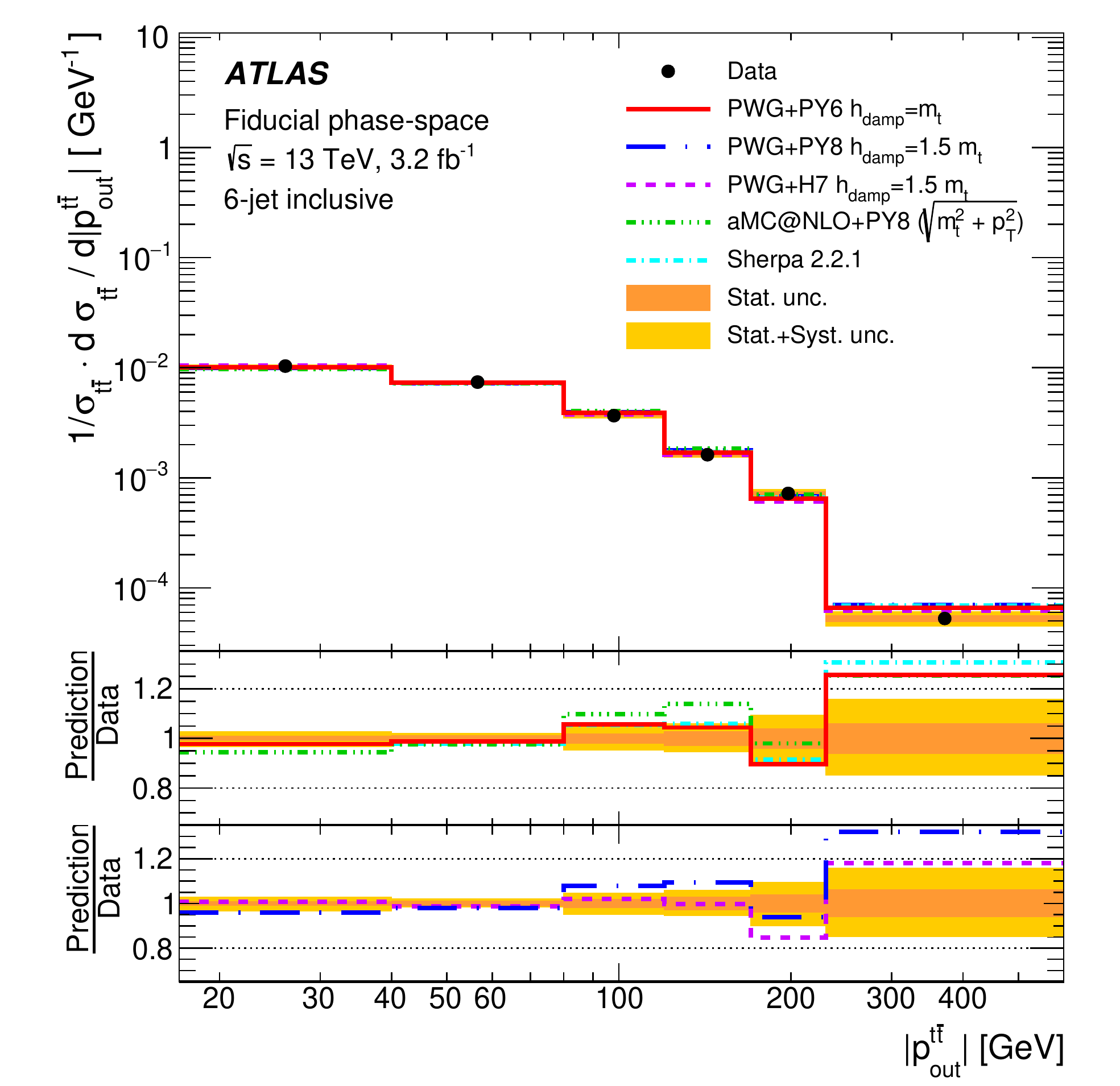}\label{fig:6ji2bi_pout_rel}}
\subfigure[]{\includegraphics[width=0.45\textwidth]{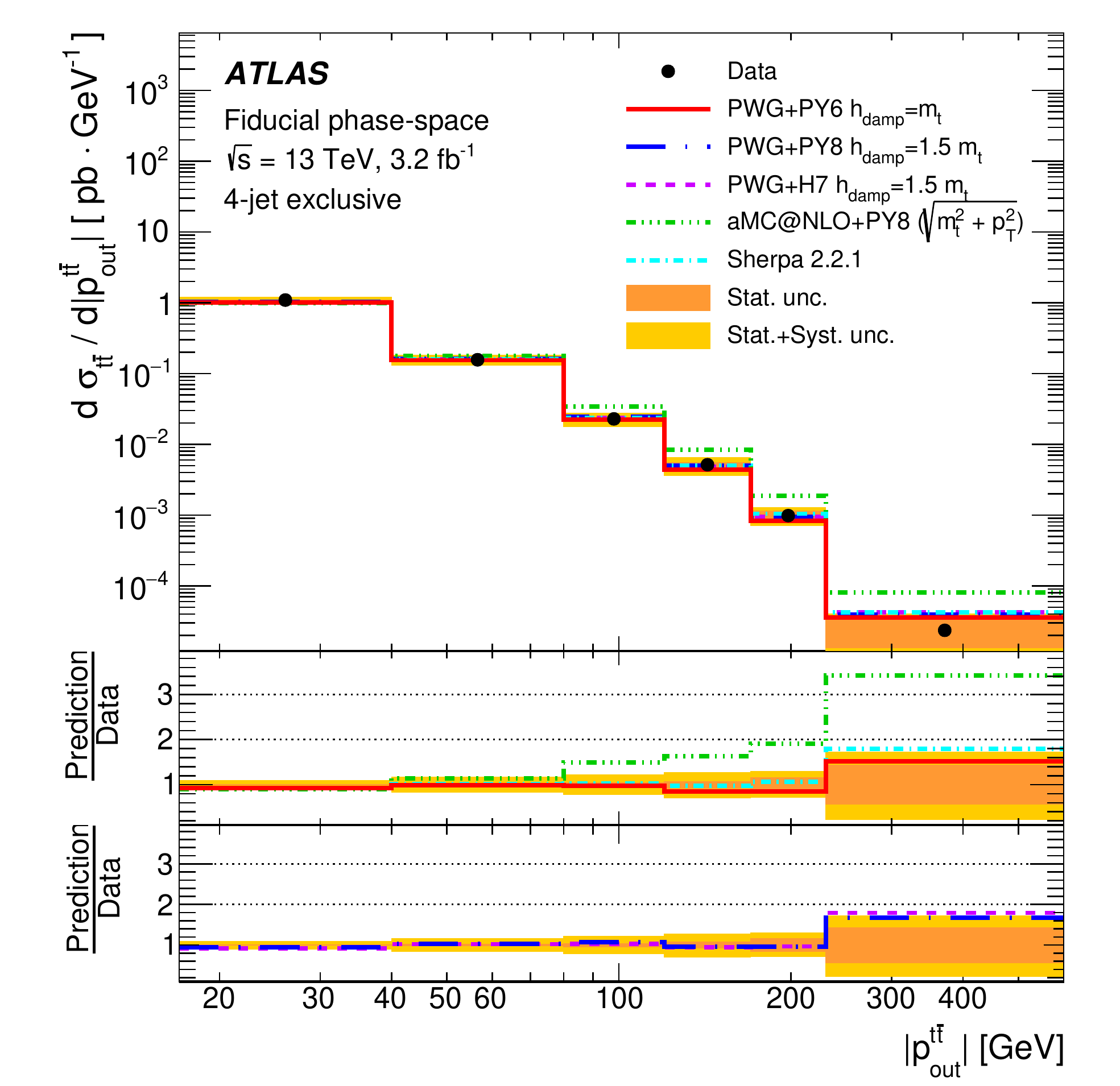}\label{fig:4je2bi_pout_abs}}
 \subfigure[]{\includegraphics[width=0.45\textwidth]{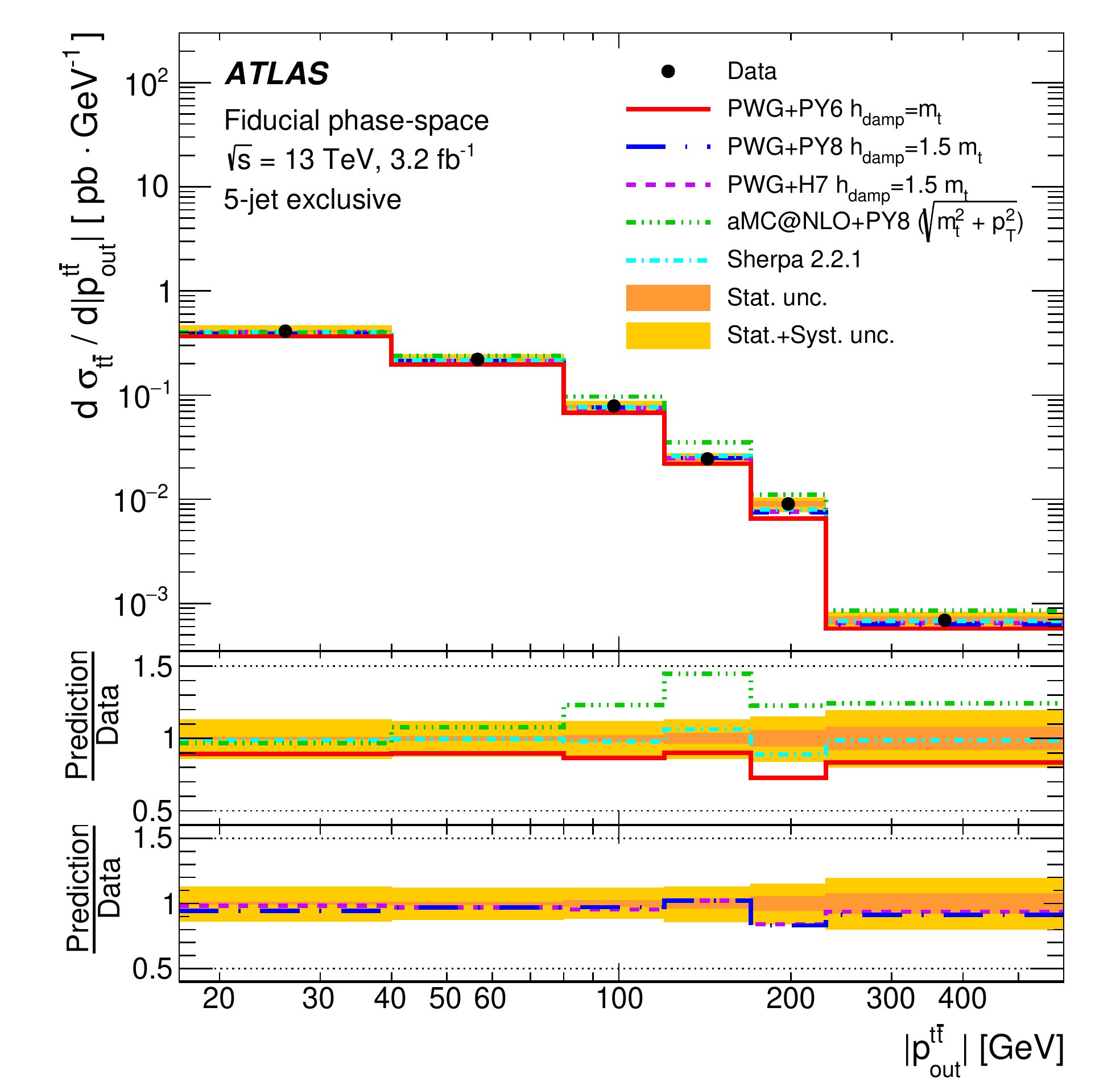}\label{fig:5je2bi_pout_abs}}
 \subfigure[]{\includegraphics[width=0.45\textwidth]{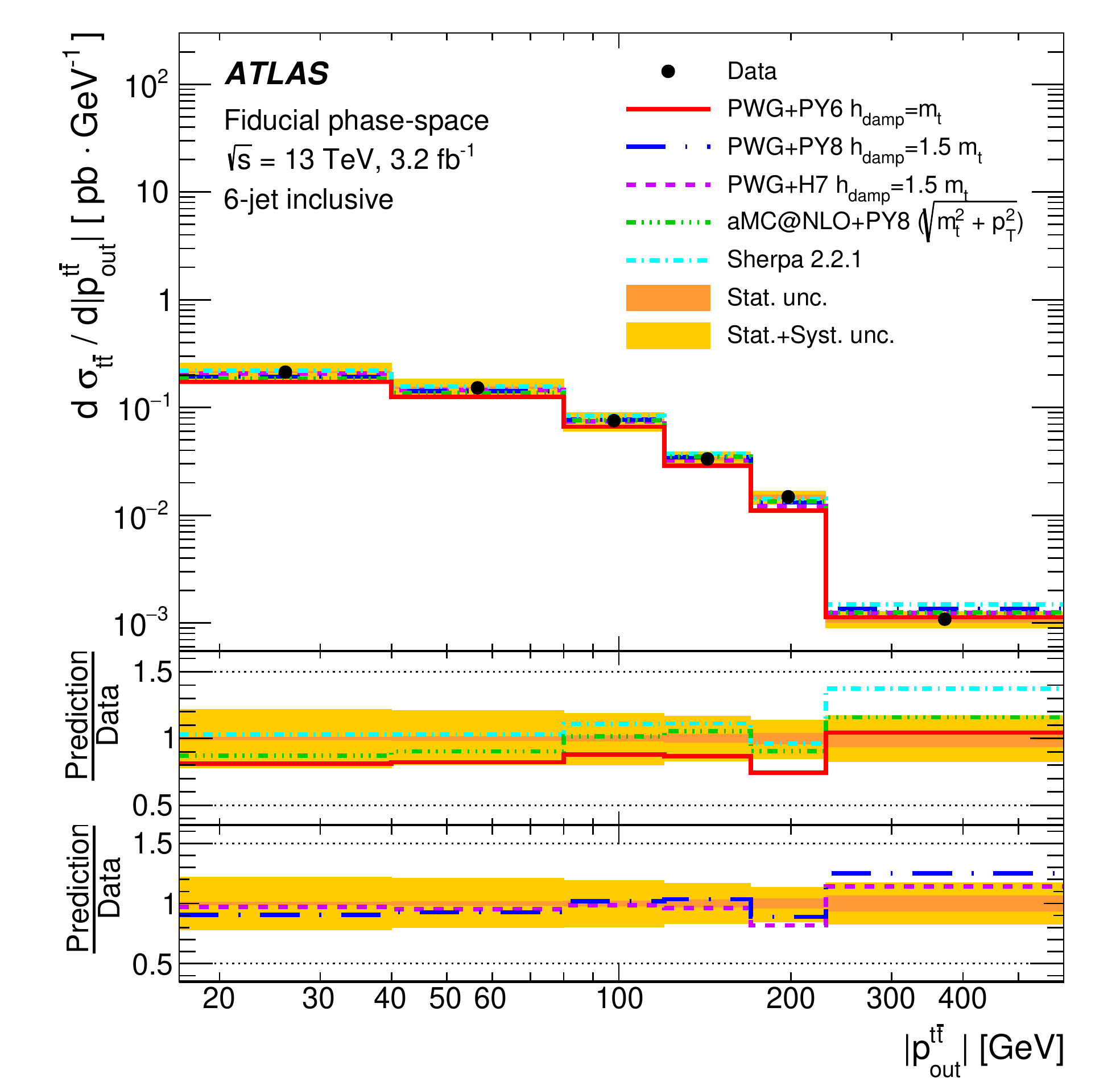}\label{fig:6ji2bi_pout_abs}}
\caption{\small {Differential cross sections in the fiducial phase space as a function of $\absPoutttbar{}$: normalised~\subref{fig:6ji2bi_pout_rel} in the 6-jet inclusive configuration, absolute~\subref{fig:4je2bi_pout_abs} in the 4-jet exclusive,~\subref{fig:5je2bi_pout_abs} 5-jet exclusive and~\subref{fig:6ji2bi_pout_abs} 6-jet inclusive configurations. The shaded area represents the total statistical and systematic uncertainties.}}
\label{fig:unfolding:particle:pout}
\end{center}
\end{figure}

%%%%%%%%%%%%%%%%%%%%%%%%%%%%
\begin{table}[!ht]
{\small \noindent\resizebox{\textwidth}{!}{
\setlength\arrayrulewidth{0.7pt}
\begin{tabular}{|l | r <{/\kern-\tabcolsep}>{\kern-\tabcolsep} l c | r <{/\kern-\tabcolsep}>{\kern-\tabcolsep} l c | r <{/\kern-\tabcolsep}>{\kern-\tabcolsep} l c|}
\hline
& \multicolumn{3}{c|}{4-jet exclusive} & \multicolumn{3}{c|}{5-jet exclusive} & \multicolumn{3}{c|}{6-jet inclusive} \\
& \multicolumn{2}{c}{$\chi^{2}$/NDF} &  ~$p$-value  & \multicolumn{2}{c}{$\chi^{2}$/NDF} &  ~$p$-value  & \multicolumn{2}{c}{$\chi^{2}$/NDF} &  ~$p$-value  \\
\hline
\rowcolor{Gray}
\Powheg{}+\PythiaSix{} & {\ } 28.9 & 18 & 0.05  &  {\ } 13.0 & 18 & 0.79  &  {\ } 13.0 & 18 & 0.79 \\
\Powheg{}+\PythiaSix{}\ (radHi) & {\ } 29.2 & 18 & 0.05  &  {\ } 14.7 & 18 & 0.68  &  {\ } 17.2 & 18 & 0.51 \\
\rowcolor{Gray}
\Powheg{}+\PythiaSix{}\ (radLo) & {\ } 32.5 & 18 & 0.02  &  {\ } 14.3 & 18 & 0.71  &  {\ } 13.9 & 18 & 0.74 \\
\Powheg{}+\PythiaEight{}\ ($h_\mathrm{damp} = m_{t}$) & {\ } 25.2 & 18 & 0.12  &  {\ } 14.7 & 18 & 0.68  &  {\ } 15.7 & 18 & 0.61 \\
\rowcolor{Gray}
\Powheg{}+\PythiaEight{}\ ($h_\mathrm{damp} =1.5\, m_{t}$)\ & {\ } 22.7 & 18 & 0.20  &  {\ } 13.3 & 18 & 0.77  &  {\ } 16.3 & 18 & 0.57 \\

\Powheg{}+\PythiaEight{}\ (Var3c Up) ($h_\mathrm{damp} = 3\,m_{t}$) & {\ } 20.0 & 18 & 0.33  &  {\ } 14.5 & 18 & 0.70  &  {\ } 23.9 & 18 & 0.16 \\
\rowcolor{Gray}
\Powheg{}+\PythiaEight{}\ (Var3c Down) ($h_\mathrm{damp} = 1.5\,m_{t}$) & {\ } 24.7 & 18 & 0.13  &  {\ } 14.7 & 18 & 0.68  &  {\ } 13.1 & 18 & 0.79 \\
\Powheg{}+\HerwigSeven{}\ & {\ } 20.8 & 18 & 0.29  &  {\ } 12.0 & 18 & 0.85  &  {\ } 12.4 & 18 & 0.82 \\
\rowcolor{Gray}
\Powheg{}+\herwigpp{} & {\ } 37.1 & 18 & \hspace{-0.1 in} <0.01  &  {\ } 27.7 & 18 & 0.07  &  {\ } 38.7 & 18 & \hspace{-0.1 in} <0.01 \\
\mgamcatnlo{}+\herwigpp{} & {\ } 25.7 & 18 & 0.11  &  {\ } 11.1 & 18 & 0.89  &  {\ } 20.3 & 18 & 0.32 \\
\rowcolor{Gray}
\mgamcatnlo{}+\PythiaEight{}\ ($H_{\mathrm T}/2$) & {\ } 22.9 & 18 & 0.19  &  {\ } 21.2 & 18 & 0.27  &  {\ } 17.7 & 18 & 0.47 \\
\mgamcatnlo{}+\PythiaEight{}\ ($\sqrt{m_{t}^{2}+p_{\mathrm T}^{2}}$) & {\ } 25.4 & 18 & 0.11  &  {\ } 19.3 & 18 & 0.37  &  {\ } 23.1 & 18 & 0.18 \\
\rowcolor{Gray}
\textsc{Sherpa 2.2.1} & {\ } 24.7 & 18 & 0.14  &  {\ } 18.3 & 18 & 0.43  &  {\ } 18.3 & 18 & 0.44 \\
\hline
\end{tabular}}
}
\caption{Comparison of the measured fiducial phase space absolute differential cross sections as a function of $\ptthad{}$ and the predictions from several MC generators in different $n$-jet configurations. For each prediction a $\chi^2$ and a $p$-value are calculated using the covariance matrix of the measured spectrum. The number of degrees of freedom (NDF) is equal to the number of bins in the distribution.}
\label{tab:modelcomparison_top_pt_abs}
\end{table}

\begin{table}[!hb]
{\small \noindent\resizebox{\textwidth}{!}{
\setlength\arrayrulewidth{0.7pt}
\begin{tabular}{|l | r <{/\kern-\tabcolsep}>{\kern-\tabcolsep} l c | r <{/\kern-\tabcolsep}>{\kern-\tabcolsep} l c | r <{/\kern-\tabcolsep}>{\kern-\tabcolsep} l c|}
\hline
& \multicolumn{3}{c|}{4-jet exclusive} & \multicolumn{3}{c|}{5-jet exclusive} & \multicolumn{3}{c|}{6-jet inclusive} \\
& \multicolumn{2}{c}{$\chi^{2}$/NDF} &  ~$p$-value  & \multicolumn{2}{c}{$\chi^{2}$/NDF} &  ~$p$-value  & \multicolumn{2}{c}{$\chi^{2}$/NDF} &  ~$p$-value  \\
\hline
\rowcolor{Gray}
\Powheg{}+\PythiaSix{} & {\ } 23.4 & 17 & 0.14  &  {\ } 14.1 & 17 & 0.66  &  {\ } 14.8 & 17 & 0.61 \\
\Powheg{}+\PythiaSix{}\ (radHi) & {\ } 23.4 & 17 & 0.14  &  {\ } 14.9 & 17 & 0.60  &  {\ } 15.9 & 17 & 0.53 \\
\rowcolor{Gray}
\Powheg{}+\PythiaSix{}\ (radLo) & {\ } 25.6 & 17 & 0.08  &  {\ } 16.5 & 17 & 0.49  &  {\ } 16.5 & 17 & 0.49 \\
\Powheg{}+\PythiaEight{}\ ($h_\mathrm{damp} = m_{t}$) & {\ } 22.7 & 17 & 0.16  &  {\ } 16.9 & 17 & 0.46  &  {\ } 18.3 & 17 & 0.37 \\
\rowcolor{Gray}
\Powheg{}+\PythiaEight{}\ ($h_\mathrm{damp} =1.5\,m_{t}$)\ & {\ } 20.4 & 17 & 0.25  &  {\ } 15.6 & 17 & 0.56  &  {\ } 18.8 & 17 & 0.34 \\
\Powheg{}+\PythiaEight{}\ (Var3c Up) ($h_\mathrm{damp} = 3\,m_{t}$) & {\ } 17.8 & 17 & 0.40  &  {\ } 16.3 & 17 & 0.50  &  {\ } 19.3 & 17 & 0.31 \\
\rowcolor{Gray}
\Powheg{}+\PythiaEight{}\ (Var3c Down) ($h_\mathrm{damp} = 1.5\,m_{t}$) & {\ } 21.1 & 17 & 0.22  &  {\ } 17.8 & 17 & 0.40  &  {\ } 17.5 & 17 & 0.42 \\
\Powheg{}+\HerwigSeven{}\ & {\ } 16.6 & 17 & 0.48  &  {\ } 12.1 & 17 & 0.80  &  {\ } 12.8 & 17 & 0.75 \\
\rowcolor{Gray}
\Powheg{}+\herwigpp{} & {\ } 19.1 & 17 & 0.33  &  {\ } 20.7 & 17 & 0.24  &  {\ } 28.1 & 17 & 0.04 \\
\mgamcatnlo{}+\herwigpp{} & {\ } 16.3 & 17 & 0.50  &  {\ } 11.5 & 17 & 0.83  &  {\ } 23.9 & 17 & 0.12 \\
\rowcolor{Gray}
\mgamcatnlo{}+\PythiaEight{}\ ($H_{\mathrm T}/2$) & {\ } 20.3 & 17 & 0.26  &  {\ } 21.9 & 17 & 0.19  &  {\ } 22.5 & 17 & 0.17 \\
\mgamcatnlo{}+\PythiaEight{}\ ($\sqrt{m_{t}^{2} + p_{\mathrm T}^{2}}$) & {\ } 20.7 & 17 & 0.24  &  {\ } 18.1 & 17 & 0.38  &  {\ } 28.5 & 17 & 0.04 \\
\rowcolor{Gray}
\textsc{Sherpa 2.2.1} & {\ } 21.8 & 17 & 0.19  &  {\ } 20.0 & 17 & 0.28  &  {\ } 17.5 & 17 & 0.42 \\
\hline
\end{tabular}}
}
\caption{Comparison of the measured fiducial phase space normalised
  differential cross sections as a function of $\ptthad{}$ and the predictions from several MC
  generators in different $n$-jet configurations. For each prediction a $\chi^2$ and a $p$-value are
  calculated using the covariance matrix of the measured spectrum. The number of degrees of freedom (NDF) is equal to the number of bins in the distribution minus one.}
\label{tab:modelcomparison_top_pt_rel}
\end{table}

%%%%%%%%%%%%%%%%%%%%%%%%%%%%
\begin{table}[!ht]
{\small \noindent\resizebox{\textwidth}{!}{
\setlength\arrayrulewidth{0.7pt}
\begin{tabular}{|l | r <{/\kern-\tabcolsep}>{\kern-\tabcolsep} l c | r <{/\kern-\tabcolsep}>{\kern-\tabcolsep} l c | r <{/\kern-\tabcolsep}>{\kern-\tabcolsep} l c|}
\hline
& \multicolumn{3}{c|}{4-jet exclusive} & \multicolumn{3}{c|}{5-jet exclusive} & \multicolumn{3}{c|}{6-jet inclusive} \\
& \multicolumn{2}{c}{$\chi^{2}$/NDF} &  ~$p$-value  & \multicolumn{2}{c}{$\chi^{2}$/NDF} &  ~$p$-value  & \multicolumn{2}{c}{$\chi^{2}$/NDF} &  ~$p$-value  \\
\hline
\rowcolor{Gray}
\Powheg{}+\PythiaSix{} & {\ } 7.9 & 6 & 0.25  &  {\ } 6.0 & 6 & 0.43  &  {\ } 6.4 & 6 & 0.38 \\
\Powheg{}+\PythiaSix{}\ (radHi) & {\ } 15.9 & 6 & 0.01  &  {\ } 5.8 & 6 & 0.45  &  {\ } 36.2 & 6 & \hspace{-0.1 in} <0.01 \\
\rowcolor{Gray}
\Powheg{}+\PythiaSix{}\ (radLo) & {\ } 4.9 & 6 & 0.56  &  {\ } 5.8 & 6 & 0.45  &  {\ } 6.5 & 6 & 0.37 \\
\Powheg{}+\PythiaEight{}\ ($h_\mathrm{damp} = m_{t}$) & {\ } 7.3 & 6 & 0.29  &  {\ } 5.7 & 6 & 0.45  &  {\ } 8.0 & 6 & 0.24 \\
\rowcolor{Gray}
\Powheg{}+\PythiaEight{}\ ($h_\mathrm{damp} =1.5\,m_{t}$)\ & {\ } 7.6 & 6 & 0.27  &  {\ } 3.3 & 6 & 0.77  &  {\ } 12.3 & 6 & 0.06 \\
\Powheg{}+\PythiaEight{}\ (Var3c Up) ($h_\mathrm{damp} = 3\,m_{t}$) & {\ } 13.9 & 6 & 0.03  &  {\ } 3.2 & 6 & 0.78  &  {\ } 54.8 & 6 & \hspace{-0.1 in} <0.01 \\
\rowcolor{Gray}
\Powheg{}+\PythiaEight{}\ (Var3c Down) ($h_\mathrm{damp} = 1.5\,m_{t}$) & {\ } 5.5 & 6 & 0.49  &  {\ } 5.0 & 6 & 0.55  &  {\ } 6.6 & 6 & 0.36 \\
\Powheg{}+\HerwigSeven{}\ & {\ } 10.2 & 6 & 0.12  &  {\ } 5.1 & 6 & 0.53  &  {\ } 5.0 & 6 & 0.54 \\
\rowcolor{Gray}
\Powheg{}+\herwigpp{} & {\ } 8.2 & 6 & 0.23  &  {\ } 25.8 & 6 & \hspace{-0.1 in} <0.01  &  {\ } 20.8 & 6 & \hspace{-0.1 in} <0.01 \\

\mgamcatnlo{}+\herwigpp{} & {\ } 98.3 & 6 & \hspace{-0.1 in} <0.01  &  {\ } 8.6 & 6 & 0.20  &  {\ } 12.4 & 6 & 0.05 \\
\rowcolor{Gray}
\mgamcatnlo{}+\PythiaEight{}\ ($H_{\mathrm T}/2$) & {\ } 41.2 & 6 & \hspace{-0.1 in} <0.01  &  {\ } 34.5 & 6 & \hspace{-0.1 in} <0.01  &  {\ } 22.8 & 6 & \hspace{-0.1 in} <0.01 \\
\mgamcatnlo{}+\PythiaEight{}\ ($\sqrt{m_{t}^{2} + p_{\mathrm T}^{2}}$) & {\ } 46.7 & 6 & \hspace{-0.1 in} <0.01  &  {\ } 31.4 & 6 & \hspace{-0.1 in} <0.01  &  {\ } 18.6 & 6 & \hspace{-0.1 in} <0.01 \\
\rowcolor{Gray}
\textsc{Sherpa 2.2.1} & {\ } 13.3 & 6 & 0.04  &  {\ } 1.8 & 6 & 0.94  &  {\ } 21.7 & 6 & \hspace{-0.1 in} <0.01 \\
\hline
\end{tabular}}
}
\caption{Comparison of the measured fiducial phase space absolute
  differential cross sections as a function of $\ptttbar{}$ and the predictions from several MC
  generators in different $n$-jet configurations. For each prediction a $\chi^2$ and a $p$-value are
  calculated using the covariance matrix of the measured spectrum. The number of degrees of freedom (NDF) is equal to the number of bins in the distribution.}
\label{tab:modelcomparison_tt_pt_abs}
\end{table}

\begin{table}[!hb]
{\small \noindent\resizebox{\textwidth}{!}{
\setlength\arrayrulewidth{0.7pt}
\begin{tabular}{|l | r <{/\kern-\tabcolsep}>{\kern-\tabcolsep} l c | r <{/\kern-\tabcolsep}>{\kern-\tabcolsep} l c | r <{/\kern-\tabcolsep}>{\kern-\tabcolsep} l c|}
\hline
& \multicolumn{3}{c|}{4-jet exclusive} & \multicolumn{3}{c|}{5-jet exclusive} & \multicolumn{3}{c|}{6-jet inclusive} \\
& \multicolumn{2}{c}{$\chi^{2}$/NDF} &  ~$p$-value  & \multicolumn{2}{c}{$\chi^{2}$/NDF} &  ~$p$-value  & \multicolumn{2}{c}{$\chi^{2}$/NDF} &  ~$p$-value  \\
\hline
\rowcolor{Gray}
\Powheg{}+\PythiaSix{} & {\ } 4.3 & 5 & 0.51  &  {\ } 3.0 & 5 & 0.70  &  {\ } 3.9 & 5 & 0.56 \\
\Powheg{}+\PythiaSix{}\ (radHi) & {\ } 5.2 & 5 & 0.40  &  {\ } 6.3 & 5 & 0.28  &  {\ } 9.8 & 5 & 0.08 \\
\rowcolor{Gray}
\Powheg{}+\PythiaSix{}\ (radLo) & {\ } 6.2 & 5 & 0.29  &  {\ } 3.5 & 5 & 0.62  &  {\ } 5.2 & 5 & 0.39 \\
\Powheg{}+\PythiaEight{}\ ($h_\mathrm{damp} = m_{t}$) & {\ } 7.6 & 5 & 0.18  &  {\ } 4.5 & 5 & 0.48  &  {\ } 4.7 & 5 & 0.46 \\
\rowcolor{Gray}
\Powheg{}+\PythiaEight{}\ ($h_\mathrm{damp} =1.5\,m_{t}$)\ & {\ } 5.5 & 5 & 0.36  &  {\ } 3.9 & 5 & 0.57  &  {\ } 6.2 & 5 & 0.28 \\
\Powheg{}+\PythiaEight{}\ (Var3c Up) ($h_\mathrm{damp} = 3\,m_{t}$) & {\ } 6.5 & 5 & 0.26  &  {\ } 4.0 & 5 & 0.55  &  {\ } 10.5 & 5 & 0.06 \\
\rowcolor{Gray}
\Powheg{}+\PythiaEight{}\ (Var3c Down) ($h_\mathrm{damp} = 1.5\,m_{t}$) & {\ } 5.2 & 5 & 0.39  &  {\ } 5.6 & 5 & 0.35  &  {\ } 7.6 & 5 & 0.18 \\
\Powheg{}+\HerwigSeven{}\ & {\ } 10.5 & 5 & 0.06  &  {\ } 5.1 & 5 & 0.41  &  {\ } 3.1 & 5 & 0.68 \\
\rowcolor{Gray}
\Powheg{}+\herwigpp{} & {\ } 18.6 & 5 & \hspace{-0.1 in} <0.01  &  {\ } 16.2 & 5 & \hspace{-0.1 in} <0.01  &  {\ } 19.4 & 5 & \hspace{-0.1 in} <0.01 \\
\mgamcatnlo{}+\herwigpp{} & {\ } 12.8 & 5 & 0.03  &  {\ } 10.0 & 5 & 0.07  &  {\ } 9.3 & 5 & 0.10 \\
\rowcolor{Gray}
\mgamcatnlo{}+\PythiaEight{}\ ($H_{\mathrm T}/2$) & {\ } 26.8 & 5 & \hspace{-0.1 in} <0.01  &  {\ } 10.2 & 5 & 0.07  &  {\ } 8.2 & 5 & 0.14 \\
\mgamcatnlo{}+\PythiaEight{}\ ($\sqrt{m_{t}^{2} + p_{\mathrm T}^{2}}$) & {\ } 17.3 & 5 & \hspace{-0.1 in} <0.01  &  {\ } 10.0 & 5 & 0.07  &  {\ } 7.8 & 5 & 0.17 \\
\rowcolor{Gray}
\textsc{Sherpa 2.2.1} & {\ } 7.5 & 5 & 0.19  &  {\ } 1.7 & 5 & 0.89  &  {\ } 2.2 & 5 & 0.82 \\
\hline
\end{tabular}}
}
\caption{Comparison of the measured fiducial phase space normalised
  differential cross sections as a function of $\ptttbar{}$ and the predictions from several MC
  generators in different $n$-jet configurations. For each prediction a $\chi^2$ and a $p$-value are
  calculated using the covariance matrix of the measured spectrum. The number of degrees of freedom (NDF) is equal to the number of bins in the distribution minus one.}
\label{tab:modelcomparison_tt_pt_rel}
\end{table}

%%%%%%%%%%%%%%%%%%%%%%%%%%%%
\begin{table}[!ht]
{\small \noindent\resizebox{\textwidth}{!}{
\setlength\arrayrulewidth{0.7pt}
\begin{tabular}{|l | r <{/\kern-\tabcolsep}>{\kern-\tabcolsep} l c | r <{/\kern-\tabcolsep}>{\kern-\tabcolsep} l c | r <{/\kern-\tabcolsep}>{\kern-\tabcolsep} l c|}
\hline
& \multicolumn{3}{c|}{4-jet exclusive} & \multicolumn{3}{c|}{5-jet exclusive} & \multicolumn{3}{c|}{6-jet inclusive} \\
& \multicolumn{2}{c}{$\chi^{2}$/NDF} &  ~$p$-value  & \multicolumn{2}{c}{$\chi^{2}$/NDF} &  ~$p$-value  & \multicolumn{2}{c}{$\chi^{2}$/NDF} &  ~$p$-value  \\
\hline
\rowcolor{Gray}
\Powheg{}+\PythiaSix{} & {\ } 4.1 & 6 & 0.67  &  {\ } 10.0 & 6 & 0.12  &  {\ } 10.2 & 6 & 0.12 \\
\Powheg{}+\PythiaSix{}\ (radHi) & {\ } 7.1 & 6 & 0.32  &  {\ } 7.4 & 6 & 0.28  &  {\ } 14.4 & 6 & 0.03 \\
\rowcolor{Gray}
\Powheg{}+\PythiaSix{}\ (radLo) & {\ } 2.5 & 6 & 0.87  &  {\ } 10.2 & 6 & 0.12  &  {\ } 14.8 & 6 & 0.02 \\
\Powheg{}+\PythiaEight{}\ ($h_\mathrm{damp} = m_{t}$) & {\ } 3.0 & 6 & 0.81  &  {\ } 9.7 & 6 & 0.14  &  {\ } 10.1 & 6 & 0.12 \\
\rowcolor{Gray}
\Powheg{}+\PythiaEight{}\ ($h_\mathrm{damp} =1.5\,m_{t}$)\ & {\ } 3.1 & 6 & 0.80  &  {\ } 7.3 & 6 & 0.29  &  {\ } 10.7 & 6 & 0.10 \\
\Powheg{}+\PythiaEight{}\ (Var3c Up) ($h_\mathrm{damp} = 3\,m_{t}$) & {\ } 5.4 & 6 & 0.49  &  {\ } 7.4 & 6 & 0.29  &  {\ } 24.8 & 6 & \hspace{-0.1 in} <0.01 \\
\rowcolor{Gray}
\Powheg{}+\PythiaEight{}\ (Var3c Down) ($h_\mathrm{damp} = 1.5\,m_{t}$) & {\ } 2.4 & 6 & 0.88  &  {\ } 8.2 & 6 & 0.22  &  {\ } 9.2 & 6 & 0.16 \\
\Powheg{}+\HerwigSeven{}\ & {\ } 4.6 & 6 & 0.59  &  {\ } 6.4 & 6 & 0.38  &  {\ } 12.3 & 6 & 0.06 \\
\rowcolor{Gray}
\Powheg{}+\herwigpp{} & {\ } 8.0 & 6 & 0.24  &  {\ } 28.7 & 6 & \hspace{-0.1 in} <0.01  &  {\ } 37.6 & 6 & \hspace{-0.1 in} <0.01 \\
\mgamcatnlo{}+\herwigpp{} & {\ } 59.9 & 6 & \hspace{-0.1 in} <0.01  &  {\ } 10.0 & 6 & 0.12  &  {\ } 22.1 & 6 & \hspace{-0.1 in} <0.01 \\
\rowcolor{Gray}
\mgamcatnlo{}+\PythiaEight{}\ ($H_{\mathrm T}/2$) & {\ } 41.0 & 6 & \hspace{-0.1 in} <0.01  &  {\ } 38.1 & 6 & \hspace{-0.1 in} <0.01  &  {\ } 10.3 & 6 & 0.11 \\
\mgamcatnlo{}+\PythiaEight{}\ ($\sqrt{m_{t}^{2} + p_{\mathrm T}^{2}}$) & {\ } 41.0 & 6 & \hspace{-0.1 in} <0.01  &  {\ } 40.9 & 6 & \hspace{-0.1 in} <0.01  &  {\ } 10.5 & 6 & 0.10 \\
\rowcolor{Gray}
\textsc{Sherpa 2.2.1} & {\ } 3.5 & 6 & 0.74  &  {\ } 5.7 & 6 & 0.46  &  {\ } 12.8 & 6 & 0.05 \\
\hline
\end{tabular}}
}
\caption{Comparison of the measured fiducial phase space absolute
  differential cross sections as a function of $\absPoutttbar$ and the predictions from several MC
  generators in different $n$-jet configurations. For each prediction a $\chi^2$ and a $p$-value are
  calculated using the covariance matrix of the measured spectrum. The number of degrees of freedom (NDF) is equal to the number of bins in the distribution.}
\label{tab:modelcomparison_pout_abs}
\end{table}

\begin{table}[!hb]
{\small \noindent\resizebox{\textwidth}{!}{
\setlength\arrayrulewidth{0.7pt}
\begin{tabular}{|l | r <{/\kern-\tabcolsep}>{\kern-\tabcolsep} l c | r <{/\kern-\tabcolsep}>{\kern-\tabcolsep} l c | r <{/\kern-\tabcolsep}>{\kern-\tabcolsep} l c|}
\hline
& \multicolumn{3}{c|}{4-jet exclusive} & \multicolumn{3}{c|}{5-jet exclusive} & \multicolumn{3}{c|}{6-jet inclusive} \\
& \multicolumn{2}{c}{$\chi^{2}$/NDF} &  ~$p$-value  & \multicolumn{2}{c}{$\chi^{2}$/NDF} &  ~$p$-value  & \multicolumn{2}{c}{$\chi^{2}$/NDF} &  ~$p$-value  \\
\hline
\rowcolor{Gray}
\Powheg{}+\PythiaSix{} & {\ } 2.1 & 5 & 0.84  &  {\ } 5.1 & 5 & 0.41  &  {\ } 8.0 & 5 & 0.15 \\
\Powheg{}+\PythiaSix{}\ (radHi) & {\ } 5.2 & 5 & 0.40  &  {\ } 5.7 & 5 & 0.34  &  {\ } 11.6 & 5 & 0.04 \\
\rowcolor{Gray}
\Powheg{}+\PythiaSix{}\ (radLo) & {\ } 1.2 & 5 & 0.95  &  {\ } 5.1 & 5 & 0.40  &  {\ } 8.6 & 5 & 0.13 \\
\Powheg{}+\PythiaEight{}\ ($h_\mathrm{damp} = m_{t}$) & {\ } 1.4 & 5 & 0.93  &  {\ } 6.7 & 5 & 0.25  &  {\ } 9.0 & 5 & 0.11 \\
\rowcolor{Gray}
\Powheg{}+\PythiaEight{}\ ($h_\mathrm{damp} =1.5\,m_{t}$)\ & {\ } 1.0 & 5 & 0.96  &  {\ } 6.2 & 5 & 0.29  &  {\ } 11.9 & 5 & 0.04 \\
\Powheg{}+\PythiaEight{}\ (Var3c Up) ($h_\mathrm{damp} = 3\,m_{t}$) & {\ } 2.9 & 5 & 0.71  &  {\ } 7.5 & 5 & 0.18  &  {\ } 14.3 & 5 & 0.01 \\
\rowcolor{Gray}
\Powheg{}+\PythiaEight{}\ (Var3c Down) ($h_\mathrm{damp} = 1.5\,m_{t}$) & {\ } 0.5 & 5 & 0.99  &  {\ } 6.6 & 5 & 0.26  &  {\ } 10.5 & 5 & 0.06 \\
\Powheg{}+\HerwigSeven{}\ & {\ } 2.3 & 5 & 0.80  &  {\ } 4.6 & 5 & 0.46  &  {\ } 5.3 & 5 & 0.38 \\
\rowcolor{Gray}
\Powheg{}+\herwigpp{} & {\ } 7.3 & 5 & 0.20  &  {\ } 15.0 & 5 & 0.01  &  {\ } 10.4 & 5 & 0.07 \\
\mgamcatnlo{}+\herwigpp{} & {\ } 36.3 & 5 & \hspace{-0.1 in} <0.01  &  {\ } 10.2 & 5 & 0.07  &  {\ } 6.7 & 5 & 0.24 \\
\rowcolor{Gray}
\mgamcatnlo{}+\PythiaEight{}\ ($H_{\mathrm T}/2$) & {\ } 47.9 & 5 & \hspace{-0.1 in} <0.01  &  {\ } 28.9 & 5 & \hspace{-0.1 in} <0.01  &  {\ } 16.2 & 5 & \hspace{-0.1 in} <0.01 \\
\mgamcatnlo{}+\PythiaEight{}\ ($\sqrt{m_{t}^{2} + p_{\mathrm T}^{2}}$) & {\ } 46.5 & 5 & \hspace{-0.1 in} <0.01  &  {\ } 30.7 & 5 & \hspace{-0.1 in} <0.01  &  {\ } 15.7 & 5 & \hspace{-0.1 in} <0.01 \\
\rowcolor{Gray}
\textsc{Sherpa 2.2.1} & {\ } 1.5 & 5 & 0.92  &  {\ } 4.6 & 5 & 0.46  &  {\ } 8.2 & 5 & 0.15 \\
\hline
\end{tabular}}
}
\caption{Comparison of the measured fiducial phase space normalised
  differential cross sections as a function of $\absPoutttbar$ and the predictions from several MC
  generators in different $n$-jet configurations. For each prediction a $\chi^2$ and a $p$-value are
  calculated using the covariance matrix of the measured spectrum. The number of degrees of freedom (NDF) is equal to the number of bins in the distribution minus one.}
\label{tab:modelcomparison_pout_rel}
\end{table}

%%%%%%%%%%%%%%%%%%%%%%%%%%%%

\begin{figure}[htb]
\begin{center}
\includegraphics[width=0.9\textwidth]{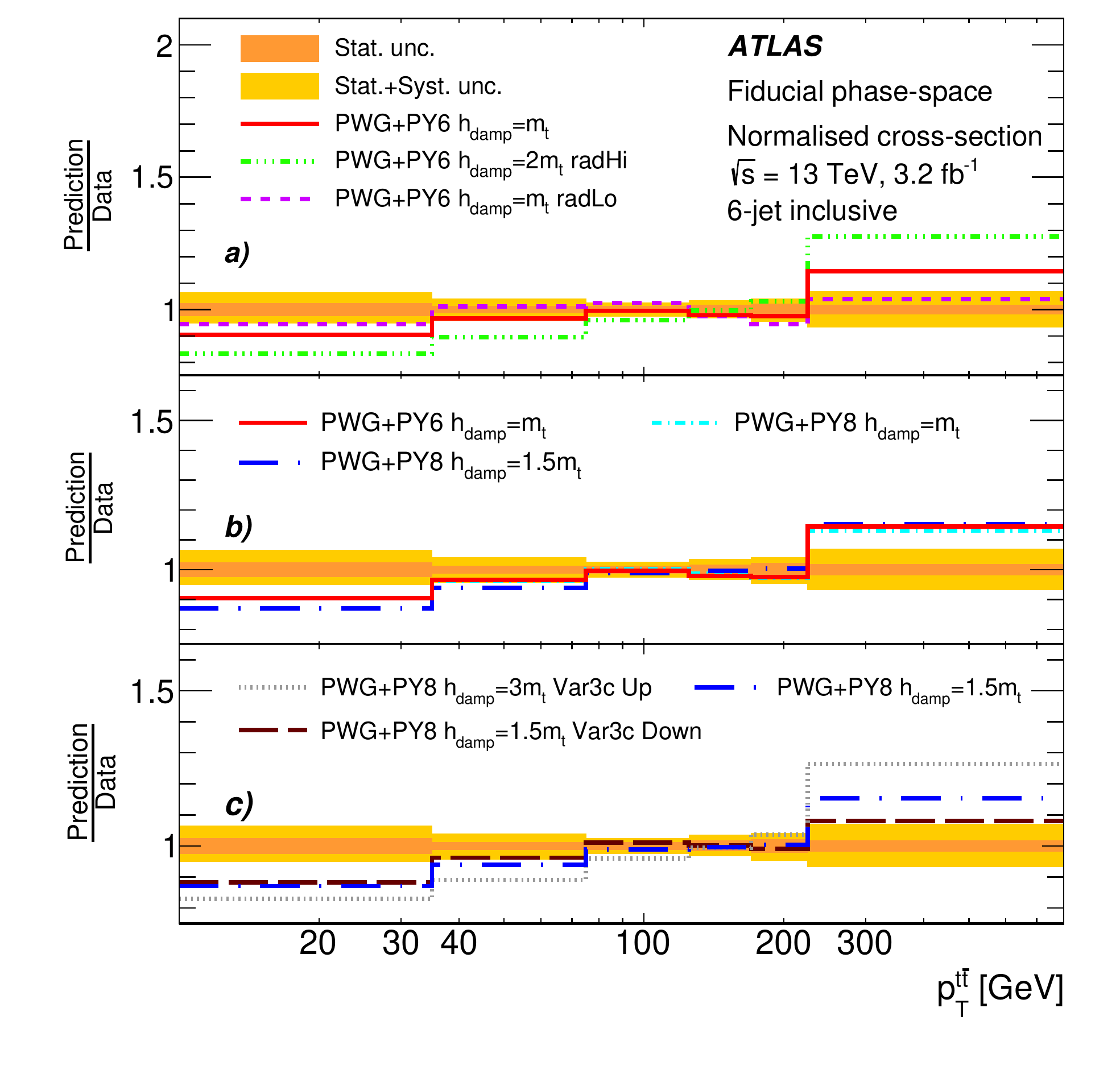}
\caption{\small{Normalised differential cross sections as a function of $\ptttbar{}$ in the 6-jet inclusive configuration in the fiducial phase space. The dark shaded area is the statistical uncertainty and the light shaded area represents the total uncertainty.}}
\label{fig:unfolding:special:hdamp}
\end{center}
\end{figure}

\begin{figure}[htb]
\begin{center}
\subfigure[]{\includegraphics[width=0.8\textwidth]{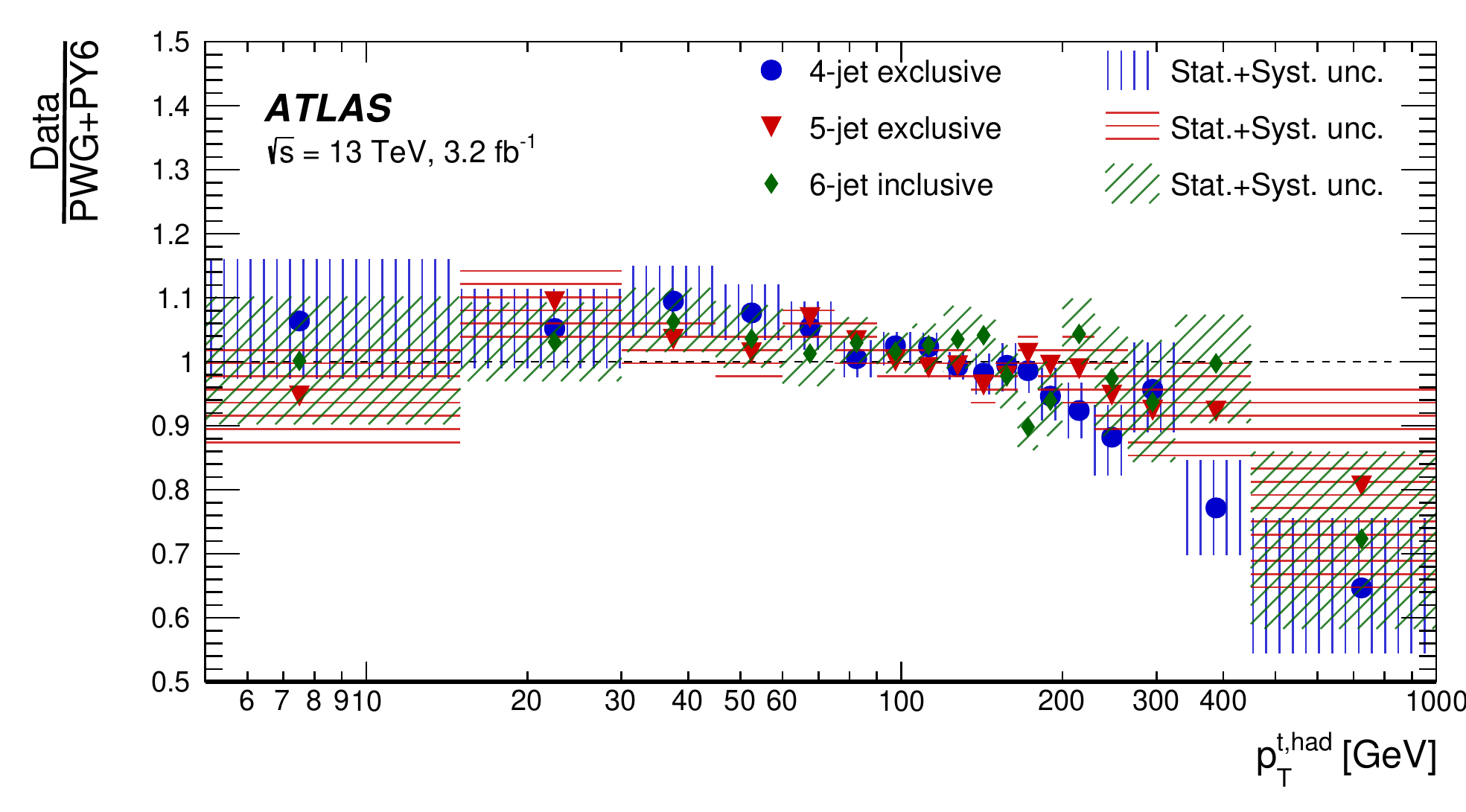}\label{fig:ratio_top}}
\subfigure[]{\includegraphics[width=0.8\textwidth]{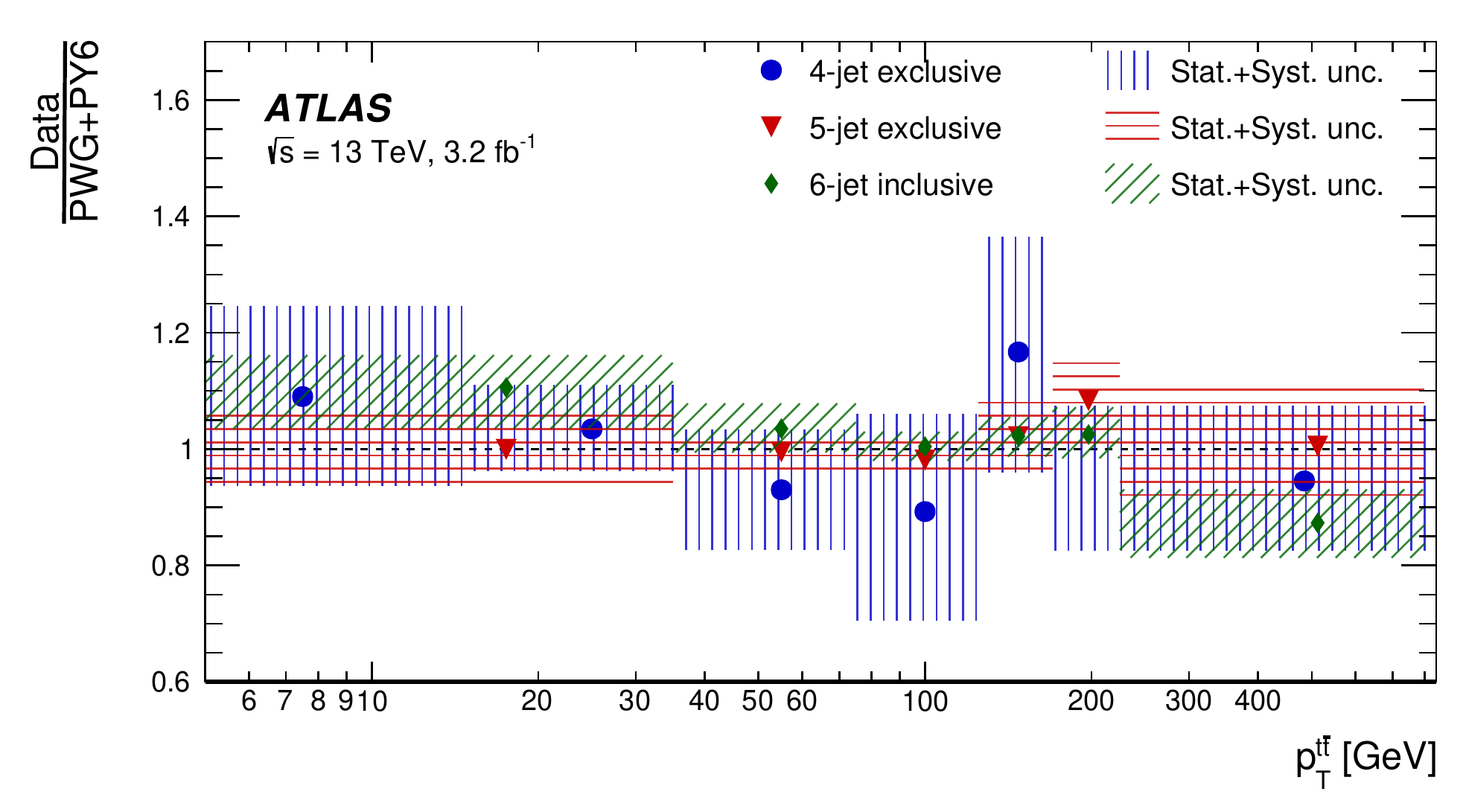}\label{fig:ratio_tt}}

\caption{\small{Normalised~\subref{fig:ratio_top} ratio of data to the nominal prediction as a function of $\ptthad{}$ and~\subref{fig:ratio_tt} as a function of $\ptttbar$ in the 4-jet exclusive, 5-jet exclusive and 6-jet inclusive configurations. 
}}
\label{fig:fit}
\end{center}
\end{figure}

\FloatBarrier
\section{Conclusions}
\label{sec:conclusions}

Measurements of differential cross sections for top quark pair production in association with jets are presented using data from the 13 \TeV{} $pp$ collisions collected by the ATLAS detector at the LHC in 2015, corresponding to an integrated luminosity of 3.2 fb$^{-1}$. Both the absolute and normalised differential cross sections are measured as functions of the top quark transverse momentum, the transverse momentum of the top quark pair system and the out-of-plane transverse momentum. The top quark pair events are selected in the lepton (electron or muon) + jets channel and three mutually exclusive configurations are defined according to the number of additional jets reconstructed in each event. Regions of phase space sensitive to the effects of gluon radiation are identified.
The predictions of several Monte Carlo calculations are compared to the measurements. Differences between the data and some of the predictions are observed. The measured $\absPoutttbar$ and $\ptttbar$ distributions in the 6-jet configuration disfavour several predictions. The measured $\ptthad$ distribution in the 4-jet configuration is underestimated by the predictions at low values and overestimated at high values; this tendency of the predictions is reduced at higher jet multiplicity. Overall, the measurements presented here improve the discriminating power of previous ATLAS results and the data have the potential to further constrain the MC models used to describe the top quark pair production.

% Acknowledgements for papers with collision data
% Version 14-Feb-2018

% Standard acknowledgements start here
%----------------------------------------------
\section*{Acknowledgements}

We thank CERN for the very successful operation of the LHC, as well as the
support staff from our institutions without whom ATLAS could not be
operated efficiently.

We acknowledge the support of ANPCyT, Argentina; YerPhI, Armenia; ARC, Australia; BMWFW and FWF, Austria; ANAS, Azerbaijan; SSTC, Belarus; CNPq and FAPESP, Brazil; NSERC, NRC and CFI, Canada; CERN; CONICYT, Chile; CAS, MOST and NSFC, China; COLCIENCIAS, Colombia; MSMT CR, MPO CR and VSC CR, Czech Republic; DNRF and DNSRC, Denmark; IN2P3-CNRS, CEA-DRF/IRFU, France; SRNSFG, Georgia; BMBF, HGF, and MPG, Germany; GSRT, Greece; RGC, Hong Kong SAR, China; ISF, I-CORE and Benoziyo Center, Israel; INFN, Italy; MEXT and JSPS, Japan; CNRST, Morocco; NWO, Netherlands; RCN, Norway; MNiSW and NCN, Poland; FCT, Portugal; MNE/IFA, Romania; MES of Russia and NRC KI, Russian Federation; JINR; MESTD, Serbia; MSSR, Slovakia; ARRS and MIZ\v{S}, Slovenia; DST/NRF, South Africa; MINECO, Spain; SRC and Wallenberg Foundation, Sweden; SERI, SNSF and Cantons of Bern and Geneva, Switzerland; MOST, Taiwan; TAEK, Turkey; STFC, United Kingdom; DOE and NSF, United States of America. In addition, individual groups and members have received support from BCKDF, the Canada Council, CANARIE, CRC, Compute Canada, FQRNT, and the Ontario Innovation Trust, Canada; EPLANET, ERC, ERDF, FP7, Horizon 2020 and Marie Sk{\l}odowska-Curie Actions, European Union; Investissements d'Avenir Labex and Idex, ANR, R{\'e}gion Auvergne and Fondation Partager le Savoir, France; DFG and AvH Foundation, Germany; Herakleitos, Thales and Aristeia programmes co-financed by EU-ESF and the Greek NSRF; BSF, GIF and Minerva, Israel; BRF, Norway; CERCA Programme Generalitat de Catalunya, Generalitat Valenciana, Spain; the Royal Society and Leverhulme Trust, United Kingdom.

The crucial computing support from all WLCG partners is acknowledged gratefully, in particular from CERN, the ATLAS Tier-1 facilities at TRIUMF (Canada), NDGF (Denmark, Norway, Sweden), CC-IN2P3 (France), KIT/GridKA (Germany), INFN-CNAF (Italy), NL-T1 (Netherlands), PIC (Spain), ASGC (Taiwan), RAL (UK) and BNL (USA), the Tier-2 facilities worldwide and large non-WLCG resource providers. Major contributors of computing resources are listed in Ref.~\cite{ATL-GEN-PUB-2016-002}.
%----------------------------------------------

\clearpage
\label{app:References}
\printbibliography

\clearpage

\clearpage

\clearpage 
% ATLAS Collaboration author list
% Reference date of TOPQ-2017-01 is 2017-08-10
% Author list last updated on date 23-OCT-18
% Data extracted on 23-Oct-2018 for paper reference TOPQ-2017-01
% at 2:33pm
 
\begin{flushleft}
{\Large The ATLAS Collaboration}

\bigskip

M.~Aaboud$^\textrm{\scriptsize 34d}$,    
G.~Aad$^\textrm{\scriptsize 99}$,    
B.~Abbott$^\textrm{\scriptsize 125}$,    
O.~Abdinov$^\textrm{\scriptsize 13,*}$,    
B.~Abeloos$^\textrm{\scriptsize 129}$,    
S.H.~Abidi$^\textrm{\scriptsize 165}$,    
O.S.~AbouZeid$^\textrm{\scriptsize 143}$,    
N.L.~Abraham$^\textrm{\scriptsize 153}$,    
H.~Abramowicz$^\textrm{\scriptsize 159}$,    
H.~Abreu$^\textrm{\scriptsize 158}$
% ...
\bigskip
\\

% ...
$^{13}$Institute of Physics, Azerbaijan Academy of Sciences, Baku; Azerbaijan.\\
% ...
$^{34}$$^{(a)}$Facult\'e des Sciences Ain Chock, R\'eseau Universitaire de Physique des Hautes Energies - Universit\'e Hassan II, Casablanca;$^{(b)}$Centre National de l'Energie des Sciences Techniques Nucleaires (CNESTEN), Rabat;$^{(c)}$Facult\'e des Sciences Semlalia, Universit\'e Cadi Ayyad, LPHEA-Marrakech;$^{(d)}$Facult\'e des Sciences, Universit\'e Mohamed Premier and LPTPM, Oujda;$^{(e)}$Facult\'e des sciences, Universit\'e Mohammed V, Rabat; Morocco.\\
% ...
$^{99}$CPPM, Aix-Marseille Universit\'e, CNRS/IN2P3, Marseille; France.\\
% ...
$^{125}$Homer L. Dodge Department of Physics and Astronomy, University of Oklahoma, Norman OK; United States of America.\\
% ...
$^{129}$LAL, Universit\'e Paris-Sud, CNRS/IN2P3, Universit\'e Paris-Saclay, Orsay; France.\\
% ...
$^{143}$Santa Cruz Institute for Particle Physics, University of California Santa Cruz, Santa Cruz CA; United States of America.\\
% ...
$^{153}$Department of Physics and Astronomy, University of Sussex, Brighton; United Kingdom.\\
% ...
$^{158}$Department of Physics, Technion, Israel Institute of Technology, Haifa; Israel.\\
$^{159}$Raymond and Beverly Sackler School of Physics and Astronomy, Tel Aviv University, Tel Aviv; Israel.\\
% ...
$^{165}$Department of Physics, University of Toronto, Toronto ON; Canada.\\
% ...
$^{a}$ Also at  Department of Physics, University of Malaya, Kuala Lumpur; Malaysia.\\
$^{b}$ Also at Borough of Manhattan Community College, City University of New York, NY; United States of America.\\
$^{c}$ Also at Centre for High Performance Computing, CSIR Campus, Rosebank, Cape Town; South Africa.\\
$^{d}$ Also at CERN, Geneva; Switzerland.\\
$^{e}$ Also at CPPM, Aix-Marseille Universit\'e, CNRS/IN2P3, Marseille; France.\\
% ...
$^{*}$ Deceased

\end{flushleft}

% Created with Glance <Atlas.Glance@cern.ch>

\clearpage

\end{document}